\documentclass[structabstract]{aa}  
\usepackage{graphicx}
\usepackage{color}
\usepackage{txfonts}
\usepackage{longtable,lscape}
\usepackage{natbib}
\begin{document}
%
%===================================================================================================

   \title{HCOOCH$\rm_{3}$ as a probe of temperature and structure of Orion-KL\thanks{Based on observations carried out with the IRAM Plateau de Bure Interferometer. IRAM is supported by INSU/CNRS (France), MPG (Germany) and IGN (Spain).}\thanks{A fits image of the HCOOCH$\rm_{3}$ integrated intensity map (Fig. 4) is only available in electronic form
at the CDS via anonymous ftp to cdsarc.u-strasbg.fr (130.79.128.5) or via http://cdsweb.u-strasbg.fr/cgi-bin/qcat?J/A+A/ . All spectra can be obtained upon request to the authors.}
}

 %  \subtitle{}

   \author{C. Favre
          \inst{1,2}
          \and
          D. Despois\inst{1,2}
          \and 
          N. Brouillet\inst{1,2}
          \and 
          A. Baudry\inst{1,2}
          \and
          F. Combes\inst{3}
          \and
          M. Gu\'elin\inst{4}          
          \and
          A. Wootten\inst{5}
          \and
          G. Wlodarczak\inst{6}
          }

   \institute{ Universit\'e de Bordeaux, OASU,  2 rue de l'Observatoire, BP 89, 33271 Floirac Cedex, France\\
              \email{favre@phys.au.dk, despois, brouillet, baudry@obs.u-bordeaux1.fr}
         \and
            CNRS, UMR 5804, Laboratoire dÕAstrophysique de Bordeaux, 2 rue de lÕObservatoire, BP 89, 33271 Floirac Cedex, France
         \and
             Observatoire de Paris, LERMA, 61 Av. de l'Observatoire, 75014 Paris, France\\
             \email{francoise.combes@obspm.fr}       
          \and
             Institut de Radioastronomie Millim\'etrique, Domaine Universitaire, 300 rue de la piscine, St Martin d'H\`eres, 38400, France\\
             \email{guelin@iram.fr}        
          \and
             National Radio Astronomy Observatory, 520 Edgemont Road, Charlottesville, VA 22903-2475, USA\\
             \email{awootten@nrao.edu}  
           \and
             Laboratoire de Physique des Lasers, Atomes et Mol\'ecules, Universit\'e de Lille1, UMR 8523, 59655 Villeneuve d'Ascq Cedex, France\\
             \email{georges.wlodarczak@univ-lille1.fr}            
             }

   \date{Received July 07, 2010; accepted March 04, 2011}

%===================================================================================================

%-----------------------------------------------------------------------------------------------------------------------------
%--------------------------------------------ABSTRACT -------------------------------------------------
%-----------------------------------------------------------------------------------------------------------------------------

% \abstract{}{}{}{}{} 
% 5 {} token are mandatory
 
  \abstract
  % context heading (optional)
  % {} leave it empty if necessary  
   {The Orion Kleinmann-Low nebula (Orion-KL) is a complex region of star formation. Whereas its proximity allows studies on a scale of a few hundred AU, spectral confusion makes it difficult to identify molecules with low abundances.}
  % aims heading (mandatory)
   {We studied an important oxygenated molecule, HCOOCH$\rm_{3}$, to characterize the physical conditions, temperature and density of the different molecular source components. Methyl formate presents strong close rotational transitions covering a large range of energy and its emission in Orion-KL is not contaminated by the emission of N-bearing molecules. This study will help in the future 1) to constrain chemical models for the formation of methyl formate in gas phase or on grain mantles, 2)  to search for more complex or prebiotic molecules.}
  % methods heading (mandatory)
   {To reduce the spectral confusion we used high resolution observations from the IRAM Plateau de Bure Interferometer in order to better isolate the molecular emission regions. We used twelve data sets with a spatial resolution down to 1.8$\arcsec$ $\times$ 0.8$\arcsec$. Continuum emission was subtracted by selecting apparently line free channels.}
  % results heading (mandatory)
   {We identify 28 methyl formate emission peaks throughout the 50$\arcsec$ field of observations. The two strongest peaks, named MF1 and MF2, are in the Compact Ridge and in the South West of the Hot Core respectively. From a comparison with single dish observations, we estimate that we miss less than 15\% of the flux and that spectral confusion is still prevailing as half of the expected transitions are blended over the region. Assuming that the transitions are thermalized, we derive the temperature at the five main emission peaks. At the MF1 position in the Compact Ridge we find a temperature of 80~K in a 1.8$\arcsec$ $\times$ 0.8$\arcsec$ beam size and 120~K on a larger scale (3.6$\arcsec$ $\times$ 2.2$\arcsec$), suggesting an external source of heating, whereas the temperature is about 130~K at the MF2 position on both scales. Transitions of methyl formate in its first torsionally excited state are detected as well and the good agreement of the positions on the rotational diagrams between the ground state and the v$\rm_{t}$=1 transitions suggests a similar temperature. The LSR velocity of the gas is between 7.5 and 8.0~km~s$^{-1}$ depending on the positions and column density peaks vary from 1.6$\times$10$^{16}$ to 1.6$\times$10$^{17}$~cm$^{-2}$. 
A second velocity component is observed around 9-10~km~s$^{-1}$ in a North-South structure stretching from the Compact Ridge up to the BN object; ; this component is warmer at the MF1 peak.
The two other C$\rm_{2}$H$\rm_{4}$O$\rm_{2}$ isomers are not detected and the derived upper limit for the column density is $\leq$3$\times$10$^{14}$~cm$^{-2}$ for glycolaldehyde and $\leq$2$\times$10$^{15}$~cm$^{-2}$ for acetic acid. 
From the 223~GHz continuum map, we identify several dust clumps with associated gas masses in the range 0.8 to 5.8~M$\rm_{\sun}$.
Assuming that the methyl formate is spatially distributed as the dust, we find relative abundances of methyl formate in the range $\le$0.1$\times$10$^{-8}$ to 5.2$\times$10$^{-8}$. 
 We suggest a relation between the methyl formate distribution and shocks as traced by 2.12~$\mu m$ H$\rm_2$ emission.}
 % conclusions heading (optional), leave it empty if necessary 
   {}


   \maketitle

%===================================================================================================

%-----------------------------------------------------------------------------------------------------------------------------
%---------------------------------------------- INTRODUCTION---------------------------------------------
%-----------------------------------------------------------------------------------------------------------------------------

\section{Introduction}

The Orion Nebula  is one of the most studied regions in the sky. It contains remarkable groups of bright, visible stars and, at the same time, it is an extremely rich region of star formation, up to very high masses. In addition, it is the closest high--mass star formation region to the Sun  \citep[414$\pm$7~pc,][]{Menten:2007}. It is thus the best source to investigate with very high spatial resolution the processes leading to star formation. However, Orion cannot be considered as the prototypical star-forming region in our Galaxy because it exhibits some unique characteristics which need to be understood before any generalization is made.

The Orion Kleinmann-Low nebula is an atypical region of the Orion Molecular Cloud 1 (OMC-1), which harbours one of the most luminous embedded IR sources of this region \citep[luminosity $\sim$$10^{5}~L_{\sun}$, ][]{Wynn-Williams:1984}.
Several molecular components (referred to as the Hot Core, Compact and Extended Ridges) and several IR sources or radio sources are associated with Orion-KL. The nature of the source(s) responsible for this IR emission is still poorly known and much debated. Many young stellar objects are still embedded in the dusty gas, including radio source I, a deeply embedded, high-mass young stellar object, which drives a bipolar outflow along a northeast-southwest axis \citep{Beuther:2005, Goddi:2009, Plambeck:2009} and presents a disk perpendicular to it \citep{Matthews:2010}.

Another specificity of the Orion protocluster region is the presence of high speed shocks generated in the center of Orion-KL, reminiscent of an explosive event. From VLA velocity measurements, it was proposed that a very unique phenomenon had taken place some 500-1000 years ago: the close encounter, or collision, of two or more rather massives stars. The objects involved in such a dynamical interaction could have included the Becklin-Neugebauer object (BN) and sources I and n \citep{Gomez:2005,Rodriguez:2005,Goddi:2010}. We might thus be observing a very recent and energetic event at the heart of the nebula, providing unique conditions for the study of a rich interstellar chemistry. With this scheme in mind, many molecules could have thus been released from the grain mantles because of dust heating or multiple shocks.

The search for complex or prebiotic molecules is difficult because of their relatively low abundance and line intensity. Especially in a very rich molecular source such as Orion-KL, high spectral confusion makes it difficult to detect the weakest lines. Among them,  glycine (NH$\rm_{2}$CH$\rm_{2}$COOH, the simplest amino acid) has been searched for in the interstellar medium by many groups but has not been detected and only upper limits are given by the most sensitive studies \citep[e.g.] []{ Combes:1996, Guelin:2008, Snyder:2005}. On the other hand, observing abundant complex molecules is necessary to characterize the gas temperature and density of the various Orion-KL molecular source components. Observations of O- and N-bearing molecules show a clearly different spatial distribution. N-bearing molecules tend  to peak to the north of the Hot Core while O-bearing molecules cover the Hot Core and the Compact Ridge \citep{Guelin:2008,Beuther:2005,Friedel:2008}.

Rotational temperature maps have been derived from interferometric observations of CH$\rm_{3}$OH \citep{Beuther:2005}, NH$\rm_{3}$ \citep{Wilson:2000}, both essentially focused on the Hot Core, and CH$\rm_{3}$CN \citep{Wang:2010}; the intense lines of these abundant species  require however to correct for opacity effects.
Several single dish observations \citep[e.g.] []{Blake:1987,Comito:2005} were also used  to derive rotational temperatures towards the Hot Core and the Compact Ridge, which are supposed to be differentiated from their velocity structure. The spatial structure is lacking in these data, however. 

In this paper we investigate the  spatial structure and temperature distribution of Orion-KL from observations of the O-bearing methyl formate (HCOOCH$\rm_{3}$) molecule, which is relatively abundant in several interstellar hot cores and corinos, and especially abundant in Orion-KL \citep{Blake:1987, Kobayashi:2007, Friedel:2008}. 
In Sect. 2 we present our observations and the methyl formate frequency data base used in this work and we briefly describe the data reduction methodology. The results of our maps are presented in Sect. 3 together with details on the various molecular emission peaks identified in our maps. In Sect. 4, the temperature and molecular abundance across the entire V-shaped molecular structure linking IRc2 to BN are deduced from the rotational diagrams of the methyl formate molecule. In Sect. 5, upper limits on the abundance of methyl formate isomers are given. The dust and total mass and the mean gas density of the most prominent dust emission peaks are estimated from our continuum data in Section 6. The methyl formate fractional abundance is also estimated and briefly discussed in this same Section. We compare our results to previous studies in Section 7.  In Sect. 8, the methyl formate distribution is discussed in the light of the complex structure of the Orion nebula. Conclusions are presented in the last Section.

%-----------------------------------------------------------------------------------------------------------------------------
%---------------OBSERVATIONS AND METHYL FORMATE FREQUENCIES-------------------
%-----------------------------------------------------------------------------------------------------------------------------

\section{Observations and methyl formate frequencies}
 \label{sec:Observations}

%-------------------------------------------------------------
%----------Observations-----------------------
%-------------------------------------------------------------
\subsection{Observations}
\begin{figure}[h!]
   \centering
 	\resizebox{6.cm}{!}{\includegraphics[angle=270]{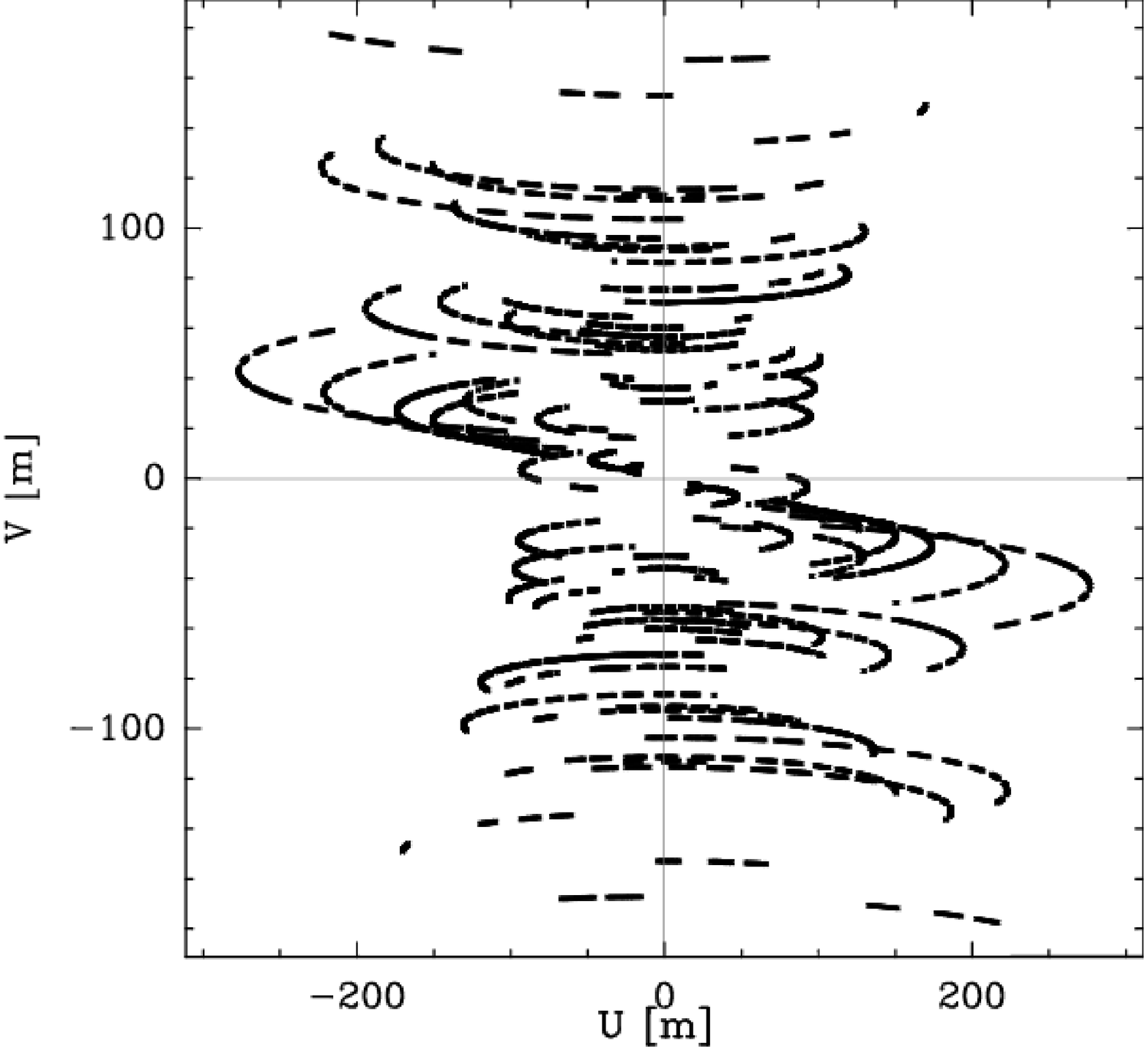}}
	\resizebox{6.cm}{!}{\includegraphics[angle=270]{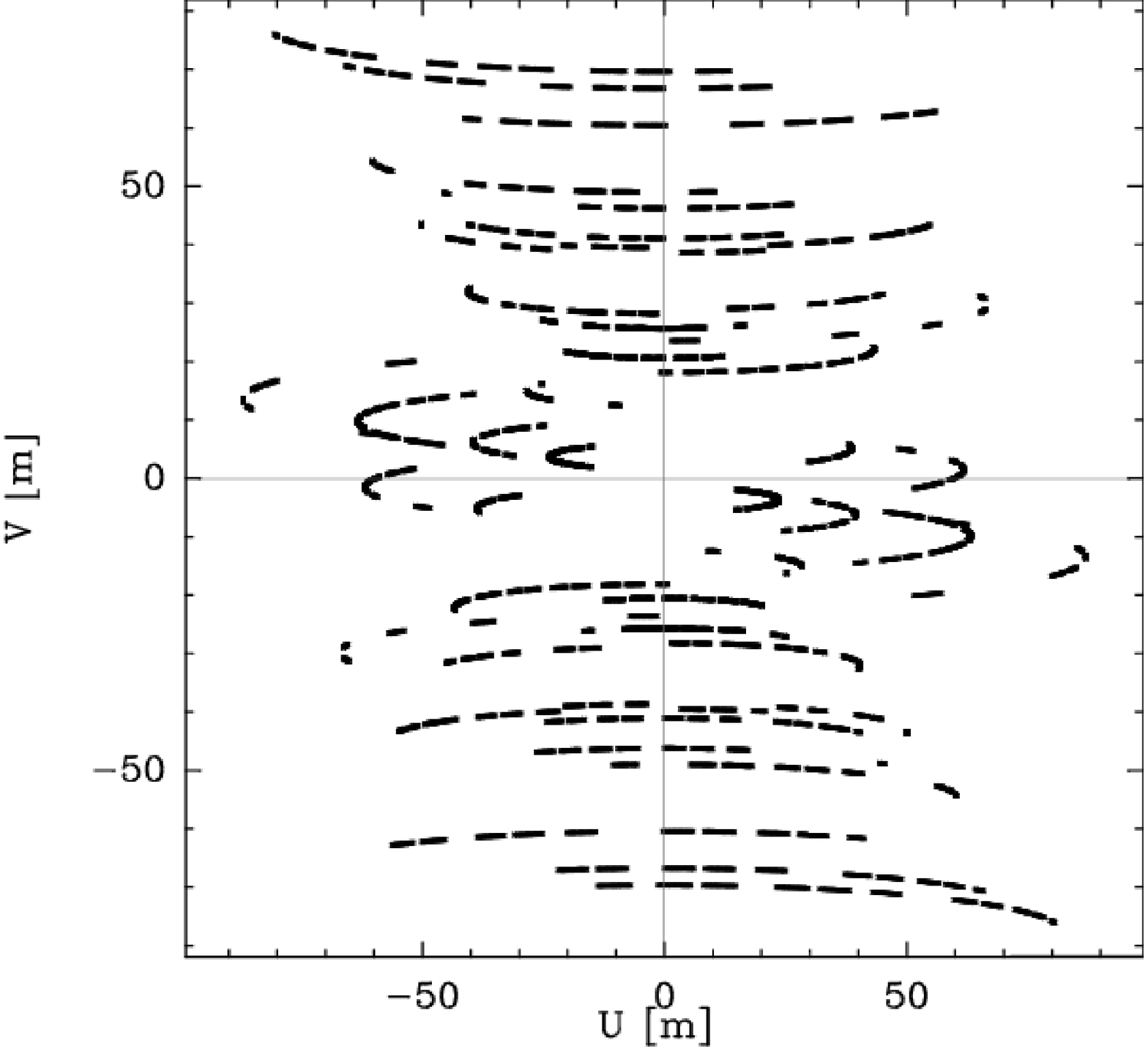}}
   \caption{Resulting \textit{uv} coverage from the six tracks at 223~GHz (top panel) and four tracks at 225~GHz (bottom panel).}
              \label{uvco}%
                 \end{figure}

We used twelve observational data sets obtained with the IRAM Plateau de  Bure Interferometer (PdBI) towards the IRc2 region and its surroundings between 1996 and 2007. The bulk of the observations were obtained between 1999 and 2007. Table \ref{Table.dataset_parameters} lists the different parameters for each data set. The highest spatial resolution (1.79$\arcsec$ $\times$ 0.79$\arcsec$) was achieved in the period 2003 to 2007. Most observations used five antennas which were equiped with two SIS receivers operated simultaneously at 3mm and 1mm until 2006, then independently. 0420-014, 0458-020, 0528+134, 0605-085 and 0607-157 were used as phase and amplitude calibrators. The short--term atmospheric phase fluctuations were corrected at 1.3~mm on line. 
Tropospheric and radiometric phase corrections have been used since 1995 and October 2001, respectively, at the IRAM PdBI, based on the measured water line wings in the continuum emission observed with the 1.3~mm receiver and on the 22~GHz water line radiometers.
The six units of the correlator allowed us to observe several data subsets with different bandwidths and spectral resolution which are also given in Table \ref{Table.dataset_parameters}. For the highest spatial resolution the spectral resolution was around 0.84~km~s$^{-1}$ while some lower spatial resolution maps were obtained with 0.42~km~s$^{-1}$ spectral resolution. The \textit{uv} coverage of these two data sets is shown in figure \ref{uvco}.

%-------------------
%TABLE 1
%-------------------
\begin{table*}
\begin{minipage}[t]{17cm}
\caption{Main parameters of IRAM Plateau de Bure interferometer observations.}             
\label{Table.dataset_parameters}      
\centering          
\small\addtolength{\tabcolsep}{-1pt} 
\renewcommand{\footnoterule}{}  % to avoid a line before footnotes
\begin{tabular}{c c c c c c c c c c c }
\hline\hline       
Set \footnote{Sets 3 and 9 were centered on coordinates  ($\alpha_{J2000}$ = 05$^{h}$35$^{m}$14$\fs$20, $\delta_{J2000}$ = -05$\degr$22$\arcmin$36$\farcs$00). Sets 1, 2, 4, 5, 6, 7, 8, 10, 11 and 12 were centered on coordinates ($\alpha_{J2000}$ = 05$^{h}$35$^{m}$14$\fs$46, $\delta_{J2000}$ = -05$\degr$22$\arcmin$30$\farcs$59). The observation velocity was 6 ~km~s$^{-1}$ for sets 1, 2, 6, 7, 8 and 8 ~km~s$^{-1}$ for the other sets.} & Bandwidth & Date  of & Confi- & HPBW & \multicolumn{2}{c}{Spectral resolution}  & Flux conversion  & \multicolumn{2}{c}{Synthesized beam}  & Pixel size \\
 & (GHz) & observation\footnote{Format: Month-Month/Year.} & guration & ($\arcsec$)  & (MHz) & (~km~s$^{-1}$) & (1 Jy~beam$^{-1}$) & ($\arcsec$ $\times$ $\arcsec$) & PA ($\degr$) & ($\arcsec$ $\times$ $\arcsec$) \\
(1) & (2) & (3) & (4) &  (5) &  (6) &  (7) &  (8) &  (9) &  (10) &  (11) \\ 
\hline                    
1 & 80.502 - 80.574 & 10-11/1999 & CD & 60 & 0.625 & 2.33 & 4.6~K & 7.63 $\times$ 5.35 & 15 & 0.80 $\times$ 0.80 \\ 
2 & 80.582 - 80.654 & 10-11/1999 &  CD & 60 & 0.625 & 2.33 & 4.7~K & 7.62 $\times$ 5.23 & 16 & 0.96 $\times$ 0.96  \\ 
3 & 101.178 - 101.717 & \footnote{Dates of observation: 12/2003, 01-12/2004, 03/2005, 01/2006.}   &  BC & 50  & 0.625 & 1.85 & 15.8~K & 3.79 $\times$ 1.99 & 22 & 0.50 $\times$ 0.50\\ 
4 & 105.655 - 105.726 & 08-11/2005 &  D & 48  & 0.3125 & 0.89 & 2.9~K & 7.13 $\times$ 5.36 &  9 & 1.40 $\times$ 1.40 \\ 
5 & 105.785 - 105.856 & 08-11/2005 &  D & 48 & 0.3125 & 0.89 & 2.9~K & 7.12 $\times$ 5.36 & 9 & 1.40 $\times$ 1.40 \\ 
6 & 110.125 - 110.181 & 10/1996 & C &  46  & 0.15625 & 0.43 & 8.5~K & 5.85 $\times$ 2.02 & 8 & 0.54 $\times$ 0.54 \\ 
7 & 203.331 - 203.403 &  10-11/1999 & CD & 25 & 0.625 & 0.92 & 7.0~K & 2.94 $\times$ 1.44 & 27 & 0.38 $\times$ 0.38  \\
8 & 203.411 - 203.483 & 10-11/1999 & CD  & 25  & 0.625 & 0.92 & 7.0~K & 2.94 $\times$ 1.44 & 27 & 0.32 $\times$ 0.32 \\
9 & 223.402 - 223.941 & \footnote{Dates of observation: 12/2003, 01-12/2004, 03/2005, 01/2006, 11-12/2007.}  & BC  & 23  & 0.625 & 0.84 & 17.3~K & 1.79 $\times$ 0.79 & 14 & 0.25 $\times$ 0.25 \\ 
10 & 225.675 - 225.747 &  09-11/2005 & D  & 22 & 0.3125 & 0.42 & 2.9~K &  3.63 $\times$ 2.26 & 12 & 0.70 $\times$ 0.70  \\
11 & 225.805 - 225.942 &  09-11/2005 & D & 22  & 0.3125 & 0.42 & 2.9~K  & 3.63 $\times$ 2.26 & 12 & 0.70 $\times$ 0.70 \\
12 & 225.990 - 226.192 &  09-11/2005 & D  & 22 & 0.3125 & 0.42 & 2.9~K & 3.63 $\times$ 2.25 & 12 & 0.70 $\times$ 0.70  \\ 
\hline                  
\end{tabular}
\end{minipage}
\end{table*}
%-------------------

%-------------------------------------------------------------
%----------Data reduction---------------------
%-------------------------------------------------------------
\subsection{Data reduction}

\begin{figure}[h!]
   \centering
   \renewcommand{\footnoterule}{}  
 \includegraphics[width=8.8cm]{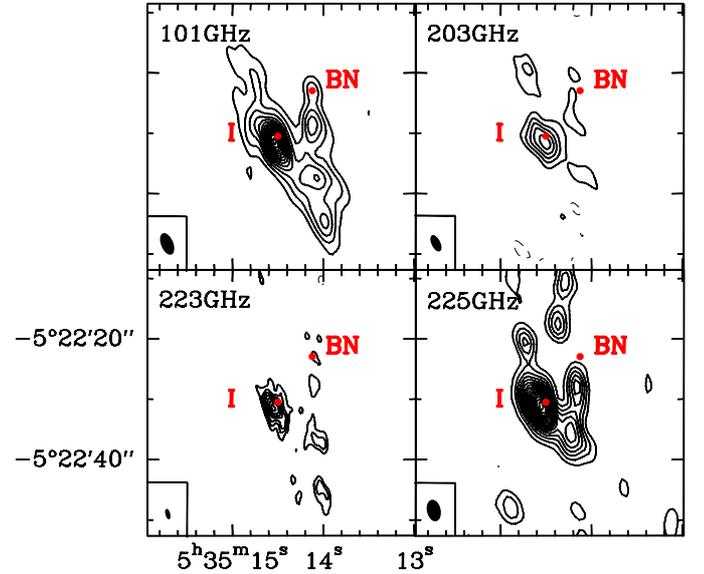}
\caption{Continuum maps obtained toward Orion-KL with the Plateau de Bure Interferometer. The clean beam is shown in the bottom left corner of each map. The first contour and the level step are 15~mJy~beam$^{-1}$, 170~mJy~beam$^{-1}$, 70~mJy~beam$^{-1}$ and 100~mJy~beam$^{-1}$ at 101.45~GHz,  203.41~GHz, 223.67~GHz and 225.90~GHz respectively. In each map, positions of the BN object ($\alpha_{J2000}$ = 05$^{h}$35$^{m}$14$\fs$1094, $\delta_{J2000}$ = -05$\degr$22$\arcmin$22$\farcs$724) and the radio source I  ($\alpha_{J2000}$ = 05$^{h}$35$^{m}$14$\fs$5141, $\delta_{J2000}$ = -05$\degr$22$\arcmin$30$\farcs$575) \citep{Goddi:2010} are indicated. Note that the maps have not been observed with the same center (see table \ref{Table.dataset_parameters}). }
              \label{Fig.continuum}%
    \end{figure}

We used the GILDAS package\footnote{http://www.iram.fr/IRAMFR/GILDAS} for data reduction. The continuum emission was subtracted in the data cubes by selecting line-free channels, discarding any contaminated (or doubtful) channels. Continuum emission in Orion-KL is essentially due to the thermal emission from the dust and to a weaker contribution from the free-free radiation of the gas.
However, the contribution from the pseudo-continuum resulting from the superposition of many weak lines cannot be easily removed as it may depend on the spectral range analysed. 
The continuum emission is strong and spatially extended, varies with frequency, and reveals a clumpy structure at the highest spatial resolutions (see Fig.\ref{Fig.continuum} and Fig.\ref{Fig.dust223GHz}). As the center of the observations is not the same for all data sets, the continuum emission is better mapped toward the South at 223~GHz and toward the North at 225~GHz. Mapping the continuum emission is not obvious because of the high density of lines. It is difficult to isolate channels devoid of molecular lines. Nevertheless, we have been able to identify several channels (typically 4 to 22 by data sets) without line emission in our observations. All of these channels have been averaged and the average subtracted from each channel in the data cubes. 

Finally, we have cleaned the data cubes, channel by channel, using the Clark method. Columns 9, 10 and 11 in Table \ref{Table.dataset_parameters} summarize the synthesized beam parameters for each dataset.

%-------------------
%FIGURE 1
%-------------------
%-------------------
%
%-------------------
%FIGURE 2
%-------------------
%-------------------

%-------------------------------------------------------------
%---Interest of methyl formate observations--
%-------------------------------------------------------------
\subsection{Interest of methyl formate observations}

Methyl formate, one of three [C$\rm_{2}$H$\rm_{4}$O$\rm_{2}$] isomers, is relatively abundant towards Orion-KL, in view of its complexity.
High angular resolution allows us 1) to isolate the different emission peaks of methyl formate and to investigate the gas structure, 2) and with large scale maps, to understand the overall Orion nebula structure (as discussed later in Section 8 of this work). 
High angular resolution also  lowers  the risk of spectral contamination by line-rich species such as  C$\rm_{2}$H$\rm_{5}$CN, because the methyl formate  distribution is different from that of  N-bearing molecules \citep{Blake:1987, Kobayashi:2007, Friedel:2008}.

\begin{figure}[h!]
  \centering
   \includegraphics[width=5cm, angle=270]{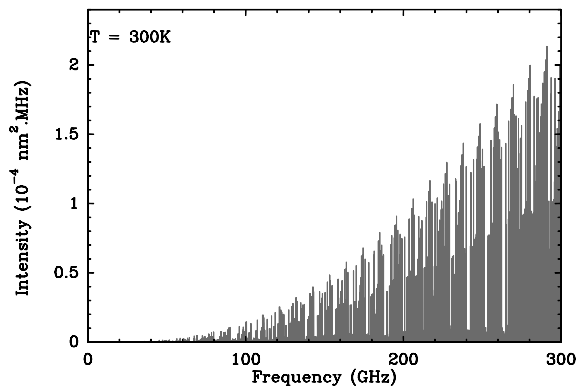}
   \caption{Methyl formate (HCOOCH$\rm_{3}$) line intensities assuming LTE and a temperature of  300~K (sources : JPL database and Ilyushin et al. 2009).}
              \label{Fig.structuMF}%
    \end{figure}

Methyl formate presents close rotational transitions (see Fig.\ref{Fig.structuMF}) with strong line strengths (S$\mu^{2}$ up to 50.2~D$^{2}$ in our selection, see Tables \ref{MF1} and \ref{MF5}). Several transitions  covering an important range of energy can be  observed at the same time. Comparison of these data obtained with identical spatial and spectral resolutions will allow an optimal estimate of the molecular gas physical conditions. The observed lines are all optically thin, which ease greatly their interpretation.

From a chemical point of view, formation of methyl formate on grain surfaces or in gas phase is still debated, as for many other complex species. The study of its spatial distribution, and its relative abundance with respect to other O-bearing molecules like methanol CH$_3$OH, dimethyl ether CH$_3$OCH$_3$ and the two methyl formate isomers, glycolaldehyde CH$_2$OHCHO and acetic acid CH$_3$COOH, could help evaluate the different chemical models \citep{Charnley:2005,Bisschop:2007,Garrod:2006,Garrod:2008}.

%-------------------
%FIGURE 3
%-------------------
%-------------------    
    
%-------------------------------------------------------------
%--Methyl formate and isomers frequencies--
%-------------------------------------------------------------
\subsection{Methyl formate and isomers frequencies}

Recently, \citet{Ilyushin:2009} recalculated the rotational transitions of HCOOCH$\rm_{3}$ for the ground and first excited torsional states of this molecule. The JPL database\footnote{http://spec.jpl.nasa.gov/home.html} \citep{Pickett:1991,Pickett:1998} is now updated (since April 2009) to include both the newly recalculated methyl formate frequencies and their own predictions (Drouin, B.J.). The methyl group, because of its three-fold internal rotation, leads to a series of thermally populated torsional levels which are split into two torsional substates with A and E symmetries. Previous data separated these two states while current data treat both substates simultaneously. In our study, we use the measured and predicted transitions coming from both Ilyushin's table and the JPL database up to E$\rm_{upper}$ $\la$ 650~K. Assuming HCOOCH$\rm_{3}$ lines are thermalized, we take into account transitions in their fundamental and first excited torsional states v$\rm_{t}$ = 0 and v$\rm_{t}$ = 1, respectively. Our observations confirm the detection of transitions of methyl formate in the first torsionally excited state \citep{Kobayashi:2007}.

We also searched for the methyl formate isomers (see Sect. \ref{sec:isomers}) and used \citet{Ilyushin:2008} for the acetic acid and the CDMS database \footnote{http://www.astro.uni-koeln.de/cdms} \citep{Muller:2001,Muller:2005} for the glycolaldehyde transitions.

%-----------------------------------------------------------------------------------------------------------------------------
%-----MAPPING METHYL FORMATE EMISSION WITH THE IRAM INTERFEROMETER----
%-----------------------------------------------------------------------------------------------------------------------------

\section{Mapping methyl formate emission (HCOOCH$\rm_{3}$) with the IRAM interferometer}
\label{sec:MF} 

\begin{figure}[h!]
  \centering
   \includegraphics[width=8.5cm]{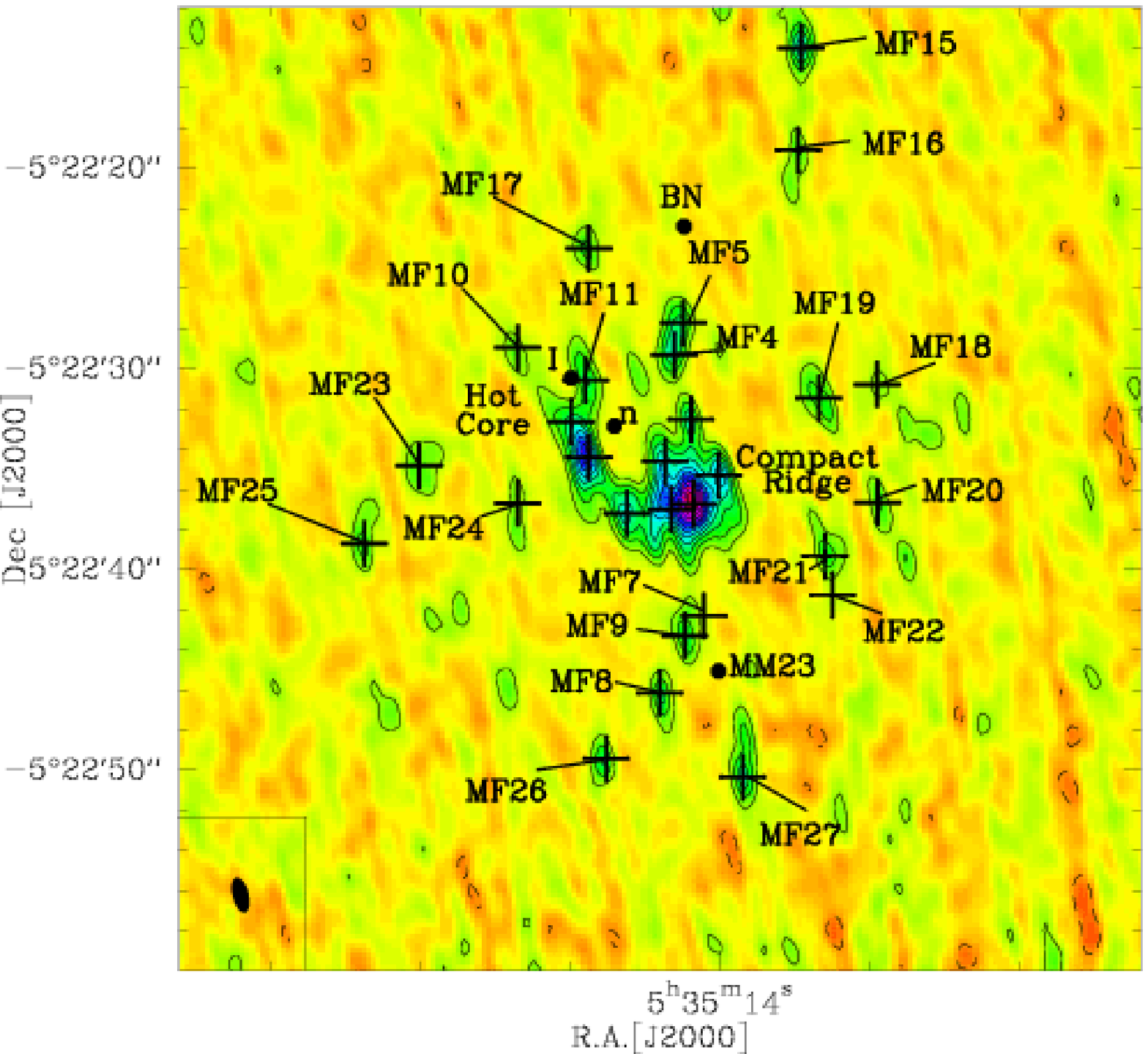}
      \includegraphics[width=8cm]{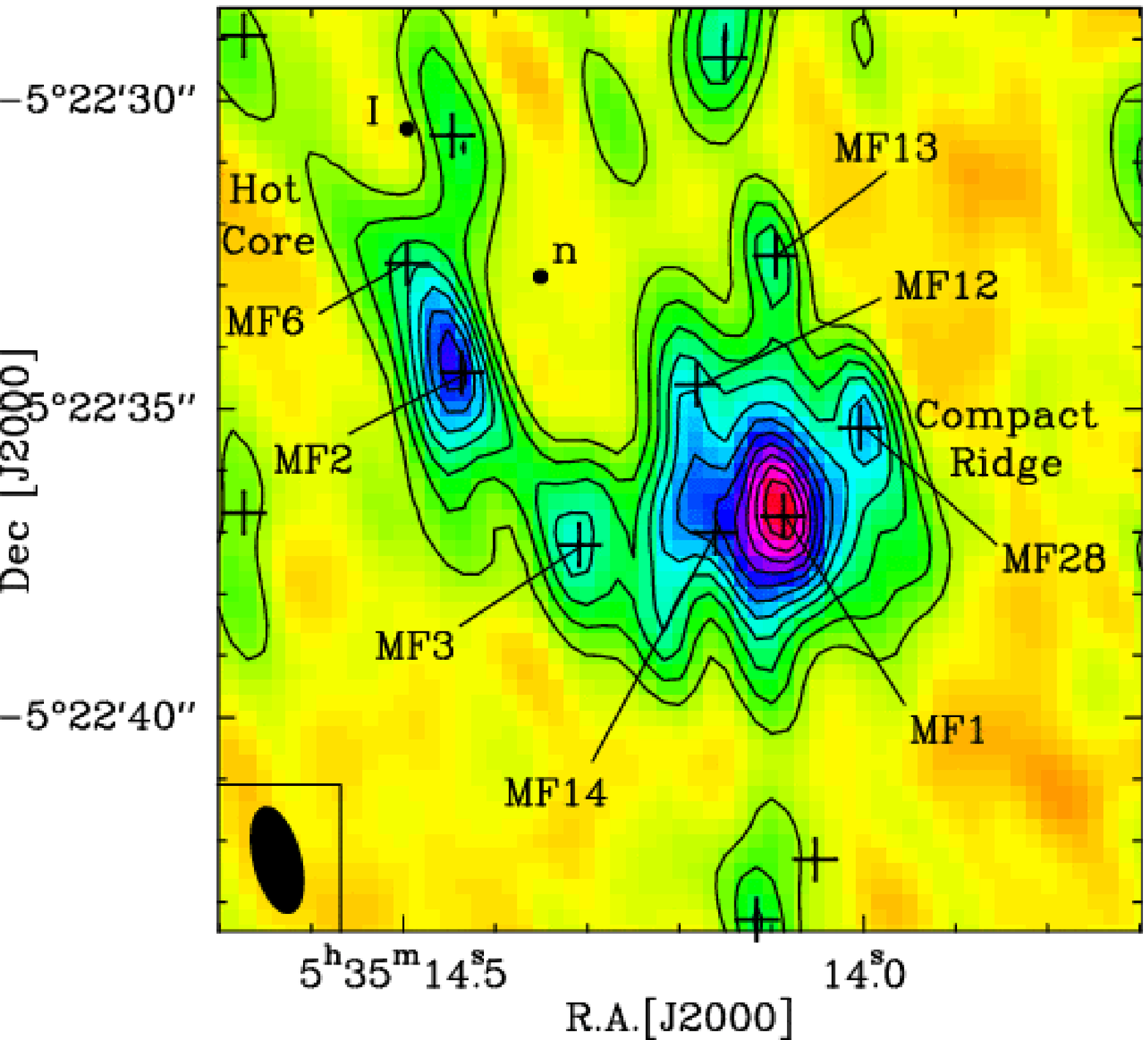}
   \caption{Methyl formate integrated intensity maps  obtained with the Plateau de Bure Interferometer (sum of  emission at 223465.340~MHz, 223500.463~MHz and 223534.727~MHz between 5 and 12~km~s$^{-1}$). The bottom image is a blowup of the Hot Core/Compact Ridge map area. The beam is 1.79$\arcsec$ $\times$ 0.79$\arcsec$; the level step and first contour are 3.2~K~km~s$^{-1}$. The BN object position is ($\alpha_{J2000}$ = 05$^{h}$35$^{m}$14$\fs$1094, $\delta_{J2000}$ = -05$\degr$22$\arcmin$22$\farcs$724), the radio source I position is ($\alpha_{J2000}$ = 05$^{h}$35$^{m}$14$\fs$5141, $\delta_{J2000}$ = -05$\degr$22$\arcmin$30$\farcs$575) and the IR source n position is ($\alpha_{J2000}$ = 05$^{h}$35$^{m}$14$\fs$3571, $\delta_{J2000}$ = -05$\degr$22$\arcmin$32$\farcs$719) \citep{Goddi:2010}. The position of the millimetre source MM23  \citep{Eisner:2008} is also indicated. The main different HCOOCH$\rm_{3}$ emission peaks identified on channel maps are marked by a cross and labelled MF$\rm_{NUMBER}$. }
              \label{Fig.vshape}%
    \end{figure}

The methyl formate molecule, HCOOCH$\rm_{3}$, allows us to trace the spatial distribution of one major oxygenated molecule in Orion-KL, especially in the direction of the Compact Ridge where high spatial resolution data are missing. We have adopted the Hot Core and Compact Ridge positions labelled in \citet{Beuther:2005}. Figure \ref{Fig.vshape} shows our large scale, 1.79$\arcsec$ $\times$ 0.79$\arcsec$ beam size interferometric map of the HCOOCH$\rm_{3}$ line emission integrated over the line profile and on three different transitions: 11$\rm_{4,8}$-10$\rm_{3,7}$ A and E in v$\rm_{t}$ =0 and 18$\rm_{5,14}$-17$\rm_{5,13}$ E in v$\rm_{t}$ =1 (at 223465.340~MHz, 223500.463~MHz and 223534.727~MHz respectively) to improve the signal to noise ratio. It shows: (a) several main emission peaks labeled MF1 to MF28, and (b) an extended, V-shaped molecular emission linking the radio source I to the BN object. 
%-------------------  
%FIGURE 4
%-------------------  
%-------------------  
%
%-------------------  
%FIGURE 5
%-------------------  
   \begin{figure}[h!]
  \centering
   \includegraphics[width=8.5cm,angle=270]{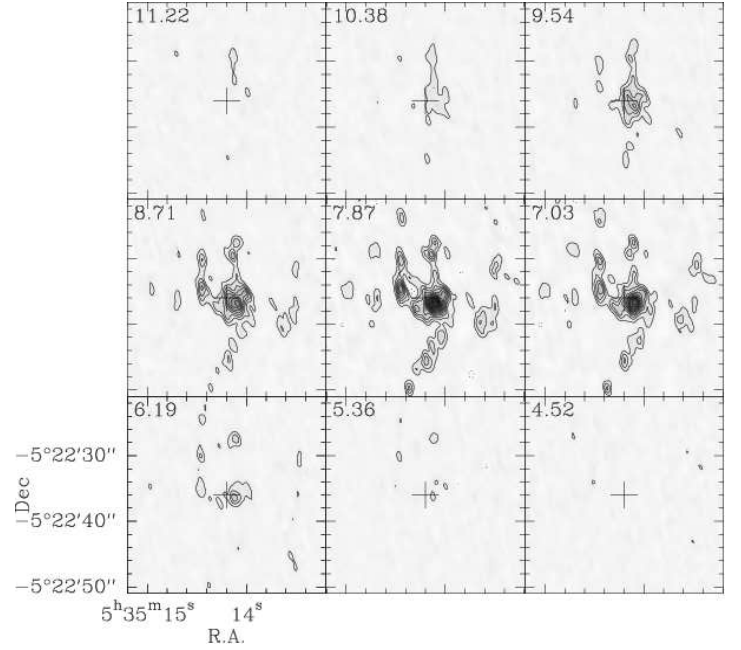}
   \caption{Methyl formate channel maps of emission at 223465.340~MHz. The beam is 1.79$\arcsec$ $\times$ 0.79$\arcsec$ and the first contour and level step are 50~mJy~beam$^{-1}$. The v$\rm_{LSR}$ velocity is indicated on each plot. }
              \label{Fig.channel}%
    \end{figure}
%

%-----------------------------------------
%--Molecular emission peaks-- 
%-----------------------------------------
\subsection{Molecular emission peaks}

We used the whole data set obtained with different frequencies and angular resolutions to model and investigate the properties of the methyl formate emission. The molecular emission peaks (MF1 to MF28) were determined from our highest angular resolution data (1.79$\arcsec$ $\times$ 0.79$\arcsec$) on the channel maps when they are present on \textit{at least three adjacent channels}; they are listed in Table \ref{Table.position-clumps}. The individual positions can vary by one pixel (0.25$\arcsec$) depending on the HCOOCH$\rm_{3}$ transition. Two main molecular peaks, MF1 and MF2,  are located toward the Compact Ridge and the Hot Core-SW  respectively \citep[e.g references taken from][]{Beuther:2005,Friedel:2008}.  Emission peaks are present throughout the observed field and exhibit more or less extended structures. At some frequencies a molecular emission is also detected at the position of the millimeter source MM23 \citep{Eisner:2008}. 

%-------------------  
%TABLE 2
%-------------------  
\begin{table}
\caption{Position of the main HCOOCH$\rm_{3}$ emission peaks observed with the Plateau de Bure Interferometer toward Orion-KL.}             
\label{Table.position-clumps}      
\centering        
\begin{tabular}{l c c}
\hline\hline       
Position name & R.A (J2000) & Dec (J2000) \\
 & 05$^{h}$35$^{m}$ & -05$\degr$22$\arcmin$ \\
\hline                    
MF1  & 14$\fs$09  & 36$\farcs$7 \\ 
MF2 & 14$\fs$44  & 34$\farcs$4 \\  
MF3  & 14$\fs$31 & 37$\farcs$2 \\ 
MF4  & 14$\fs$15 & 29$\farcs$3 \\ 
MF5  & 14$\fs$12 & 27$\farcs$7 \\ 
MF6 & 14$\fs$50 & 32$\farcs$6 \\ 
MF7  & 14$\fs$05 & 42$\farcs$3 \\ 
MF8  & 14$\fs$20 & 46$\farcs$1 \\  
MF9  & 14$\fs$11 & 43$\farcs$3 \\ 
MF10  & 14$\fs$68 & 28$\farcs$9 \\ 
MF11  & 14$\fs$45 & 30$\farcs$6 \\ 
MF12  & 14$\fs$18 & 34$\farcs$6 \\ 
MF13  & 14$\fs$09 & 32$\farcs$5 \\ 
MF14  & 14$\fs$16 & 37$\farcs$0 \\ 
MF15  & 13$\fs$72 & 14$\farcs$0 \\ 
MF16  & 13$\fs$73 & 19$\farcs$1 \\ 
MF17  & 14$\fs$44 & 24$\farcs$0 \\ 
MF18  & 13$\fs$46 & 30$\farcs$8 \\ 
MF19  & 13$\fs$66 & 31$\farcs$5 \\ 
MF20  & 13$\fs$46 & 36$\farcs$6 \\ 
MF21  & 13$\fs$64 & 39$\farcs$3 \\ 
MF22  & 13$\fs$61 & 41$\farcs$3 \\ 
MF23  & 15$\fs$01 & 34$\farcs$8 \\ 
MF24  & 14$\fs$68 & 36$\farcs$7 \\ 
MF25  & 15$\fs$20 & 38$\farcs$7 \\ 
MF26  & 14$\fs$38 & 49$\farcs$4 \\ 
MF27  & 13$\fs$92 & 50$\farcs$3 \\ 
MF28  & 14$\fs$00 & 35$\farcs$3 \\ 
\hline                  
\end{tabular}
\end{table}
%-------------------  

The molecular emission peaks MF1, MF2, MF4-5 and MM23 are associated with the 4 main dust peaks identified at  223~GHz  (see Fig. \ref{Fig.dust223GHz} in Sect.  \ref{sec:dust}). However the continuum emission does not coincide exactly with the associated molecular line peak ($\Delta\alpha$ $\sim$ 0.1$\arcsec$ to 1.2$\arcsec$ and  $\Delta\delta$ $\sim$ 0.03$\arcsec$ to 1.8$\arcsec$, except for MF2 where $\Delta\alpha$ = 1.8$\arcsec$ and $\Delta\delta$ = 2.9$\arcsec$, for a beam size of 1.8$\arcsec$ $\times$ 0.8$\arcsec$). Since both continuum and line maps come from the same data sets the observed spatial separations cannot be ascribed to instrumental effects. \citet{de-Vicente:2002} report the same effect from their HC$\rm_{3}$N (10-9) lines observed in different vibrational levels. We conclude that even our best resolution is probably not sufficient to determine exactly the relative positions of molecular and dust emissions.

%--------------------------------------------------
%-------------Velocity structure-------
%--------------------------------------------------
\subsection{Velocity structure}

Figure \ref{Fig.channel} presents the channel maps of the 223465.340~MHz line. Most of the peaks appear between 6 and 9~km~s$^{-1}$. However a linear structure is seen mainly toward the North at higher velocities (see discussion on MF1 and MF4 peaks in Sect. \ref{sec:observedpeaks}); it goes from MF5 to MF8 . There is no methyl formate emission at the Hot Core position at the usual 5~km~s$^{-1}$ velocity.

%-----------------------------------------
%------Critical density---------- 
%-----------------------------------------
\subsection{Critical density}
\label{sec:density} 
We assumed that the HCOOCH$\rm_{3}$ transitions are thermalized at the five molecular emission peaks MF1 to MF5. From the H$\rm_{2}$ column densities measured at the continuum peaks associated to those positions (see Sect. \ref{sec:dust} and Table \ref{Table.continuum-clumps-peak}) we find indeed n$\rm_{H_{2}}$~$\gg$~n$\rm_{cr}$~$\sim$~10$^{4-7}$~cm$^{-3}$  for the detected lines at frequencies between 80~GHz and 226~GHz. To compute n$\rm_{cr}$, we have adopted the collision coefficients of H$\rm_{2}$O \citep{Faure:2008}  because of a similar dipole moment to HCOOCH$\rm_{3}$ ($\mu$ = 1.8~D), and we have taken into account a multiplicative factor between 1 and 10 because HCOOCH$_{3}$ is a larger molecule. Even if n$\rm_{cr}$ would reach 10$^{7}$~cm$^{-3}$ the critical density remains well below n$\rm_{H_{2}}$ (see Table \ref{Table.continuum-clumps-peak}).

%-----------------------------------------
%---Missing flux estimation----
%-----------------------------------------
\subsection{Missing flux estimation}
\label{sec:missingflux} 

To estimate whether there is some flux missing in our data due to spatial filtering with the PdBI interferometer we have compared our PdBI spectra at 3~mm and 1~mm with IRAM 30-m telescope spectra. The PdBI spectra have been convolved with a Gaussian beam similar to the 30-m beam (25$\arcsec$ at 3~mm and 11$\arcsec$ at 1~mm). For all data sets the convolution was performed at the positions observed with the 30-m telescope. Error bars arising from the calibration and the pointing of the 30-m antenna fall within the typical range of 5 --10 $\%$.

\begin{figure}[h!]
  \centering
  \includegraphics[width=6.2cm,angle=270]{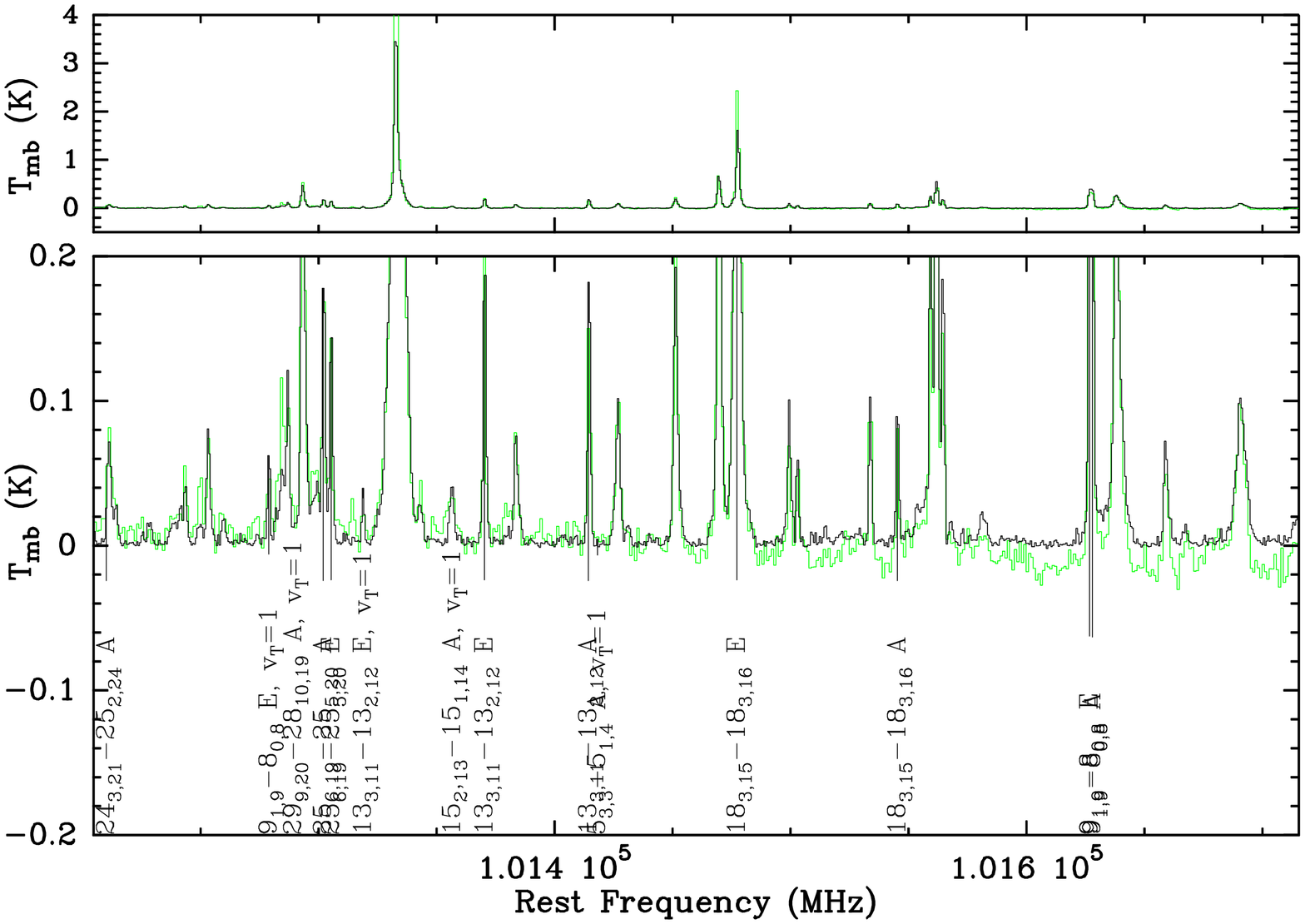}
     \caption{Spectra of molecular emission at 3~mm toward the Compact Ridge. The lower panel is a blow up of the upper one. Green lines show IRAM 30-m telescope data of \citet{Combes:1996} and black lines illustrate PdBI data (set n$^{\circ}$3, see table \ref{Table.dataset_parameters}) convolved with the 30-m beam.  A and E transitions of HCOOCH$\rm_{3}$ are marked on the figure.} 
              \label{Fig.fluxlost}%
    \end{figure}

In the Compact Ridge, comparison of our PdBI observations (which have a spatial resolution of 3.79$\arcsec$ $\times$ 1.99$\arcsec$ at the relevant frequencies) with the 30-m observations made at 101~GHz \citep{Combes:1996} indicates a ratio of missing flux of 3$\%$ and 15$\%$ measured from lines at 101370.505~MHz (13$\rm_{3,11}$-13$\rm_{2,12}$, E) and 101414.746~MHz (13$\rm_{3,11}$-13$\rm_{2,12}$, A) respectively.
 At 101477.421~MHz the 18$\rm_{3,15}$-18$\rm_{3,16}$E line is strongly blended with an intense H$\rm_{2}$CS transition (see Fig \ref{Fig.fluxlost}). 

 Concerning the Hot Core position, we also compared our PdBI observations (spatial resolution: 1.79$\arcsec$ $\times$ 0.79$\arcsec$) with the 30-m observations made at 223~GHz (J. Cernicharo, private communication). 
 Chemically, this region exhibits an important molecular diversity implying a very high confusion level. All HCOOCH$\rm_{3}$ lines are blended either with another strong molecule or with a multitude of weaker lines, which makes it difficult to measure the exact missing flux. 
Using the 223500.463~MHz (11$\rm_{4,8}$-10$\rm_{3,7}$, A) line, we find that about 20$\%$ of the flux is missed. At 3~mm we estimate that we can miss between 4$\%$ and 15$\%$ of methyl formate emission flux ; this is determined from the lines at 101302.159~MHz (25$\rm_{6,19}$-25$\rm_{5,20}$, A) and 101370.505~MHz (13$\rm_{3,11}$-13$\rm_{2,12}$, E).

%-------------------  
%FIGURE 6
%-------------------  
%-------------------  

%--------------------------------------------------------------------------------------
%--Observed parameters of the different HCOOCH3 peaks----
%--------------------------------------------------------------------------------------
\subsection{Observed parameters of the different HCOOCH$\rm_{3}$ peaks}
\label{sec:observedpeaks} 

Supposing that the gas  rotational temperature does not exceed a few hundred Kelvin, we have searched for the 64 methyl formate transitions with E$\rm_{upper}$ $\la$ 650~K. We have used the MAPPING GILDAS software to map and model the methyl formate source sizes, and the CLASS GILDAS software to determine the line parameters. The line frequencies,  spectroscopic  and observed parameters (v, $\Delta$v$\rm_{1/2}$, T$\rm_{B}$, W) at the MF1 to MF5 emission peaks are given in Tables \ref{MF1} to \ref{MF5}. The observational results are briefly summarized below for the main molecular peaks shown in Figure \ref{Fig.vshape}.

\begin{figure}[h!]
  \centering
   \includegraphics[width=6cm,angle=270]{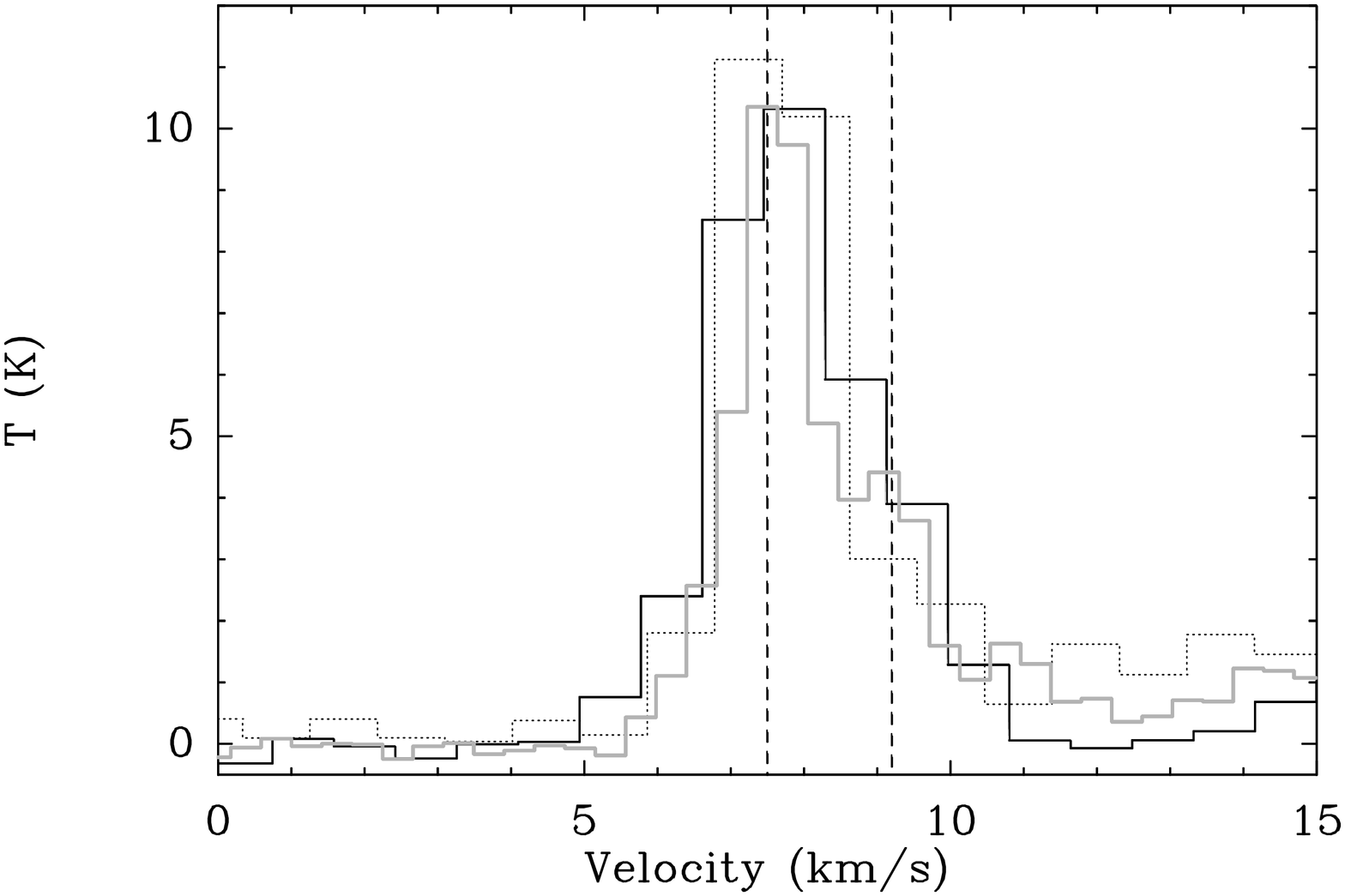}
   \caption{Two HCOOCH$\rm_{3}$ components observed on the emission peak MF1 toward the Compact Ridge. The black spectrum shows the 223465.340~MHz line, the grey spectrum the 225855.505~MHz transition and the dotted line spectrum the 203435.554~MHz transition. The intensity of the 2 latter spectra is multiplied by a factor 1.5 and 2 respectively. Dash lines mark v$\rm_{LSR}$ = 7.5 and 9.2~km~s$^{-1}$.}
              \label{Fig.decompositionMF1}%
    \end{figure}

%------
%MF1
%------
\paragraph{\textbf{MF1}}

is the main HCOOCH$\rm_{3}$ emission peak toward the Compact Ridge. It is present in all maps of detected  HCOOCH$\rm_{3}$ transitions. The source sizes, estimated at half-peak flux density (see Table \ref{MF1}), increase with a lower spatial resolution from 3.0$\arcsec$ $\times$ 2.0$\arcsec$ at 223~GHz to 7.0$\arcsec$ $\times$ 10.0$\arcsec$ at 80~GHz. The line width clearly depends on the spatial and spectral resolutions. For instance, with the same beam ($\approx$ 3$\arcsec$ $\times$ 2$\arcsec$) but using two different spectral resolutions, 1.85~km~s$^{-1}$ and 0.42~km~s$^{-1}$, we obtain $\Delta$v $\approx$ 3.6~km~s$^{-1}$ and 1.1~km~s$^{-1}$, respectively. Table \ref{MF1} summarizes the line parameters for all detected, blended or not detected transitions in all data sets. We detected 20 lines, and observed 2 partially blended and 33 blended lines; 9 transitions were too faint to be detected. At both high spatial and spectral resolutions the 203~GHz, 223~GHz and 225~GHz methyl formate spectra display two components, one around 7.5~km~s$^{-1}$ and the other around 9~km~s$^{-1}$ (see fig. \ref{Fig.decompositionMF1}). The 2-velocity fit parameters are presented in Table \ref{Table.MF1bis}.  Both components have a narrow average line width of  $\Delta$v$\rm_{1/2}$ $\approx$ 1.7~km~s$^{-1}$. For the transitions observed with less resolution the second velocity component cannot be directly separated and the line widths are broader ($\Delta$v$\rm_{1/2}$ $\approx$ 3.6~km~s$^{-1}$). 

%-------------------  
%FIGURE 7
%-------------------  
%-------------------  

%-------------------  
%FIGURE 8
%-------------------  
   \begin{figure}[h!]
  \centering
   \includegraphics[width=6cm,angle=270]{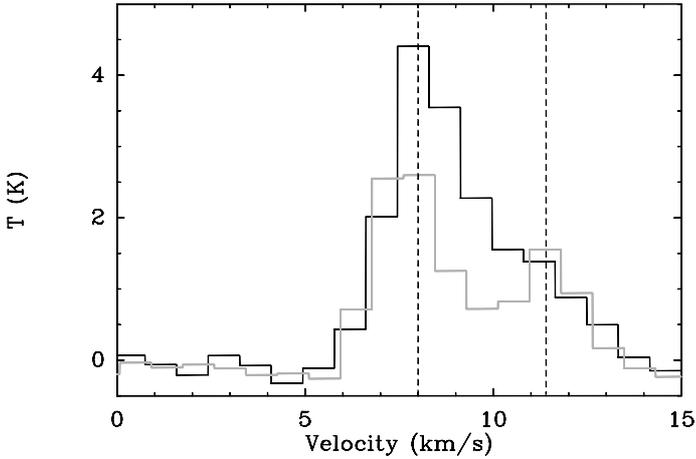}
   \caption{Two HCOOCH$\rm_{3}$ components observed toward the emission peak MF4. The black spectrum shows the 223465.340~MHz transition and the grey spectrum the 223500.463~MHz transition. Dash lines mark v$\rm_{LSR}$~=~8.0 and 11.4~km~s$^{-1}$.}
              \label{Fig.decompositionMF4}%
    \end{figure}
%-------------------  

%-------------------  
%TABLE 3
%-------------------  
\begin{table*}
\caption{Two-component analysis of methyl formate transitions observed with the IRAM Plateau de Bure Interferometer toward position MF1 in the Compact Ridge. For the definition of the parameters see Table \ref{MF1}}.            
\label{Table.MF1bis}
\small\addtolength{\tabcolsep}{-2pt} 
\begin{center}     
\begin{tabular}{llccc|cccc|cccc}
\hline\hline
Frequency &Transition & E$\rm_{up}$ & S$\mu$$^{2}$ & sigma & \multicolumn{4}{c}{Component 1}& \multicolumn{4}{c}{Component 2} \\ 
 & & &  & & v$\rm_{1}$ & ($\Delta$v$\rm_{1/2}$)$\rm_{1}$ &T$\rm_{B1}$ & W$\rm_{1}$ & v$\rm_{2}$ & ($\Delta$v$\rm_{1/2}$)$\rm_{2}$ &T$\rm_{B2}$ & W$\rm_{2}$ \\
  (MHz) &  & (K) & (D$^{2}$) & (K) & (~km~s$^{-1}$) & (~km~s$^{-1}$)  & (K) & (K~km~s$^{-1}$)& (~km~s$^{-1}$)& (~km~s$^{-1}$) & (K)  & (K~km~s$^{-1}$) \\
\hline
203435.554 & $19_{8,11} - 19_{7,12}$ A & 155 & 5.3 & 0.09 & 7.6(0.9) & 1.6(0.9) & 6.40 & 10.6(1.6) & 9.3(0.9) & 2.2(0.9) & 0.88 & 2.1(1.6) \\
203471.840 & $19_{8,11} - 19_{7,12}$ E & 155 & 4.9 & 0.09 & 7.5(0.1)& 1.6(0.1)& 5.28& 9.2(2.0)& 9.0(0.7)& 3.1(1.0)& 1.29 & 4.3(2.2) \\
223465.340 & $11_{4,8} - 10_{3,7}$ E & 50 & 2.1 & 0.17 & 7.6(0.1)& 2.0(0.2)& 10.74 &22.5(1.9) & 9.5(0.2)&1.5(0.3) & 3.17 &5.0(1.7) \\
223500.463 & $11_{4,8} - 10_{3,7}$ A & 50 & 1.8 & 0.17 & 7.4(0.1)&1.8(0.1) & 11.49 &21.8(1.7) & 9.2(0.3) & 1.9(0.4) & 2.85 & 5.8(1.2) \\
223534.727 & $18_{5,14} - 17_{5,13}$ E, (v$\rm_{t}$=1) & 305 & 41.6 & 0.17 & 7.4(0.1) & 1.7(0.1)& 10.25 & 18.5(0.8)& 8.9(0.1)& 2.9(0.2) & 4.16 & 13.0(8.2) \\
225855.505 & $6_{6,1} - 5_{5,1}$ E & 36 & 3.1 &0.10 & 7.6(0.1)&1.2(0.1) &6.84 &9.1(0.1) &9.2(0.1) & 1.1(0.1) & 2.56 &2.9(0.1)\\
225999.145 & $30_{7,23} - 29_{8,22}$ E & 312 & 2.2 &0.11 & 7.5(0.1)& 0.9(0.3)&0.70 & 0.6(0.2)& 9.1(0.1)&0.8(0.3) & 0.45 &0.4(0.1) \\
\hline             
\end{tabular}
\end{center}   
\end{table*}  
%-------------------  
%
%-------------------  
%TABLE 4
%-------------------  
\begin{table*}
\caption{Two-component analysis of methyl formate transitions observed with the IRAM Plateau de Bure Interferometer toward position MF4. For the definition of the parameters see Table \ref{MF4} (in Appendix A).}             
\label{Table.MF4bis}
\small\addtolength{\tabcolsep}{-2pt} 
\begin{center}     
\begin{tabular}{llccc|cccc|cccc}
\hline\hline
Frequency &Transition & E$\rm_{up}$ & S$\mu$$^{2}$ & sigma & \multicolumn{4}{c}{Component 1}& \multicolumn{4}{c}{Component 2} \\ 
 & & &  & & v$\rm_{1}$ & ($\Delta$v$\rm_{1/2}$)$\rm_{1}$ &T$\rm_{B1}$ & W$\rm_{1}$ & v$\rm_{2}$ & ($\Delta$v$\rm_{1/2}$)$\rm_{2}$ &T$\rm_{B2}$ & W$\rm_{2}$ \\
  (MHz) &  & (K) & (D$^{2}$) & (K) & (~km~s$^{-1}$) & (~km~s$^{-1}$)  & (K) & (K~km~s$^{-1}$)& (~km~s$^{-1}$)& (~km~s$^{-1}$) & (K)  & (K~km~s$^{-1}$) \\\hline
223465.340 & $11_{4,8} - 10_{3,7}$ E & 50 & 2.1 & 0.17 & 8.0(0.1)& 1.9(0.3)& 3.83 &7.7(2.5) & 10.2(0.8)&3.7(1.0) & 1.66 &6.6(2.7) \\
223500.463 & $11_{4,8} - 10_{3,7}$ A & 50 & 1.8 & 0.17 & 8.0(0.1)&2.2(0.2) & 1.90 &4.4(0.4) &11.4(0.2) & 2.2(0.3) & 1.29 & 3.0(0.4) \\
223534.727 & $18_{5,14} - 17_{5,13}$ E, (v$\rm_{t}$=1) &305 & 41.6 &0.17 &8.7(0.3)&3.0(0.4) & 1.78 &5.8(0.8) &11.0(0.4) & 5.3(0.8) & 1.34 & 7.6(0.9) \\
\hline             
\end{tabular}
\end{center}   
\end{table*}  
%-------------------  

%------
%MF2
%------
\paragraph{\textbf{MF2}}

is the main HCOOCH$\rm_{3}$ emission peak toward the Hot Core region. This peak is present in all detected HCOOCH$\rm_{3}$ transitions. The source sizes, estimated at half-peak flux density (see Table \ref{MF2} in Appendix A) increases with a lower spatial resolution from 2.5$\arcsec$ $\times$ 1.2$\arcsec$ at 223~GHz to 5.0$\arcsec$ $\times$ 2.5$\arcsec$ at 101~GHz\footnote{Note that the lower spatial resolution of the detected lines at 80565.210~MHz and 105815.953~MHz does not allow us to identify individual sources.} . This source appears to be more compact than MF1 in the Compact Ridge. Table \ref{MF2} (in Appendix A) summarizes the line parameters for all detected, blended or not detected transitions in all data sets. We detected 14 lines and observed 3 partially blended and 40 blended lines; 7 transitions were too faint to be detected. The mean velocity is around 7.7~km~s$^{-1}$. As for clump MF1 the line widths depend on the spatial and spectral resolutions. With the highest spatial resolution (1.79$\arcsec$ $\times$ 0.79$\arcsec$) we find $\Delta$v$\rm_{1/2}$ $\approx$ 2.4~km~s$^{-1}$ while for a resolution of 3.79$\arcsec$ $\times$ 1.99$\arcsec$, $\Delta$v$\rm_{1/2}$ $\approx$ 4.1~km~s$^{-1}$; we did not take into account the partially blended line at 226061.796~GHz. The linewidths observed here are slightly broader than those of components 1 and 2 at the MF1 peak; as the lines are optically thin (see Sect. \ref{sec:temperatures}) this indicates a larger velocity spread in the source.

%------
%MF3
%------
\paragraph{\textbf{MF3}}

is a HCOOCH$\rm_{3}$ emission peak located in-between the two previous clumps. It often appears as an extension of the emission as most of our data do not allow us to isolate the emission peak. As for clumps MF1 and MF2, the linewidth is narrow; at 203~GHz and 223~GHz $\Delta$v$\rm_{1/2}$ $\approx$ 1.7~km~s$^{-1}$. Table \ref{MF3} (in Appendix A) summarizes the line parameters for all detected, blended or not detected transitions in all data sets. We detected 14 lines, and observed 3 partially blended and 26 blended lines; 21 transitions were too faint to be detected. The methyl formate velocity is around 7.7~km~s$^{-1}$.

%------
%MF4
%------
\paragraph{\textbf{MF4}}

is one of the main HCOOCH$\rm_{3}$ emission peaks lying at the North of the Compact Ridge. Table \ref{MF4} (in Appendix A) summarizes the line parameters for all detected, blended or not detected transitions in all data sets. We detected 9 lines, and observed 4 partially blended and 28 blended lines; 23 transitions were too faint to be detected. As for clump MF1 the high spectral resolution line profiles show two components (see fig. \ref{Fig.decompositionMF4}). The 223~GHz methyl formate emission profiles are fitted with two components centered around 8.0~km~s$^{-1}$ and 11.4~km~s$^{-1}$ (Table \ref{Table.MF4bis}). Both components have linewidths in the range ($\Delta$v$\rm_{1/2}$ $\approx$ 1.9 - 3.7~km~s$^{-1}$).

%------
%MF5
%------
\paragraph{\textbf{MF5}}

lies near the emission peak MF4 and its size is similar to that of MF4. Two velocity components, visible with high spectral resolution only,  are present in our data. However, we have not made any precise velocity identification of these components because they are more blended than those identified in the direction of MF4 and MF1. The unresolved emission is centered around 7.7~km~s$^{-1}$. We detected 9 methyl formate transitions and observed 3 partially blended and 28 blended lines. 24 other lines were too faint to be detected. All results from all data sets are summarized in Table \ref{MF5} (in Appendix A).

%------------------------------------------------------------
%--Fraction of blended and detected lines--
%------------------------------------------------------------
 \subsection{Fraction of blended and detected lines}
We have mentioned earlier that spectral confusion is a major problem toward Orion-KL. Confusion increases with frequency and it is more prevalent in our data at 223~GHz--225~GHz than at 101~GHz. Spectral confusion is present to some extent toward all the emission peaks identified in this work (MF1 to MF28).  

We present here a rough estimate of the confusion level.
From a visual inspection of our spectra, we believe that some line frequencies are contaminated by other lines, and noted a priori as \emph{blended}, whereas at other line frequencies the spectrum appears \emph{free of contamination}. 
To determine which lines are detectable in our data set we use the  simple one-temperature  model derived in Section \ref{sec:temperatures}  
and compare the expected line intensity to the noise level. 
Of the $n_d$ \emph{detectable lines}, a number, $n_{db}$, are blended whereas $n_{df} = n_d - n_{db}$ others appear free of contamination.

For sources MF1, MF2 and MF3 all  lines expected above 3 sigma and free of contamination are detected; this is a consistency check of the LTE model. To roughly quantify the effect of confusion in these sources we compute $\eta_D = n_{df}/n_d$. We obtain the following results for temperatures derived from our rotational diagrams (see following Section):
\begin{itemize}
\item $\eta_D = 49\%$ at the MF1 peak  for T = 79~K,
\item $\eta_D = 40\%$ at the MF2 peak for T = 130~K,
\item $\eta_D = 30\%$ at the MF3 peak for T = 85~K
\end{itemize}

For sources MF4 and MF5, the presence of two velocity components makes the methyl formate lines broader and  more difficult to detect. The number of detected lines is thus more imprecise, as for example some broad lines may be classified as blended instead of broadened by the dynamics.
We get  $\eta_D = 13-20\%$ which is coherent with an increase of the confusion for broader lines. Some of the fainter lines expected to be detectable from the LTE model are not seen, especially in the data cubes around 225~GHz, which we attribute to an increased interferometric filtering of the  quasi vertical MF4-MF5 structure in these cubes (compare the two \textit{uv} coverages in Fig. \ref{uvco}).

The meaning of $\eta_D$ should not be overinterpreted because i) it depends on the LTE hypothesis, on the derived column density and temperature values and on the expected line profile; ii) it is not intrinsic to a molecule in a given source as it depends also on the frequency range and other observation parameters  (noise level, spectral and spatial resolution, $uv$ coverage). Note in addition that in the case of our dataset, the latter parameters vary from one datacube to another, so that $\eta_D$ is only an average value. It is however a first indication of the relative importance of the confusion problem - a  similar indicator was given by \citet{Belloche:2008a} for their search and detection of amino acetonitrile in Sgr B2(N).

%-------------------------------------------------------------------------------------------------------------------------------------------------------
%------TEMPERATURES AND ABUNDANCE OF MF TOWARDS THE WHOLE MOLECULAR V-SHAPE------
%-------------------------------------------------------------------------------------------------------------------------------------------------------

\section{Temperatures and abundance of methyl formate towards the whole molecular V-shape}
\label{sec:temperatures}

From \citet{Rohlfs:2000} the opacity at the line center can be calculated from: 
%
%------------------- 
%EQ 1
%------------------- 
      \begin{equation}
           \rm
     \tau= 
                      -ln(1-\frac{\Delta T_{mb}}{J(T_{r})-J(T_{continuum})-J(T_{bg})}) \\
      \end{equation}
%------------------- 
 where,  \[
    \begin{array}{lp{0.8\linewidth}}
    \Delta T\rm_{mb} & is the excess brightness temperature observed on spectra,     \\
    J(T\rm_{r}) & the source function of the rotational brightness temperature,  \\
   J(T\rm_{continuum}) & the source function of the continuum brightness temperature,  \\
    J(T\rm_{bg}) & the source function of the 2.7~K background brightness temperature.           \\
    \end{array}
   \]

We assumed that all the lines are optically thin (calculated opacities are generally less than 0.05 and at most 0.1) and that the local thermodynamic equilibrium is reached for all positions (see Sect. \ref{sec:density}), so that the kinetic temperature is equal to the excitation, rotational and vibrational temperatures. We estimated the rotational temperature and the column density of HCOOCH$\rm_{3}$ using the equation \citep{Turner:1991}:  
%
%------------------- 
%EQ 2
%------------------- 
      \begin{equation}
           \rm
     ln(\frac{3k_{B}W}{8\pi^{3}\nu S\mu^{2}g_{i}g_{k}}) = 
                      ln(\frac{N}{Q}) - \frac{E_{up}}{k_{b}T}\\
      \end{equation}
where,  \[
    \begin{array}{lp{0.8\linewidth}}
    W & is the integrated line intensity (K~km~s$^{-1}$),  \\
    \nu & the line frequency (Hz), \\
   S\mu^{2} & the line strength (D$^{2}$),  \\
    g_{i} & the reduced nuclear spin statistical weight, \\
    g_{k} & the K-level degeneracy, \\
    N & the total column density, \\
    Q & the partition function, \\
    E_{up} & the upper state energy, \\
    T & the excitation temperature. \\
    \end{array}
   \]
%------------------- 
   
Due to the CH$\rm_{3}$ group methyl formate is divided into two forms: A and E. \cite{Turner:1991} indicates that for A species: g$\rm_{i}$~=~2 and g$\rm_{k}$~=~1, and inversely for E species. We used  g$\rm_{i}$g$\rm_{k}$~=~2 for all the transition lines. \\

We have made the rotational diagrams for two different spatial resolutions. One is based on the data with the highest spatial resolution (synthesized beam of 1.79$\arcsec$ $\times$ 0.79$\arcsec$ at 223~GHz ). The other one includes data at 101~GHz, 203~GHz and 225~GHz where the two first data sets are smoothed to the resolution of the third one (synthesized beam of 3.63$\arcsec$ $\times$ 2.26$\arcsec$) for the MF1 emission peak. 
The rotational temperatures and column density uncertainties, estimated by a least-square fit, are only based on the statistical weight of the detection measurements\footnote{Error bars reflect only the uncertainties in the gaussian fit of the lines. Note that some points overlap because they correspond to lines with the same upper energy level E$\rm_{up}$.}\ (upper limits are not taken into account by the fit).
The derived column densities are corrected for the beam filling factor.

%------------------------------------------------
%Rotational diagram at the MF1 peak 
%------------------------------------------------
\paragraph{\textbf{Rotational diagram at the MF1 peak} \\}

%------------------- 
%FIGURE 9
%------------------- 
   \begin{figure}[h!]
  \centering
  \includegraphics[width=5.8cm,angle=270]{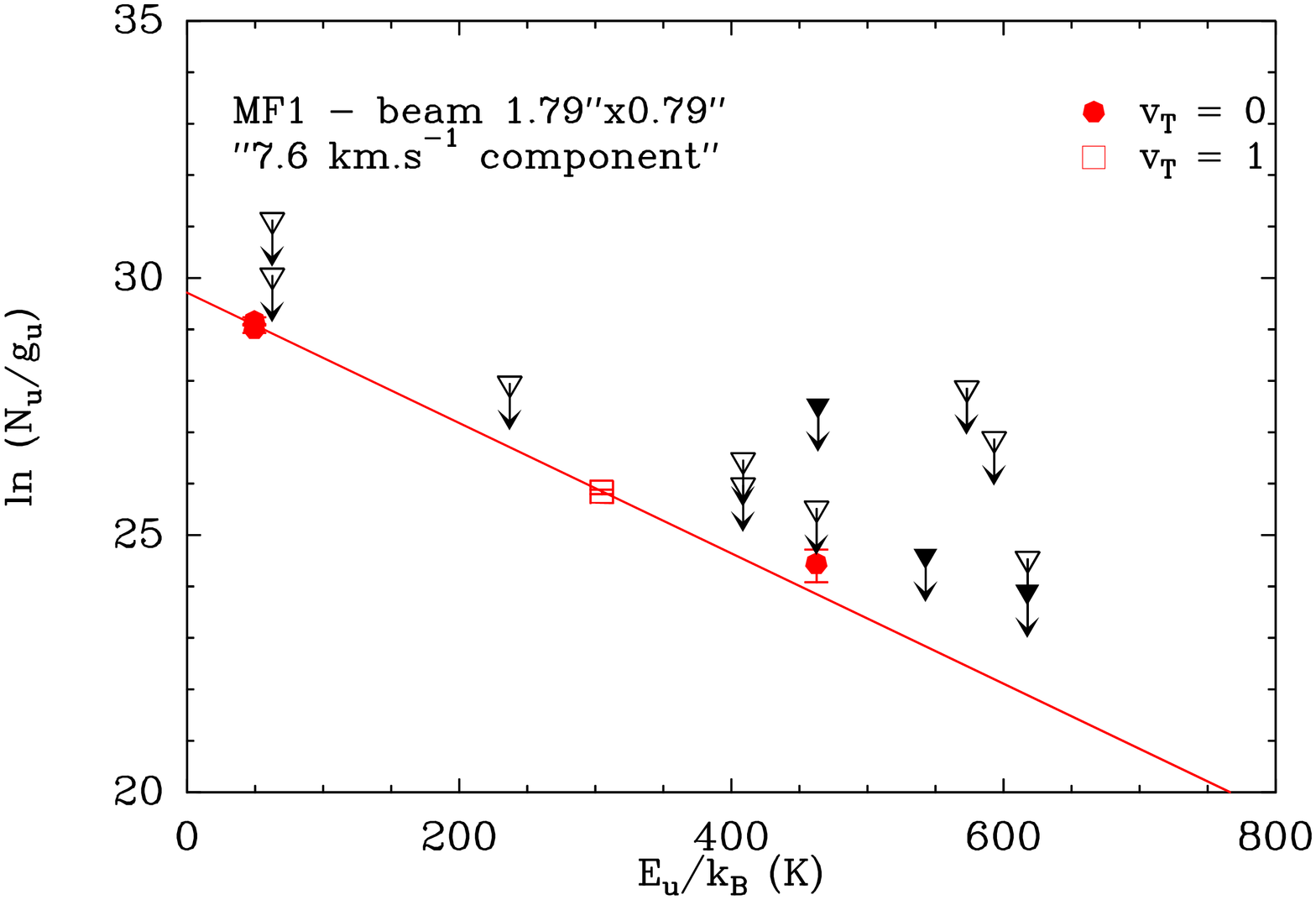}
      \includegraphics[width=5.8cm,angle=270]{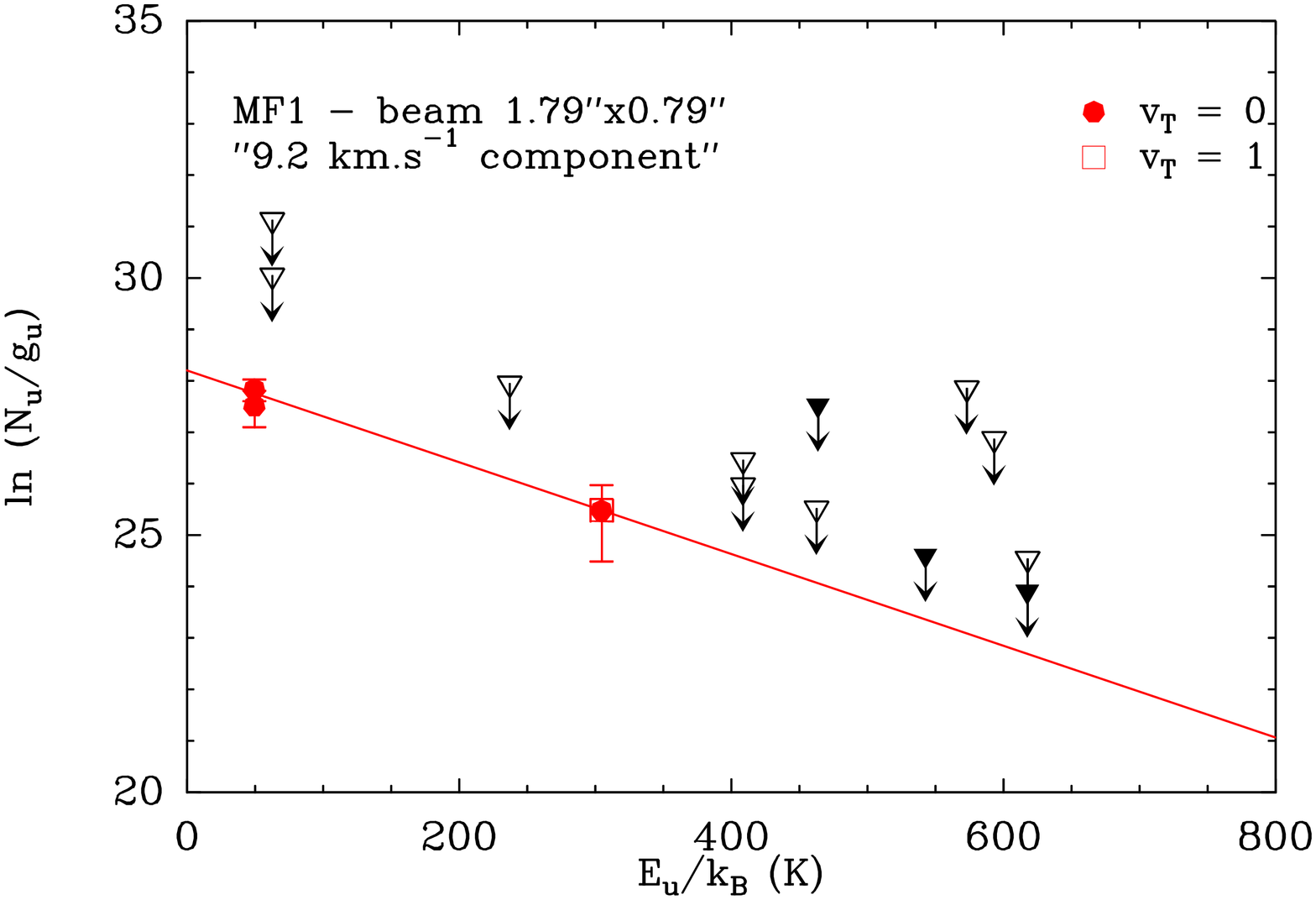}
   \caption{Rotational diagram of the first (top) and the second (bottom) components  toward the MF1 peak, based only on data observed with a synthesized beam of 1.79$\arcsec$ $\times$ 0.79$\arcsec$. Dots and squares with error bars mark detected and partially blended lines in the fundamental and first excited states v$\rm_{t}$ = 0 and v$\rm_{t}$ = 1, respectively, filled triangles mark undetected lines and open triangles blended lines. The red line is the fit according to the method described in Sect. \ref{sec:temperatures}. The derived temperature is 79$\pm$2~K and 112$\pm$50~K, respectively.}
              \label{Fig.DRc1pb1}%
    \end{figure}
%------------------- 
%
%------------------- 
%FIGURE 10
%------------------- 
    \begin{figure}[h!]
  \centering
  \includegraphics[width=5.8cm,angle=270]{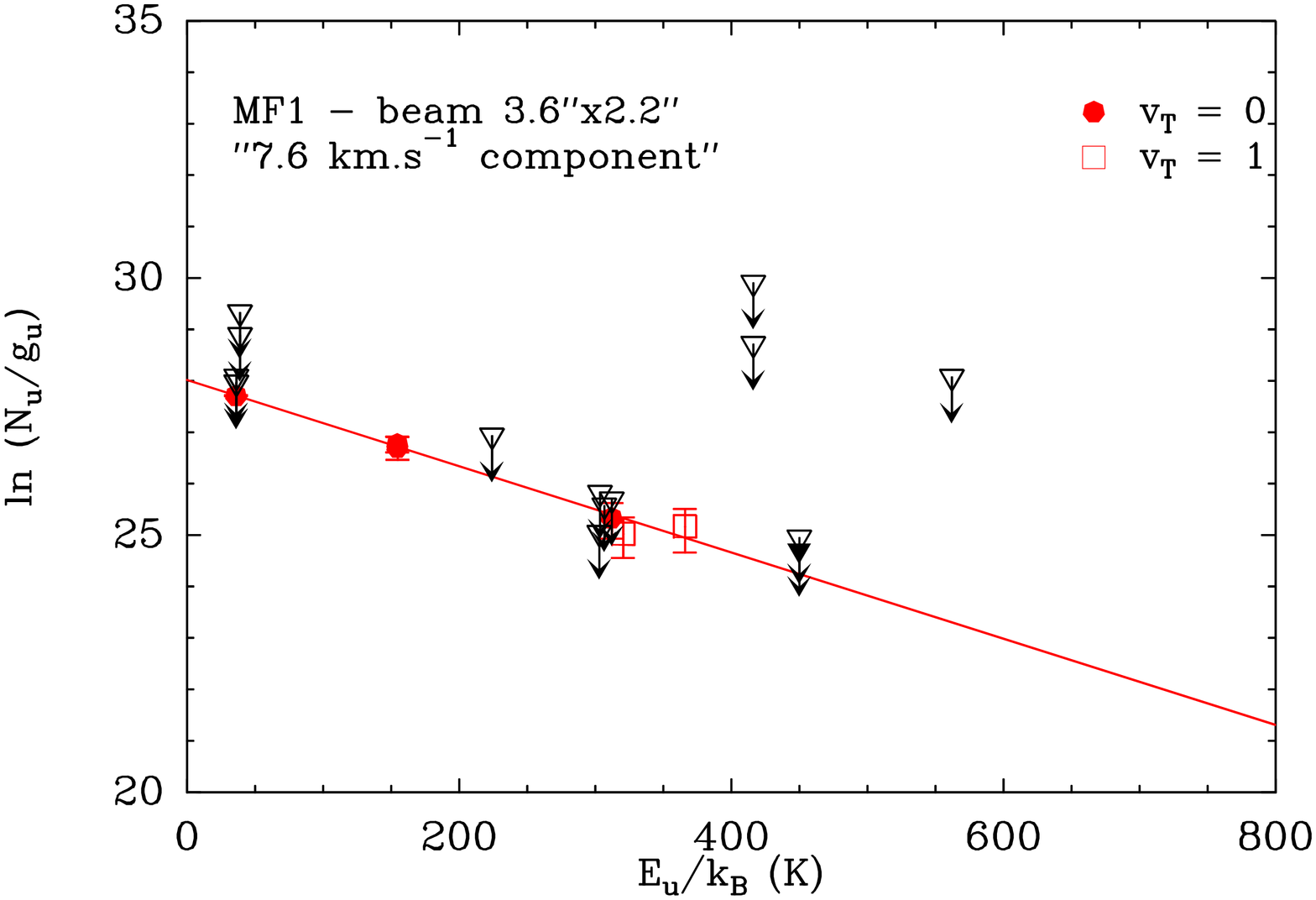}
      \includegraphics[width=5.8cm,angle=270]{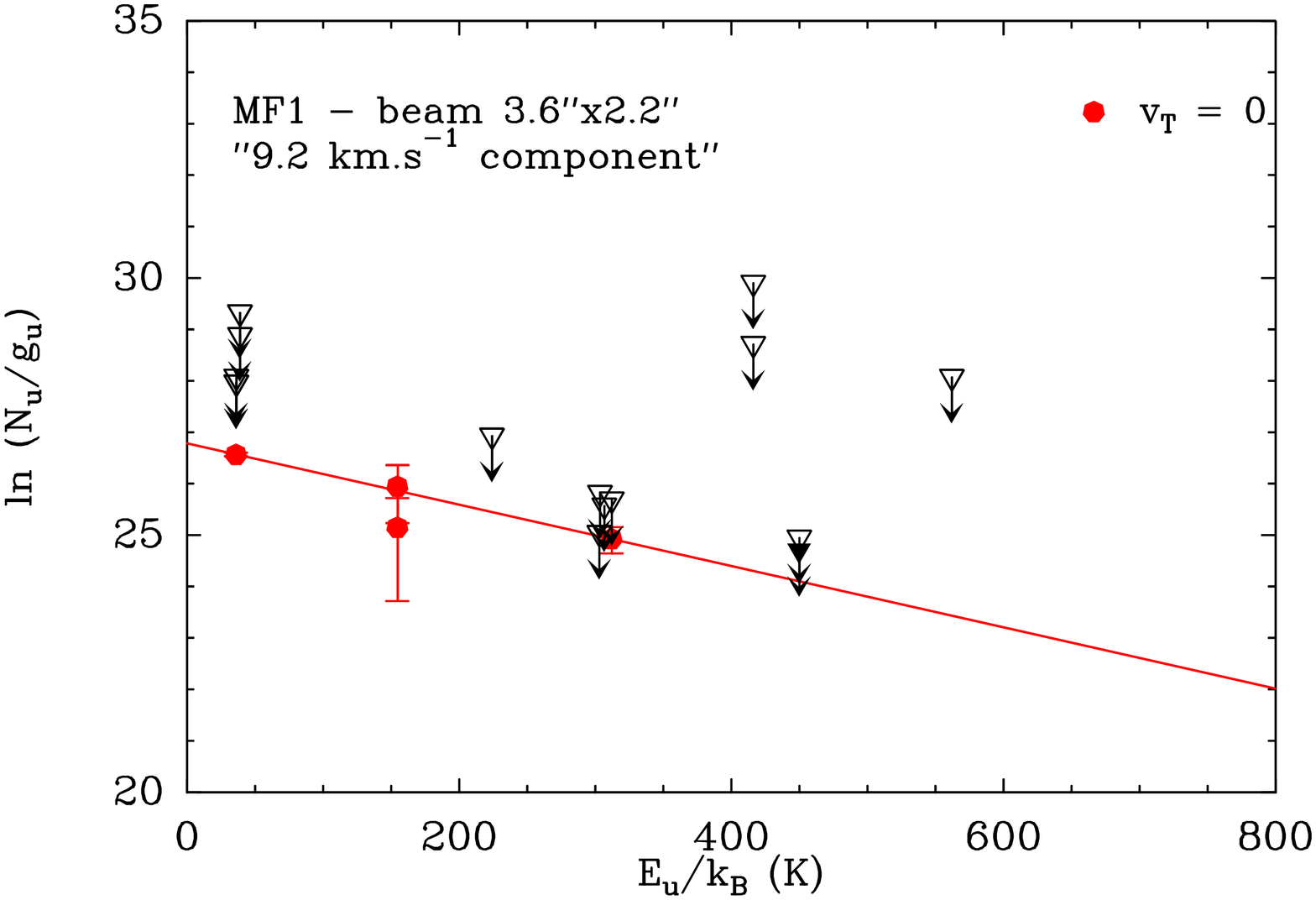}
   \caption{Rotational diagram of the first (top) and the second (bottom) components toward the MF1 peak, based only on data observed with a synthesized beam of 3.63$\arcsec$ $\times$ 2.26$\arcsec$. Dots and squares with error bars mark detected and partially blended lines in the fundamental and first excited states v$\rm_{t}$ = 0 and v$\rm_{t}$ = 1, respectively, filled triangles mark undetected lines and open triangles blended lines. The red line is the fit according to the method described in Sect. \ref{sec:temperatures}. The derived temperature is 119$\pm$10 K and 168$\pm$30 K, respectively.}
              \label{Fig.DRc1lb1}%
    \end{figure}  
%------------------- 

We have made a rotational diagram for each velocity component.
Figure \ref{Fig.DRc1pb1} displays the diagrams obtained with the data at a 1.79$\arcsec$ $\times$ 0.79$\arcsec$ resolution (see set n$^{\circ}$9 in Table \ref{Table.dataset_parameters}) and figure \ref{Fig.DRc1lb1} the diagrams obtained with the data at a 3.63$\arcsec$ $\times$ 2.26$\arcsec$ resolution (see sets 7, 8, 10, 11 and 12 in Table \ref{Table.dataset_parameters}).
They show that the temperature of the 9.2~km~s$^{-1}$ component is higher (112$\pm$50~K at high resolution and 168$\pm$30~K at a lower resolution) than that of the 7.6~km~s$^{-1}$ component (79$\pm$2~K at high resolution and 119$\pm$10~K at a lower resolution). We also note for both components that the temperatures are lower with the high spatial resolution, suggesting an external heating of the clump. The rotational diagram taking into account both components leads to an average rotational temperature of 100~K.

The derived column densities are given in Table \ref{Table.MF2345-mainresults}.
The column density derived with a high spatial resolution is larger than that derived with a lower spatial resolution.
%
%------------------- 
%FIGURE 11
%------------------- 
      \begin{figure}[h!]
  \centering
  \includegraphics[width=5.8cm,angle=270]{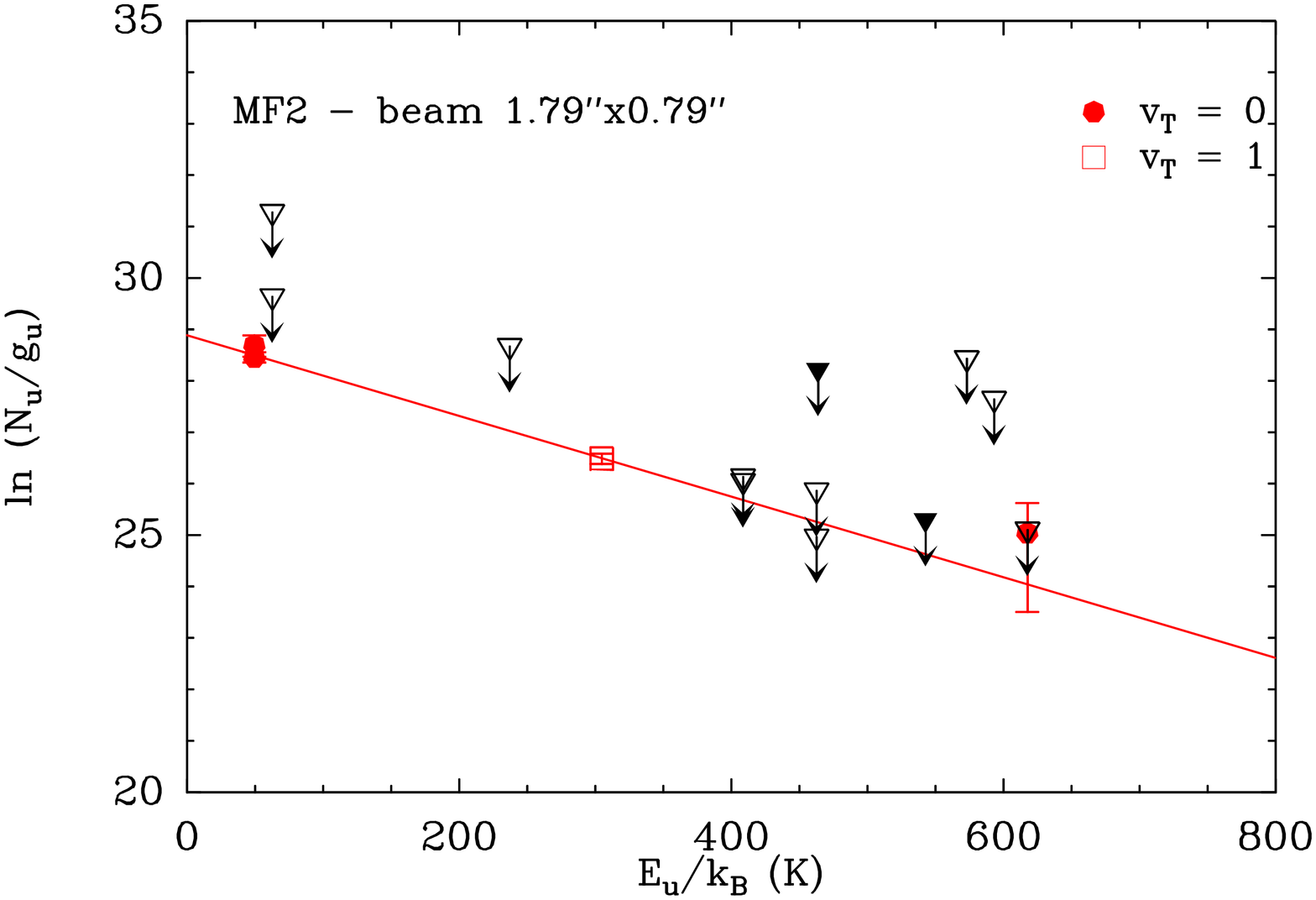}
      \includegraphics[width=5.8cm,angle=270]{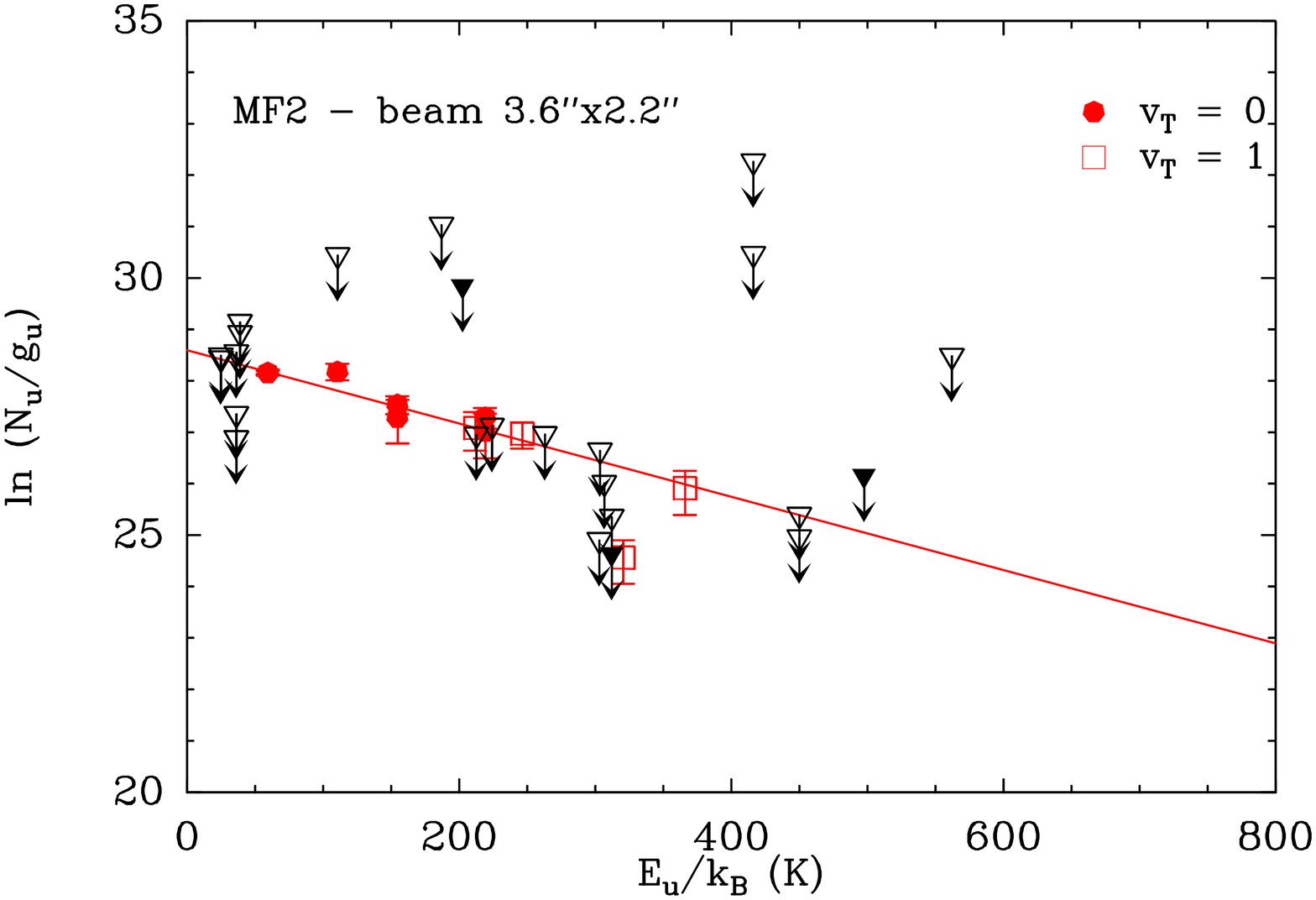}
   \caption{Rotational diagrams toward the MF2 peak based only on data observed with a beam of 1.79$\arcsec$ $\times$ 0.79$\arcsec$ (top) and a beam of 3.63$\arcsec$  $\times$ 2.26$\arcsec$ (bottom). Dots and squares with error bars mark detected and partially blended lines in the fundamental and first excited states v$\rm_{t}$ = 0 and v$\rm_{t}$ = 1, respectively, filled triangles mark undetected lines and open triangles blended lines. The red line is the fit according to the method described in Sect. \ref{sec:temperatures}. The derived rotational temperatures are 128$\pm$9~K and 140$\pm$14~K, respectively.}
              \label{Fig.DRc2pb}%
    \end{figure} 
%------------------- 
%
%------------------- 
%FIGURE 12
%-------------------     
       \begin{figure}[!h]
  \centering
   \includegraphics[width=5.8cm,angle=270]{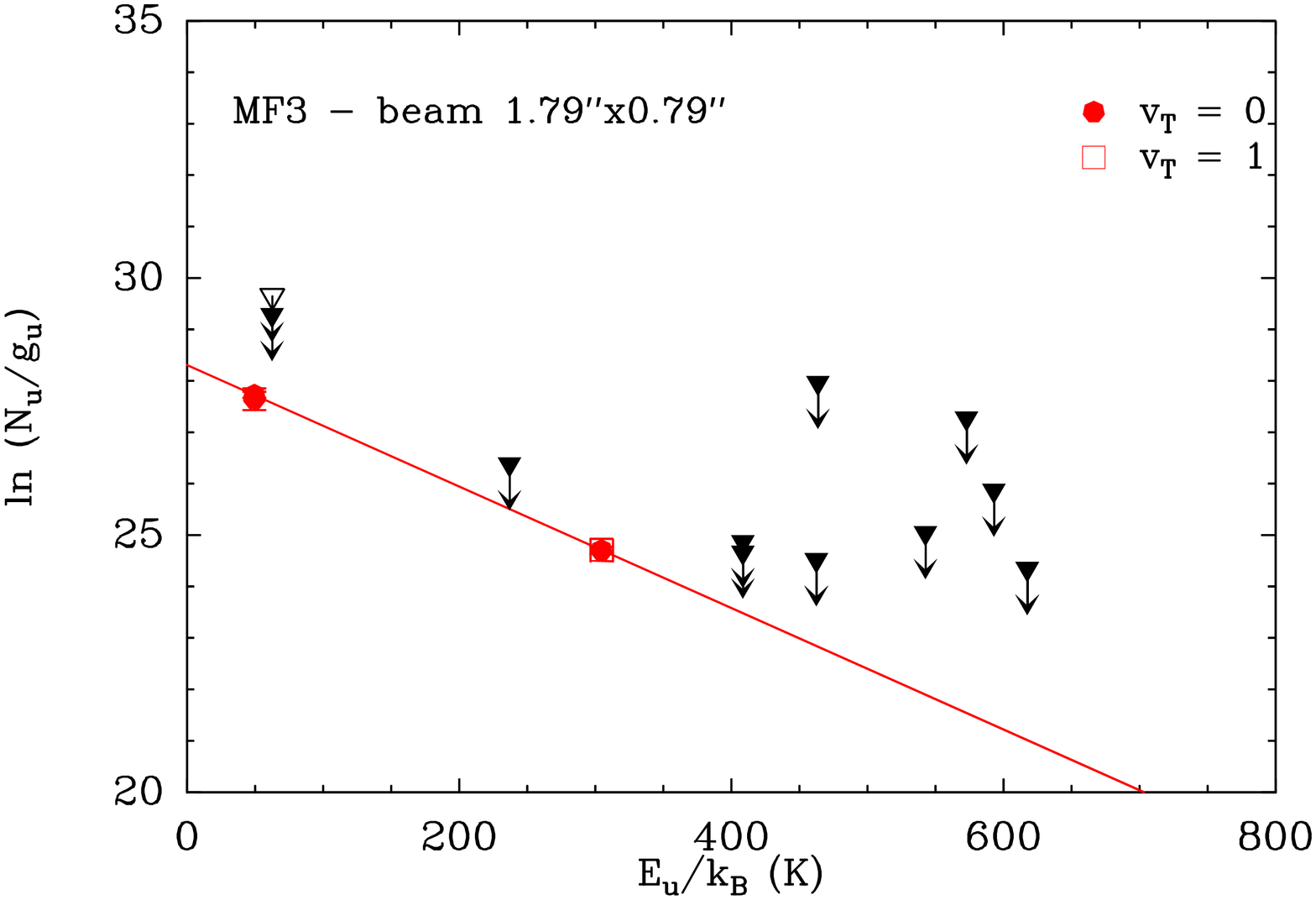}
      \includegraphics[width=5.8cm,angle=270]{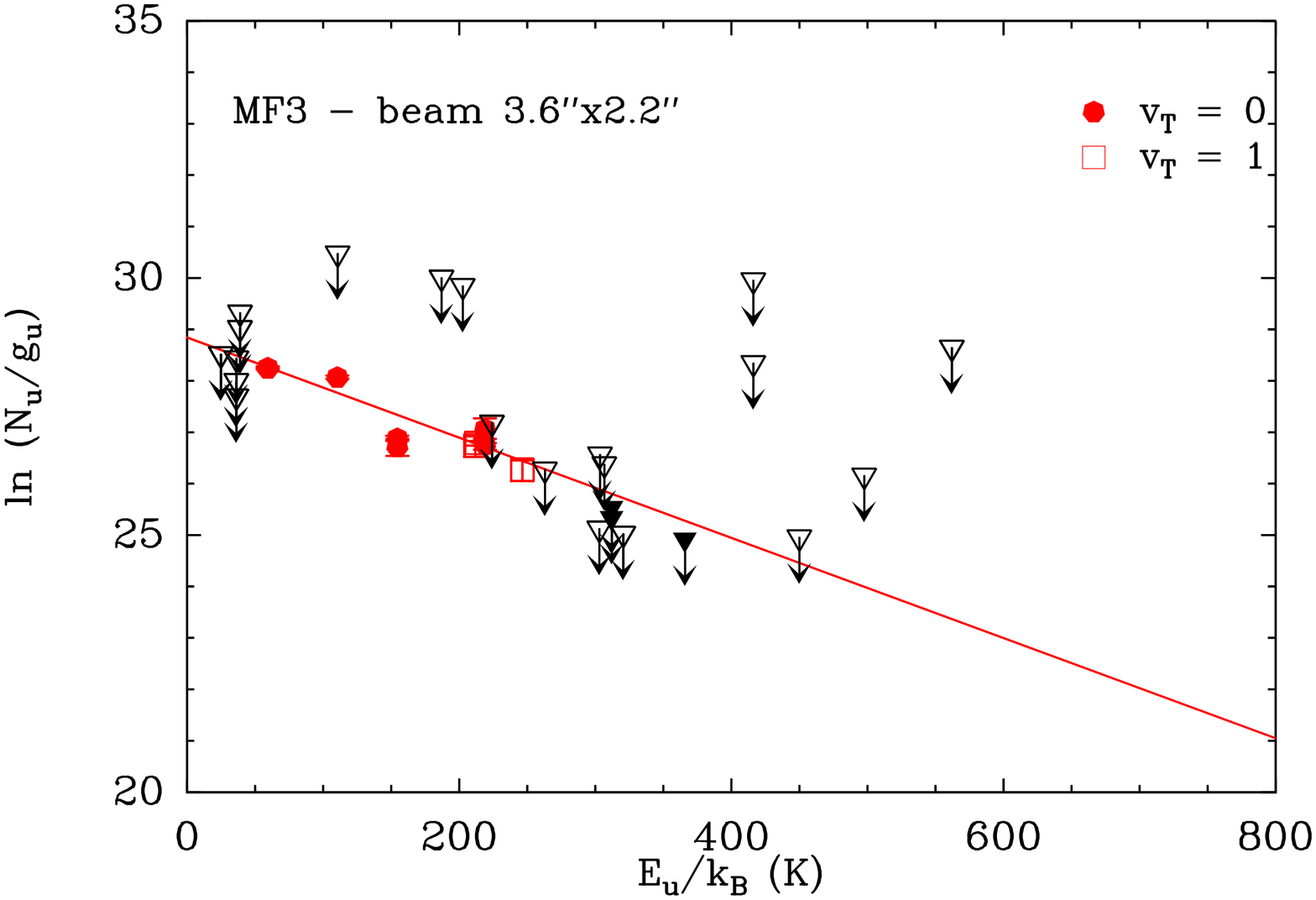}
   \caption{Rotational diagrams toward the MF3 peak based only on data observed with a  beam of 1.79$\arcsec$ $\times$ 0.79$\arcsec$ (top) and a beam of 3.63$\arcsec$  $\times$  2.26$\arcsec$ (bottom). Dots and squares with error bars mark detected and partially blended lines in the fundamental and first excited states v$\rm_{t}$ = 0 and v$\rm_{t}$ = 1, respectively, filled triangles mark undetected lines and open triangles blended lines. The red line is the fit according to the method described in Sect. \ref{sec:temperatures}. The derived rotational temperatures are 85$\pm$3~K and 103$\pm$3~K, respectively.}
              \label{Fig.DRc3pb}%
    \end{figure}
%------------------- 
%
%------------------- 
%FIGURE 13
%------------------- 
   \begin{figure}[!h]
  \centering
\includegraphics[width=5.8cm,angle=270]{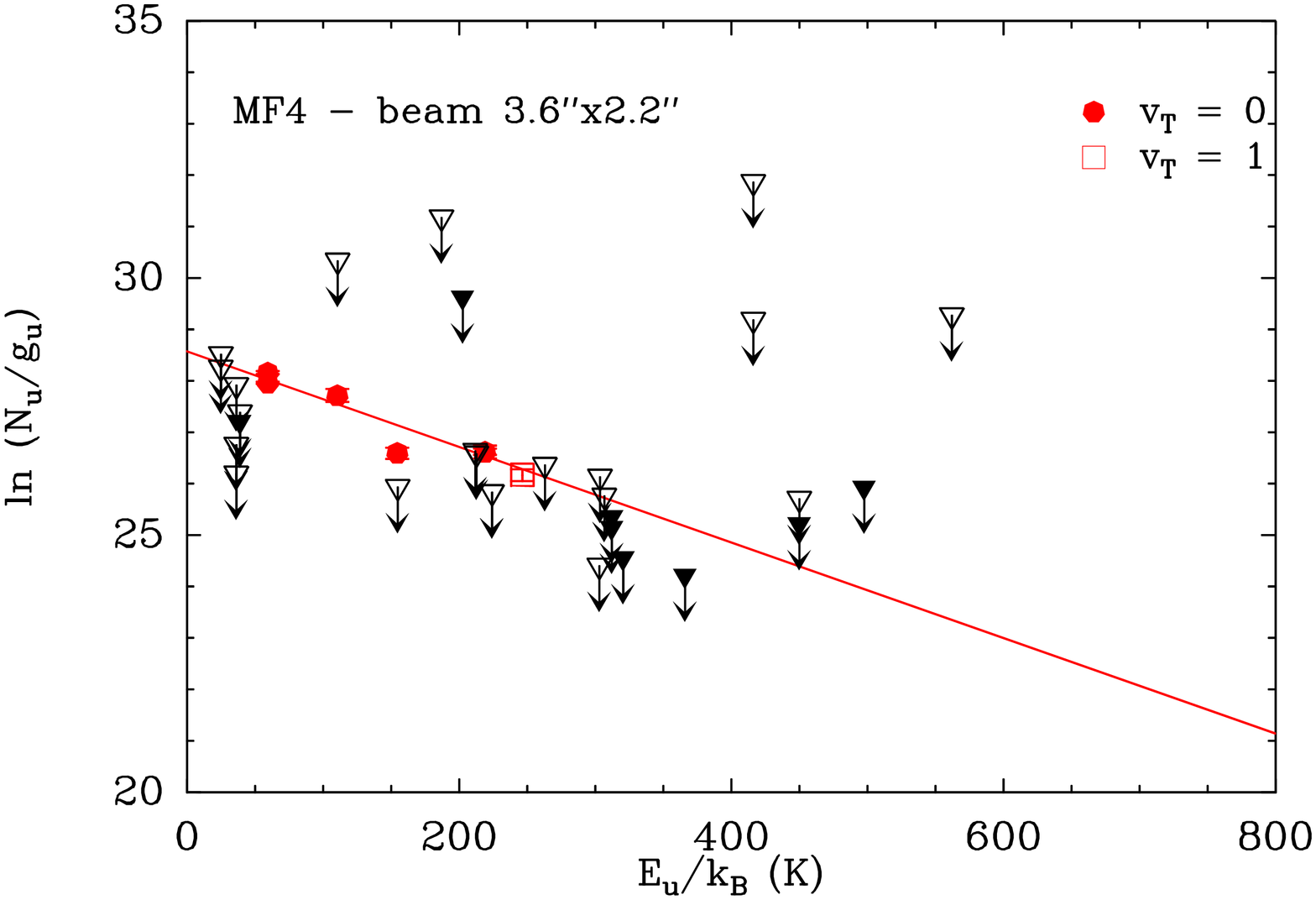}
   \caption{Rotational diagram toward the MF4 peak based only on data observed with a synthesis beam of 3.63$\arcsec$ $\times$ 2.26$\arcsec$. Dots and squares with error bars mark detected and partially blended lines in the fundamental and first excited states v$\rm_{t}$ = 0 and v$\rm_{t}$ = 1, respectively, filled triangles mark undetected lines and open triangles blended lines. The red line is the fit according to the method described in Sect. \ref{sec:temperatures}. The derived rotational temperature is 108$\pm$4~K.}
              \label{Fig.DRc4pb}%
    \end{figure}
%------------------- 
%
%------------------- 
%FIGURE 14
%-------------------     
   \begin{figure}[h!]
  \centering
   \includegraphics[width=5.8cm,angle=270]{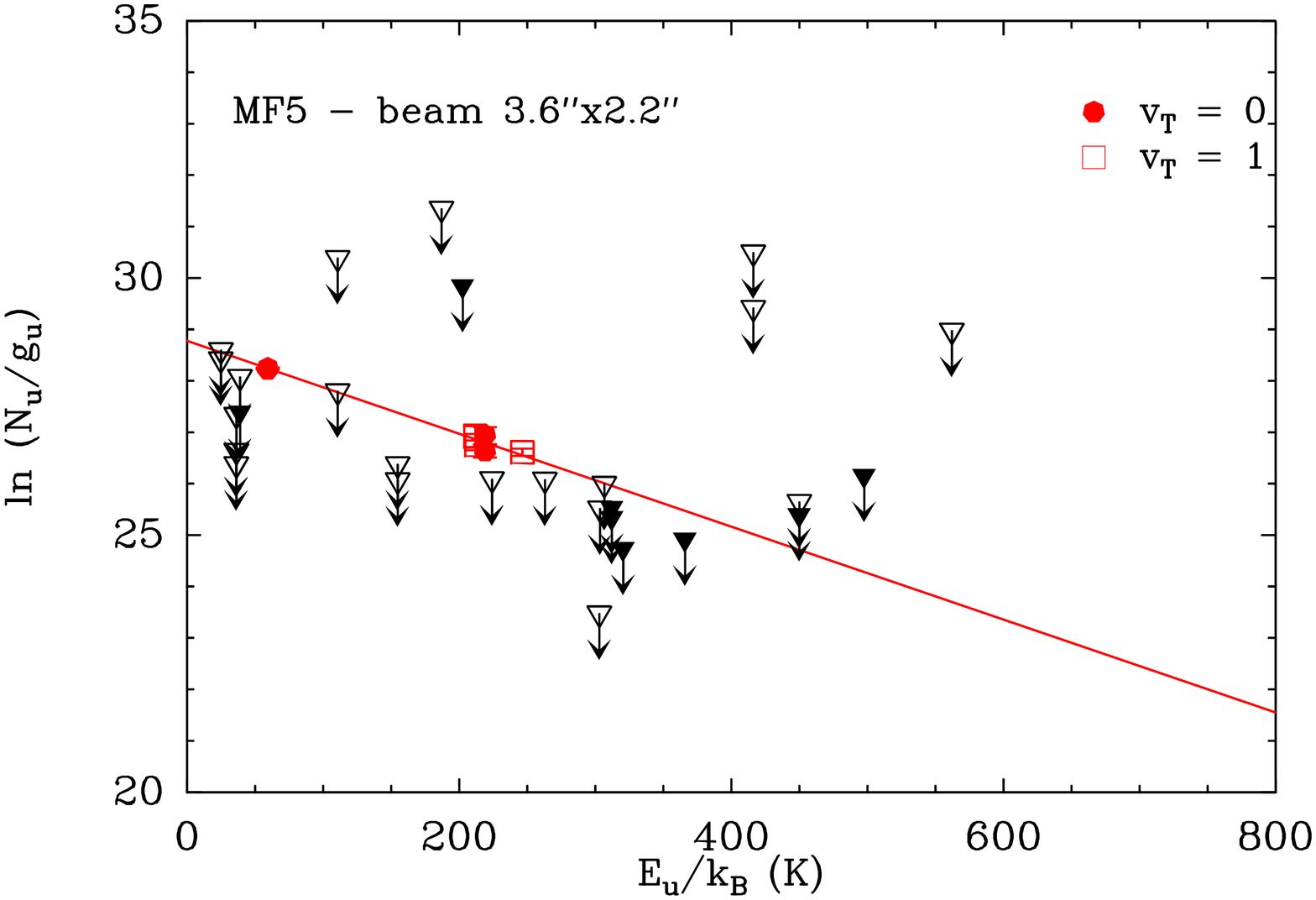}
   \caption{Rotational diagrams toward the MF5 peak based only on data observed with a synthesis beam of 3.63$\arcsec$ $\times$ 2.26$\arcsec$. Dots and squares with error bars mark detected and partially blended lines in the fundamental and first excited states v$\rm_{t}$ = 0 and v$\rm_{t}$ = 1, respectively,  filled triangles mark undetected lines and open triangles blended lines. The red line is the fit according to the method described in Sect. \ref{sec:temperatures}. The derived rotational temperature T is 111$\pm$4~K.}
              \label{Fig.DRc5pb}%
    \end{figure}
%------------------- 

%------------------------------------------------
%Rotational diagram at the MF2 peak
%------------------------------------------------
\paragraph{\textbf{Rotational diagram at the MF2 peak} \\}

Both rotational diagrams, for  angular resolutions of 1.79$\arcsec$ $\times$ 0.79$\arcsec$ and 3.63$\arcsec$ $\times$ 2.26$\arcsec$ (see sets 9 and 3, 7, 8, 10, 11, 12 in Table \ref{Table.dataset_parameters}, respectively), give comparable rotational temperatures, within the uncertainties. We obtain 128$\pm$9~K and 140$\pm$14~K, respectively (see Fig.\ref{Fig.DRc2pb}). The derived methyl formate column densities are given in Table \ref{Table.MF2345-mainresults}.

%------------------------------------------------
%Rotational diagram at the MF3 peak
%------------------------------------------------
\paragraph{\textbf{Rotational diagram at the MF3 peak} \\}

The rotational temperature derived at a higher resolution is smaller (85$\pm$3~K) than that at a lower resolution (103$\pm$3~K) (see Fig.\ref{Fig.DRc3pb}), suggesting external heating like for the MF1 peak. The derived methyl formate column densities are given in Table \ref{Table.MF2345-mainresults}.

%------------------- 
%TABLE 5
%------------------- 
\begin{table*}
\begin{minipage}[t]{16cm}
\caption{HCOOCH$\rm_{3}$ beam averaged column densities derived towards the emission peaks MF1 to MF5 for two angular resolutions.}            
\label{Table.MF2345-mainresults}      
\centering        
\renewcommand{\footnoterule}{}
\begin{tabular}{c c c | c  c | c}
\hline\hline       
Emission peaks & v\footnote{Towards MF1 and MF4 two different velocity components are identified but separate rotational diagrams could be made only for MF1 (see text) -- for the other sources, velocity is a mean value of detected lines (see tables \ref{MF2}, \ref{MF3} and \ref{MF5} for details).} & \multicolumn{2}{c}{T$\rm_{rot}$\footnote{Error bars reflect only the  uncertainties in the gaussian fit of the lines}.} & \multicolumn{2}{c}{N$\rm_{HCOOCH_{3}}$$^{b}$} \\
 & (km~s$^{-1}$)& \multicolumn{2}{c}{(K)} &  \multicolumn{2}{c}{(10$^{16}~$cm$^{-2}$)}  \\
 &  & (1.79$\arcsec$ $\times$ 0.79$\arcsec$) & (3.63$\arcsec$ $\times$  2.26$\arcsec$) & (1.79$\arcsec$ $\times$ 0.79$\arcsec$) & (3.63$\arcsec$ $\times$  2.26$\arcsec$) \\
 \hline 
MF1&7.6 &79$\pm$2 &119$\pm$10&16.0$\pm$1.0 &3.0$\pm$0.1 \\
MF1 & 9.2 & 112$\pm$50& 168$\pm$30& 6.8$\pm$2.1&  1.6$\pm$0.1 \\  
\hline
MF2 & 7.7  &128$\pm$9 &  140$\pm$14 & 16.0$\pm$2.0 & 7.0$\pm$0.4\\
MF3 & 7.7 & 85$\pm$3 & 103$\pm$3 & 4.6$\pm$0.3& 5.5$\pm$0.1\\
MF4 & 8.2 & -- & 108$\pm$4 & -- & 4.5$\pm$0.2\\ 
MF5 & 7.8 & -- &111$\pm$4 & -- & 5.8$\pm$0.1\\ 
\hline
\hline                  
\end{tabular}
\end{minipage}
\end{table*}
%------------------- 

%------------------------------------------------
%Rotational diagram at the MF4 peak
%------------------------------------------------ 
 \paragraph{\textbf{Rotational diagram at the MF4 peak} \\}   
 
At 223~GHz, two velocity components are present as for the MF1 peak. Though two different upper energies are available, no rotational diagram at the resolution of 1.79$\arcsec$ $\times$ 0.79$\arcsec$ has been made because of the important error bars in the velocity decomposition.
Fig.\ref{Fig.DRc4pb} shows the rotational diagram at a 3.63$\arcsec$ $\times$ 2.26$\arcsec$ resolution. The derived gas temperature and column density are given in Table \ref{Table.MF2345-mainresults}.

%------------------------------------------------
%Rotational diagram at the MF5 peak
%------------------------------------------------
 \paragraph{\textbf{Rotational diagram at the MF5 peak} \\}  
  
As for the MF4 peak and from the same above-mentioned argument no rotational diagram has been made at the resolution of 1.79$\arcsec$ $\times$ 0.79$\arcsec$. 
Fig.\ref{Fig.DRc5pb} shows the rotational diagram at a 3.63$\arcsec$ $\times$ 2.26$\arcsec$ resolution. The derived gas temperature and column density are given in Table \ref{Table.MF2345-mainresults}. These values are close to those found at the nearby MF4 position.\\

The different temperatures obtained at different positions (MF1 to MF5) are most likely due to different physical conditions (see Sect. \ref{sec:discussion}).
%
%------------------- 
%FIGURE 15
%------------------- 
 \begin{figure}[h!]
  \centering
\includegraphics[width=8.cm]{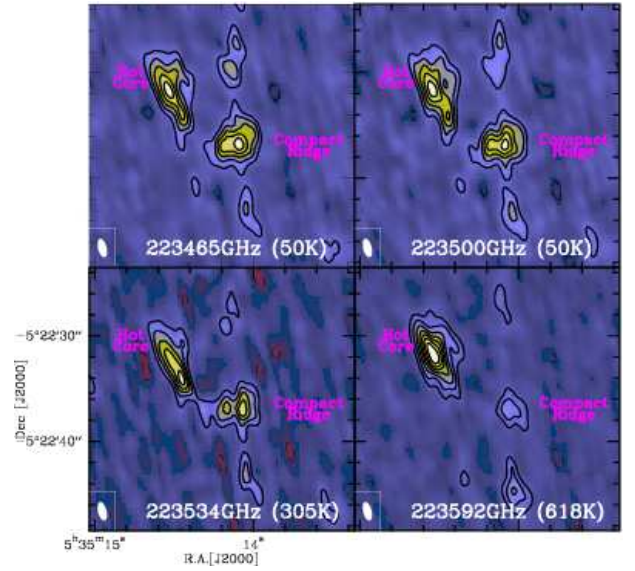}
   \caption{HCOOCH$\rm_{3}$ intensity maps integrated in velocity between 6 and 9~km~s$^{-1}$. The line frequency and the upper state energy are indicated on each plot. The methyl formate emission is stronger towards MF1 than MF2 for low upper energy transitions, while the opposite is verified for high upper state energies.}
              \label{map-eu}%
    \end{figure}
%------------------- 
%
When we compare the methyl formate emission from different energy levels (see Fig.\ref{map-eu}), it is noticeable that for low upper state energies, the emission at the MF1 position is much stronger than at MF2. The emission at MF2 becomes stronger at higher energy levels.

%-----------------------------------------------------------------------------------------------------------------------------
%------------------------COMPARISON WITH THE OTHER C2H4O2 ISOMERS-------------------
%-----------------------------------------------------------------------------------------------------------------------------

\section{Comparison with the other C$\rm_{2}$H$\rm_{4}$O$\rm_{2}$ isomers} 
\label{sec:isomers}

Methyl formate (HCOOCH$\rm_{3}$), acetic acid (CH$\rm_{3}$COOH) and glycolaldehyde (CH$\rm_{2}$OHCHO) are three isomers among which acetic acid is the most stable. In a recent paper,  \citet{Lattelais:2009} have argued that the abundance ratio of isomers could be linked to their relative stability, except for some species as methyl formate. Indeed whereas methyl formate is known to be abundant in many molecular cores, glycolaldehyde has only been detected toward Sgr B2 \citep[e.g.] []{ Hollis:2000, Halfen:2006} and acetic acid is barely detected in hot cores \citep[e.g.] []{ Shiao:2010}.

We have searched for acetic acid and glycolaldehyde in our data sets (from upper energy levels of 18~K up to 452~K and of 25~K up to 646~K, respectively), using \citet{Ilyushin:2008} for the acetic acid frequencies, but we detect neither of them.
Assuming the rotational temperatures derived from our HCOOCH$\rm_{3}$ study, we have calculated upper limits for the column densities of these isomers at the MF1 and MF2 positions (Table \ref{spec-ana-raies}). We find that the abundance of acetic acid is at least 50 times less than the methyl formate one and the abundance of glycolaldehyde at least 200-500 times less.

%------------------- 
%TABLE 6
%------------------- 
\begin{table}[h!]
\begin{minipage}[t]{8.5cm}
\renewcommand{\footnoterule}{}  
\caption{Upper limits of column densities derived for acetic acid and glycolaldehyde from our PdBI data sets.}             
\label{spec-ana-raies}     
\centering        
\begin{tabular}{c c c c}
\hline\hline       
Emission Peak  &  Temperature & Beam &  N  \\
 &  (K) & ($\arcsec$ $\times$ $\arcsec$)&  (cm$^{-2}$)  \\
\hline          
\multicolumn{4}{c}{CH$\rm_{2}$OHCHO }  \\
MF1 & 78 & 1.79 $\times$ 0.79 & $\le$2.4$\times$10$^{14}$ \\
MF1 & 120& 3.79 $\times$ 1.99 & $\le$2.8$\times$10$^{14}$ \\                  
MF2 & 128&1.79 $\times$ 0.79 & --\footnote{All the glycolaldehyde lines are blended, so that the upper limit is not significant.} \\   
MF2 & 140& 3.79 $\times$ 1.99& $\le$3.5$\times$10$^{14}$ \\   
\hline
\multicolumn{4}{c}{CH$\rm_{3}$COOH }  \\
MF1 & 78 & 1.79 $\times$ 0.79 &$\le$1.8$\times$10$^{15}$ \\
MF1 & 120&3.79 $\times$ 1.99 &$\le$1.2$\times$10$^{15}$ \\                  
MF2 & 128&1.79 $\times$ 0.79 &$\le$2.6$\times$10$^{15}$ \\   
MF2 & 140&  3.79 $\times$ 1.99 &$\le$1.4$\times$10$^{15}$ \\   
\hline
\end{tabular}
\end{minipage}
\end{table}
%------------------- 

%-----------------------------------------------------------------------------------------------------------------------------
%------------------DUST EMISSION AND CHARACTERISTICS OF MAIN CLUMPS--------------
%-----------------------------------------------------------------------------------------------------------------------------

\section{Dust emission and characteristics of main clumps}
\label{sec:dust}

\begin{figure}[h!]
  \centering
   \includegraphics[width=8.cm, angle=270]{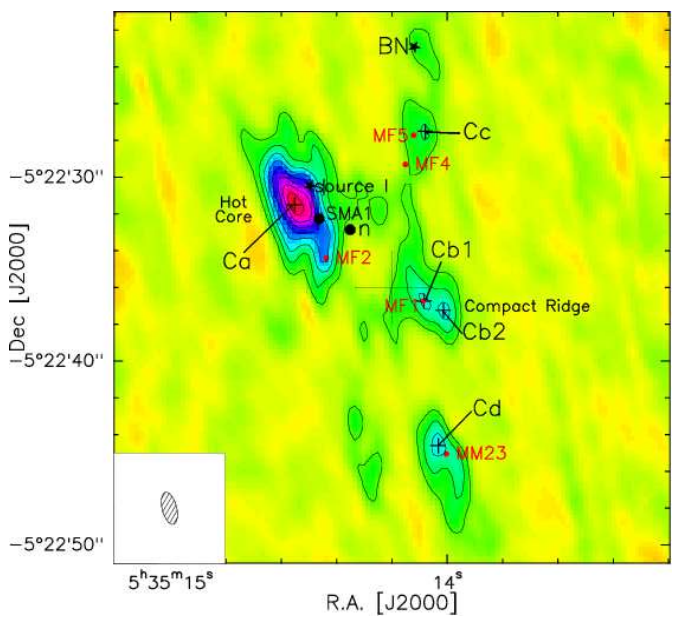}
   \caption{Continuum emission map obtained with the IRAM Plateau de Bure Interferometer at 223.67~GHz. The level step is 60~mJy~beam$^{-1}$ (3.5~$\sigma$)  for a beam size of 1.8$\arcsec$ $\times$ 0.8$\arcsec$. Dark crosses label the 4 different continuum components while points label the main molecular HCOOCH$\rm_{3}$ emission peaks.}
              \label{Fig.dust223GHz}%
    \end{figure}

Fig.\ref{Fig.continuum} shows that the continuum emission in the 101 to 225~GHz range is strong, extended and, from the highest frequency data, clumpy in places.
In this Section we use the high spatial resolution and high signal-to-noise ratio achieved at 223~GHz (see Fig.\ref{Fig.dust223GHz}) to estimate, for the main clumps of dust emission, their masses, the mean projected $\rm H\rm_{2}$ column density and the H$\rm_{2}$ volume density. Combining our line results with our N$\rm_{H\rm_{2}}$ estimates we further derive the relative abundance of the methyl formate molecule.

There are 4 main dust clumps in our 223~GHz map (Fig. \ref{Fig.dust223GHz}): two, Cb and Ca, lie in the Compact Ridge and the Hot Core near the methyl formate peaks MF1 and MF2, respectively; Cc, in the north lies near MF4 and MF5; another clump, Cd, lies in the south near MM23. Each clump exhibits a complex spatial structure which is not fully revealed even with our 1.79$\arcsec$ $\times$ 0.79$\arcsec$ spatial resolution. In the Compact Ridge for instance, we have identified toward MF1 two clear continuum maxima (labelled Cb$\rm_{1}$ and Cb$\rm_{2}$ in Figure \ref{Fig.dust223GHz}) which we consider, however, as one \textquotedblleft \textit{single clump}\textquotedblright \ (Cb). Ca, Cc and Cd are most probably not single entities as well, but we will consider them as roughly Gaussian clumps in our subsequent analysis. We first give below details on the equations used to derive the mean clump properties.

%------------------- 
%TABLE 7
%------------------- 
\begin{table*}
\begin{minipage}[t]{17cm}
\caption{Beam averaged density and methyl formate relative abundance at 223~GHz continuum and molecular peaks.}
\label{Table.continuum-clumps-peak}
\centering
\renewcommand{\footnoterule}{}  
\begin{tabular}{lcccccccl}
\hline\hline
Continuum & R.A (J2000) & Dec (J2000) & Adopted T$\rm_{d}$ & s$\rm_{\nu}$\footnote{flux density per beam or brightness temperature.}& $\langle$N$\rm_{H\rm_{2}}$$\rangle$\footnote{estimated in a beam size of 1.79$\arcsec$ $\times$ 0.79$\arcsec$.} & $\langle$n$\rm_{H\rm_{2}}$$\rangle$\footnote{estimated in a beam size of 1.79$\arcsec$ $\times$ 0.79$\arcsec$ at a distance of 414~pc \citep{Menten:2007}.}  & \multicolumn{2}{l}{$\langle$N$\rm_{ HCOOCH_{3}}$/N$\rm_{ H\rm_{2}}$$\rangle$\footnote{estimated at continuum and associated molecular peaks. The methyl formate column densities are derived from this work.}} \\
Source  & 05$^{h}$35$^{m}$ &-05$\degr$22$\arcmin$&  (K) & (Jy~beam$^{-1}$) & (10$^{24}~$cm$^{-2}$) & (10$^{8}~$cm$^{-3}$)   & \multicolumn{2}{l}{(10$^{-8}$)}  \\
\hline\hline
Ca & 14$\fs$56  & 31$\farcs$50 & 200 & 0.49 & 5.4 & 7.3 & $\le$0.6 & at continuum peak  \\
& & & 130 & 0.18 & 3.1 & 4.2 &  5.2 &at MF2 molecular peak  \\
\hline
Cb & Cb$\rm_{1}$ 14$\fs$08  & 36$\farcs$73 &  80   & 0.18 & 5.0 & 6.8 & 3.1 &at continuum peak  \\
& Cb$\rm_{2}$ 14$\fs$01  & 37$\farcs$24 &     &   & &  &   & \\
&  &    & 80   & 0.18 & 5.0 & 6.8 & 3.1 &at MF1 molecular peak  \\
\hline
Cc    & 14$\fs$08  & 27$\farcs$50 & 110 & 0.12 & 2.4 & 3.3& $\le$0.3 &at continuum peak  \\
& & & 110 & 0.08 & 1.6 & 2.2 &  $\le$0.2& at MF4 molecular peak\\
& & & 110 & 0.08 & 1.6 & 2.2 &  $\le$0.5& at MF5 molecular peak \\
\hline
Cd    & 14$\fs$03  & 44$\farcs$60   & 105 & 0.17 & 3.6 & 4.9 & $\le$0.1& at continuum peak\\
	& & & 105 & 0.17 & 3.6 & 4.9 & $\le$0.1 &at MM23 molecular peak \\
\hline
\end{tabular}
\end{minipage}
\end{table*}
%------------------- 

%------------------- 
%FIGURE 16
%------------------- 
%------------------- 

The flux density,  S$\rm_{\nu}$, from a dust cloud at frequency $\nu$ is given in the optically thin case by:

%------------------- 
%EQ 3
%------------------- 
\begin{equation}
\rm
\label{smu}
    \begin{array}{l}
       S_{\nu}  \approx
                 \tau B_{\nu}(T_{d}) \Omega \\
    \end{array}
    \end{equation}
%------------------- 

where $\tau$ is the dust opacity at frequency $\nu$, B$\rm_{\nu}$(T$\rm_{d})$ the Planck function, T$\rm_{d}$ the dust temperature and $\Omega$ the dust cloud solid angle. Estimating the dust column density and mass from the observables S$\rm_{\nu}$ and $\Omega$ requires a dust opacity model and to know the variation of the extinction efficiency with frequency. This variation strongly depends on the wavelength range, the dust composition and its temperature. \citet{Mathis:1977} suggested that the visible and UV interstellar extinction can be reproduced with a mixture of graphite and silicate grains and their model was extended to the infrared by \citet{Draine:1984}. \citet{Draine:1984} showed that a mixture of silicate and graphite explains rather well the NIR and FIR observations and they showed that the FIR opacity varies as $\lambda^{-2}$ where $\lambda$ is the wavelength. Of particular interest here is their estimate of the dust opacity at 125~$\mu$m as a function of the visual absorption or the projected H gas density N$\rm_{H}$.
They obtain $\tau=4.6\times10^{-25}$N$\rm_{H}$(cm$^{-2}$) which we extrapolate to the mm-wavelength regime, assuming that the extinction varies as $\lambda^{-2}$,

%------------------- 
%EQ 4
%------------------- 
\begin{equation}
\label{tau1}
    \rm
       \tau =
1.4\times10^{-26}\frac{N_{H_{2}}(cm^{-2})}{\lambda^{2}(mm^{2})} \\
    \end{equation}
%------------------- 
%
where we further assume the relation holds for molecular hydrogen 
with N$\rm_{H}$ $\approx$ 2N$\rm_{H_{2}}$. 
We are aware that the \citet{Draine:1984} data are rather typical of diffuse interstellar clouds and that dust coagulation as described by \citet{Ossenkopf:1994} for protostellar cores could be more appropriate. For an assumed
central luminosity and outer envelope radius, the dust mass derived from a 1--D fit of the derived fluxes to the observed spectral energy distribution of the massive protostellar object W3IRS5 is 3 times lower with the \citet{Ossenkopf:1994} opacity profile than with the \citet{Draine:1984} profile (Luis Chavarria, private communication). However, the \citet{Ossenkopf:1994} opacities may be uncertain by a factor of at least 2 due to changes in the physical conditions and,
given the additional dust to mass ratio uncertainty, derivation of the total gas mass in protostellar cores remains uncertain. 

In this work we adopt a simple approach and combine equations \ref{smu} and \ref{tau1} to derive in a straightforward manner the H$_{2}$ projected density and hence the total mass.
Using the observed peak flux density per synthesized beam, s$\rm_{\nu}$ in Jy~beam$^{-1}$ (or source brightness temperature in K), and the Rayleigh-Jeans approximation, we derive, for  the mean observing frequency of 223.65~GHz (1~Jy~beam$^{-1}$ = 17.3~K, for a HPBW of 1.79$\arcsec$ $\times$ 0.79$\arcsec$), the beam averaged projected density from
%
%------------------- 
%EQ 5
%------------------- 
\begin{equation}
\rm
\label{nh2}
        \langle N_{H_{2}}\rangle = 2.22\times10^{27}\frac{s_{\nu}(Jy~beam^{-1})}{T_{d}(K)} \\
     \end{equation}
%------------------- 

Assuming, for simplicity, that the dust clump is Gaussian with peak flux density s$\rm_{\nu}$ and size $\theta_{1} \times \theta_{2}$ determined from the 2-D Gaussian fits to the source half peak flux density, we derive the total gas mass in the entire Gaussian clump from the following equation (taking a mean molecular weight of 2.33 and adopting 414~pc as the Orion KL distance):

%------------------- 
%EQ 6
%------------------- 
\begin{equation}
\rm
\label{mgas}
           M_{gas}(M_{\sun}) \approx
188.7\times\frac{s_{\nu}(Jy~beam^{-1})}{T_{d}(K)} \times\theta_{1}(\arcsec)\times\theta_{2}(\arcsec)\\
     \end{equation}
%------------------- 

We use this equation to obtain the total gas mass present in our clumps (see Table \ref{Table.continuum-half-clumps}).

All relevant gas clump parameters are gathered in Tables \ref{Table.continuum-clumps-peak} and \ref{Table.continuum-half-clumps}. They include the positions and names of the 4 main dust clumps as well as the adopted dust temperature (see details below), the peak flux in Jy~beam$^{-1}$, the projected and volumic densities and, see last 4 columns in Table \ref{Table.continuum-half-clumps}, the clump sizes, total flux density and estimated total gas mass within the identified clumps. The methyl formate relative abundance at continuum and molecular peaks is also given in the last column of Table \ref{Table.continuum-clumps-peak}.
%
%------------------- 
%TABLE 8
%------------------- 
\begin{table*}
\begin{minipage}[t]{17cm}
\caption{Dust clump half sizes, total flux density and total gas mass within the dust clumps.}
\label{Table.continuum-half-clumps}
\centering
\renewcommand{\footnoterule}{} 
\begin{tabular}{lcccccc}
\hline\hline
Continuum & Nearby molecular  & Adopted T$\rm_{d}$ &  Adopted clump surface\footnote{taken at half size}  & M$\rm_{gas}$ & S$\rm_{total}$ \\
 source & peak & (K) &  ($\arcsec$ $\times$ $\arcsec$)  &   (M$\rm_{\sun}$) & (Jy) \\
\hline
Ca & MF2 & 200 & 5.0 $\times$ 2.5 & 5.8 & 3.9  \\
Cb\footnote{Overall structure including C$\rm_{b1}$ and C$\rm_{b2}$.} & MF1 & 80 &  4.0 $\times$ 2.5 & 4.3 & 1 \\
Cc & MF4, MF5 & 110 &  2.6 $\times$ 1.5 & 0.8 & 0.45 \\
Cd & MM23 & 105 & 3.0 $\times$ 1.5 & 1.4 & 0.8  \\
\hline
\end{tabular}
\end{minipage}
\end{table*}
%------------------- 

The dust temperature is a crucial parameter in all of the above calculations. In order to estimate T$\rm_{d}$ we assume that : (a) the gas-dust relaxation time is short \citep{Draine:2010,Tielens:2005}, and (b) the gas temperature is close to the rotational temperature T$\rm_{r}$ deduced from our methyl formate data. The gas-dust relaxation time is roughly given by 4$\times10^{16}$~s / T$\rm_{K}^{0.5}$n$\rm_{H\rm_{2}}$(cm$^{-3}$) which is $\la$ 130~years for n$\rm_{H\rm_{2}}\ga10^{6}$~cm$^{-3}$ and T$\rm_{K}$ around 100~K; therefore, dust thermalization should be well verified in Orion. T$\rm_{r}$ is rather well determined from our LTE multi-line analysis. Hence, it is reasonable to assume that both (a) and (b) are verified, thus T$\rm_{d}$ $\approx$ T$\rm_{r}$. Another complication arises because T$\rm_{d}$ is not uniform throughout Orion as demonstrated by our continum maps, and our continuum and line peaks do not match exactly. The values of T$\rm_{d}$ adopted here are given in Tables \ref{Table.continuum-clumps-peak} and \ref{Table.continuum-half-clumps}. They are taken from the rotational temperature obtained for the main gas peaks MF1, MF2, MF4, MF5 and MM23. There is also some uncertainty at the continuum peaks near both MF2 and MM23 because only 3 and 2 methyl formate lines were unanmbiguously detected in these directions (while many lines are present at the molecular peaks); in this case we used our best judgement to estimate T$\rm_{r}$ and thus T$\rm_{d}$.

The beam averaged projected density N$\rm_{H\rm_{2}}$ lies in the range 2 to 5 $\times$10$^{24}$~cm$^{-2}$ corresponding to very large visual extinctions. The latter density is used to derive the volumic H$\rm_{2}$ density across the synthesized beam projected at the 414~pc distance of Orion-KL; it lies in the range 2 to 7 $\times$10$^{8}$~cm$^{-3}$.
Our derived densities are much higher than those found in the literature \citep[e.g.][]{Irvine:1987,Persson:2007,Mezger:1990}.
However, our spatial resolution is higher and the densities are similar if we take into account the beam dilution. Moreover, we find similar H$\rm_{2}$ density values when we use the continuum map of \citet{Beuther:2004} and use their source and dust properties hypothesis.

Our individual clump masses vary from about 1 to 5 solar masses. We stress that these individual masses cannot be highly accurate because of various reasons. First, any T$\rm_{d}$ uncertainty is directly translated into a mass uncertainty in our mass equations. Second, an uncertainty in the opacity wavelength dependence also results in a mass uncertainty; at 1.3~mm there is a 30\% mass change when the extinction varies from a $\lambda^{-2}$ dependence to $\lambda^{-1}$. Third, as explained earlier our clumps might not be Gaussian and our observed spatial structure might be even more clumpy than reported here. 
Another way to estimate our mass uncertainty is to derive the total gas mass for the entire clump from the equation, $\rm M_{gas}(M_{\sun}) \approx 270\times\frac{S_{\nu}(Jy)}{T_{d}(K)}$, where S$\rm_{\nu}$ is the flux density measured in Jy in our calibrated maps (see last column in Table \ref{Table.continuum-half-clumps}). The derived masses differ from those obtained with equation \ref{mgas} by about 9$\%$ to 30$\%$. Despite these uncertainties, we estimate that the total mass derived for all clumps identified here, about 12 solar masses, is meaningful. This mass reservoir could well be used in future star forming activity.

In the last column of Table \ref{Table.continuum-clumps-peak} we also give the methyl formate relative abundance averaged over the synthesized beam in the direction of the 223~GHz dust emission peaks and of the molecular peaks. The relative abundances lie in the range $\le$0.1$\times$10$^{-8}$ to 5$\times$10$^{-8}$ and show variations from one peak to another. Toward Cc and Cd, there are not enough detected molecular lines to estimate a precise methyl formate column density and we only give upper limits of the relative abundance. Toward Cb there is no difference in abundance between the molecular and continuum peaks, while there is a marked difference in the relative abundances measured toward the Hot Core (Ca) and the associated Hot Core-SW molecular peak (MF2). The largest methyl formate relative abundances are observed toward MF2 and MF1.

%-------------------------------------------------------------------------------
%---------COMPARISON WITH PREVIOUS STUDIES-----
%-------------------------------------------------------------------------------

\section{Comparison with previous studies} 

In this section we try to relate our results to some of the most comparable and recent other studies of Orion-KL.
The large number of published works devoted to this region prevents us from being exhaustive. Therefore, we limit ourselves to the high ($<$10$\arcsec$) resolution studies of the continuum emission and methyl formate emission. There is a general agreement on space and velocity distribution and on column densities derived; however, our data show more details due to higher spatial and/or spectral resolution.

Methyl formate temperatures were often not derived  by previous authors from their interferometric studies so we quote also for reference other temperature indicators at high resolution: dust color temperature, and methanol (CH$_3$OH) and methyl cyanid (CH$_3$CN) rotational temperatures. Methanol is interesting because this O-bearing species may be formed on grains \citep[e.g.][]{van-der-Tak:2000} like methyl formate, and, as stated previously it may also be a precursor of methyl formate. It may however suffer opacity problems, and its excitation is often more complex than LTE \citep[see e.g.][]{Menten:1988,Leurini:2004}. Methyl cyanide is a well-known probe of kinetic temperature \citep{Boucher:1980}. But N-bearing species have a different distribution, so that temperatures  derived from these molecules may thus differ from those derived from methyl formate.

Much work has been done on the other hand with single dish telescopes towards Orion-KL, and large surveys allow for the determination of a temperature of the region averaged over the beam, and sometimes for a decomposition into a few spectral features identified from their velocity profiles. We quote previous results on methyl formate and methanol avoiding, however, studies most affected by optical thickness effects. The HCOOCH$_3$ temperatures we derive fall in the 60-220\ K range found in these studies.

%-----------------------------------------------------------
%-----High resolution continuum maps-----
%-----------------------------------------------------------
\subsection{High resolution continuum maps}
Our 1.3~mm continuum map differs from the Hot Core maps obtained with the SMA at 345~GHz and 690~GHz by \citet{Beuther:2004} and \citet{Beuther:2006} because we do not see any emission peak on source I and SMA1. We note that the \citet{Beuther:2004} and \citet{Beuther:2006} maps already show differences between each other. This could be due to the different spatial resolutions and/or to the different \textit{uv} coverages used in these two works. Finally, we note that the SMA 870~$\mu$m dust continuum maps of \citet{Tang:2010} is much similar to our 223~GHz map obtained with a 1.79$\arcsec$ $\times$ 0.79$\arcsec$ resolution. Our map covers a more extended region of the southern Compact Ridge but all other clumps observed by us along the Ridge up to BN are also detected in the 0.8$\arcsec$ $\times$ 0.7$\arcsec$  SMA map. As also suggested by our 1.3~mm map, the 870~$\mu$m dust emission map shows that there are at least two clumps in the Hot Core region. From their polarization study of the Hot Core and Compact Ridge \citet{Tang:2010} suggest that the magnetic field is regulating star formation at large physical scales. Hence, all clumps identified here are probably intimately related with the Orion large scale magnetic field. 

%----------------------------------------------------------------------------------------------------------------
%-----High resolution  studies of methyl formate distribution and temperature-----
%----------------------------------------------------------------------------------------------------------------
\subsection{High resolution  studies of methyl formate distribution and temperature}
Our methyl formate emission peaks MF1, MF2 and MF4-5 \footnote{The MF5 emission peak corresponds to the IRc6 position in \citet{Friedel:2008} and \citet{Beuther:2005}.} are also present in the  interferometric maps of \citet{Friedel:2008} and \citet{Beuther:2005} but the MF3 and MF4 peaks are less obvious in their data. Their maps, like ours, do not show any emission of this molecule towards the Hot Core around 5~km~s$^{-1}$.

The methyl formate observations of \citet{Friedel:2008} made with the CARMA interferometer with  a 2.5$\arcsec$ $\times$ 0.85$\arcsec$ beam size also reveal components at 7.6~km~s$^{-1}$ and 9.3~km~s$^{-1}$ towards the Compact Ridge region\footnote{Note that the Compact Ridge position defined in \citet{Friedel:2008} is not the the same as that used in this study. Their IRc5 position is closer to our Compact Ridge position (1.5$\arcsec$ away).}. However the authors do not separate the two velocity components to derive the temperature and the column density. Their results agree with ours if we do not separate the two HCOOCH$\rm_{3}$ spectral components: T$_{rot}$=101~K and N$\rm_{HCOOCH_{3}}$=1.5$\times$10$^{17}$~cm$^{-2}$. Moreover methyl formate emission towards the Compact Ridge region has also been observed with BIMA by \citet{Liu:2002}  and \citet{Remijan:2003}. Their spatial and spectral resolutions do not allow them to see the two velocity components and they derive a temperature and a column density similar to those found by \citet{Friedel:2008} and to ours when we combine the two components into one.

\citet{Hollis:2003}  also imaged methyl formate at high resolution (2.9$\arcsec$ $\times$ 1.6$\arcsec$) using the VLA at 45~GHz. The very low ($E_u=$~6~K) energy of the upper state of the transition they observed shows that emission from the Compact Ridge is dominant, thus confirming the trend shown in our Fig.\ref{map-eu}.
 Despite their very good spatial resolution, their limited spectral resolution (1.36~km~s$^{-1}$ channel separation) allows them to detect only two broad peaks beside the Compact Ridge, corresponding to our MF2 and MF4-5 peaks.
At the Compact Ridge peak (MF1), again summing up our two velocity components, our column density agrees with their value derived for a temperature of 100~K.

The interferometric map of methyl formate at 2$\arcsec$ $\times$ 1.5$\arcsec$ resolution integrated over velocities obtained with the Owens Valley interferometer by \citet{Blake:1996} shows a reasonable global agreement with ours: Hot Core-SW and MF4-5 (IRc6) peaks are displaced from ours by $<$1$\arcsec$, and their Compact Ridge peak lies 1.5$\arcsec$ North of MF1;  the emission around MF4-5 (IRc6) appears more intense and more extended, with peaks displaced by $\sim1\arcsec$ compared to ours.

%-------------------------------------------------------------------
%-----Other temperature determinations--------
%-------------------------------------------------------------------
\subsection{Other temperature determinations}

%--------------------------------------
%From dust infrared emission
%--------------------------------------
\subsubsection{From dust infrared emission}
Our temperature estimates lie in the range 80-170~K which is close to the range of (color) dust temperatures found by \citet{Wynn-Williams:1984} from their 30~$\mu$m and 20~$\mu$m measurements, but they are lower than the temperatures they derive at shorter (and more optically thick) wavelengths. 
However, the temperature variations observed by \citet{Wynn-Williams:1984} between the Hot Core, the Compact Ridge and IRc6 (MF4-5) are not ordered as in our study. This may be due to remaining opacity effects in the mid-infrared. In a more recent work \citet{Robberto:2005} derived color temperatures in Orion-KL (excepting BN) of 125-140~K from 10 and 20~$\mu$m observations, and higher values  (200-230~K) from silicate features, using a simple \textquotedblleft two-component\textquotedblright \  model for the IR emission.

%-------------------------------------------------------------------------------
%From methanol and methyl cyanide high resolution studies
%-------------------------------------------------------------------------------
\subsubsection{From methanol and methyl cyanide high resolution studies}
A comparison of our results with the temperatures derived from previous interferometric works on methanol (CH$_{3}$OH) by \citet{Beuther:2005} cannot be done toward the Compact Ridge and the Hot Core-SW because, unfortunately but not surprisingly, our two strongest methyl formate peaks correspond to areas where methanol is optically thick. It is only in the direction of MF4-MF5 (IRc6) that we observe some overlap with their temperature map: they find a somewhat higher temperature, around 180-240~K to be compared with our $\sim$ 110~K. Note that methanol emission is more extended than methyl formate and does not sample exactly
the same gas, and methanol may be formed at least in part in the gas phase  \citep[e.g.] []{Plambeck:1988}; note also that a methanol maser has been identified by the latter authors at IRc6.

As N-bearing and O-bearing molecules clearly do not share the same spatial and velocity distribution  (see e.g. their position velocity diagram showing the much broader velocity range of CH$\rm_{3}$CN emission), no simple comparison can be made of our work with the higher temperature values recently obtained by \citet{Wang:2010} using CH$\rm_{3}$CN (190-620~K in the Hot Core and 170-280~K in the Compact Ridge).

Our temperatures and column densities derived from the analysis of the MF1 emission peak suggest an external heating towards the Compact Ridge. It is interesting to mention that  recently \citet{Zapata:2010b} have also found some evidence for external heating towards the Hot Core. These two dense regions of the Orion-KL nebula seem to be associated with the dynamical event which occurred 500 years ago.

%-------------------------------------------------------------------------
%From single dish methyl formate and methanol studies
%-------------------------------------------------------------------------
\subsubsection{From single dish methyl formate and methanol studies }

From a single dish measurement of the optically thin isotopologue $\rm^{13}$CH$\rm_{3}$OH \citet{Blake:1984} derived a rotational temperature of 120~K averaged over their 30$\arcsec$ beam; this is close to the 140~K derived for CH$\rm_{3}$OH by \citet{Johansson:1984} with a similar beam. \citet{Blake:1987}, also in a 30$\arcsec$ beam, have attributed the methanol and methyl formate emissions to the Compact Ridge and derived rotational temperatures  of 146$\pm$3~K and 90$\pm$10~K respectively, whereas a temperature of 166$\pm$70~K was found in the Hot Core from their H$\rm_{2}$CO data. Within a $\sim$ 22-28$\arcsec$ beam \citet{Menten:1988} have derived from their CH$\rm_{3}$OH and $\rm^{13}$CH$\rm_{3}$OH  observations rotational temperatures of 109-147~K and 33~K respectively.
Their CH$\rm_{3}$OH line profiles are interpreted as the superposition of a broad and a narrow component (the former marginally warmer by 10-20~K). This interpretation , aimed both at thermal emission and maser excitation, suggests that infrared radiation from the dust plays an important role in the excitation of methanol in addition to collisions. From a 325-360~GHz survey made with 20$\arcsec$ resolution, \citet{Schilke:1997} found rotational temperatures of 188$\pm$3~K (methanol) and 98$\pm$3~K (methyl formate). More recently, at 350~$\mu$m and with a 11$\arcsec$ beam \citet{Comito:2005} infer  220~K for the Hot Core, and 160~K for the Compact Ridge from CH$\rm_{3}$OH lines, and an almost identical value (220~K and 155~K) from H$\rm_{2}$CO. A somewhat lower temperature, 61~K, was found in a 40$\arcsec$ beam for methyl formate by \citet{Ziurys:1993}.

Very recently Herschel observations of $\rm^{13}$CH$\rm_{3}$OH (and CH$\rm_{3}$OH) at 534~GHz and 1061~GHz  made by \citet{Wang:2011} determined T$\rm_{rot}$ = 105$\pm$2.6~K  (resp. T$\rm_{rot}$ = 125$\pm$2.8~K) in the Compact Ridge; the Compact ridge was isolated from other spatial components in the 43$\arcsec$ and 20$\arcsec$ Herschel beams thanks to its velocity profile. A more elaborated  transfer model developed for a spherical source with a temperature gradient provides a very good fit to the methanol population diagram of the Compact Ridge at both frequencies if it is externally heated. This conclusion well agrees with our findings.

%-----------------------------------------------------------------------------------------------------------------------------
%-------DISCUSSION ON METHYL FORMATE DISTRIBUTION AND ORION STRUCTURE---
%-----------------------------------------------------------------------------------------------------------------------------

\section{Discussion on methyl formate distribution and Orion structure}
\label{sec:discussion}

The ÊOrion-KL region is located about 1$\arcmin$ from the Trapezium OB stars at the heart of a large stellar cluster still in formation. Many remarkable objects Êhave been identified in this region \citep[see e.g. reviews by] []{Irvine:1987,Genzel:1989,ODell:1993,ODell:2001,ODell:2008,ODell:2009}. In this Section, we primarily try to correlate the HCOOCH$\rm_{3}$ distribution with several optical, NIR, MIR, X, radio continuum and radio line data, with the help of the ALADIN software\footnote{Centre de Donn\'ees Astronomiques de Strasbourg, see:Ê\ http://aladin.u-strasbg.fr/aladin.gml \ andÊ\ http://simbad.u-strasbg.fr/simbad/.} \citep{Bonnarel:2000} and the SIMBAD database$\rm^{8}$. We focus our discussion on the Compact Ridge and Hot Core regions as they contain the main methyl formate peaks in Orion-KL (MF1 and MF2, respectively).

%---------------------------------------------------------
%----------------Stars and YSOs------------- 
%---------------------------------------------------------
\subsection{Stars and YSOs }
Orion-KL is a complex region, with a very high density of YSOs (Young Stellar Objects) and recently formed stars.

Putting together the remarkably detailed NIR pictures of \citet{Stolovy:1998} and D. Rouan (private communication) together with the X picture \citep{Getman:2005,Grosso:2005} and previous IR work \citep{Greenhill:2004,Hillenbrand:2000,Lonsdale:1982,Muench:2002},Ê
and including the IR polarization work of \citet{Simpson:2006} we have been able to identify more than 34 probable stars or forming stars in the 0.5$\arcmin$ region of methyl formate emission in Fig. \ref{Fig.vshape}.
In this 0.058~pc wide field, this means a projected density of 156~objects~arcmin$^{-2}$ (1 object every $\sim$5$\arcsec$) or $10^{4}$~ objects~pc$^{-2}$. For comparison, \citet{Lada:2004} determined stellar projected densities for deeply embedded sources reaching up to 3000~star~pc$^{-2}$ Êsouth of Orion-KL (this would mean 10 objects in the same 0.5$\arcmin$ region of Fig. \ref{Fig.vshape}); for less embedded stars close to the Trapezium, the observed projected density is twice higher, up to 6000~star~pc$^{-2}$. With such a high density, one should Êalways keep in mind the possibility of chance projection effects Êin the following spatial comparisons.

\begin{figure}[h!]
\centering
\includegraphics[width=8cm,angle=270]{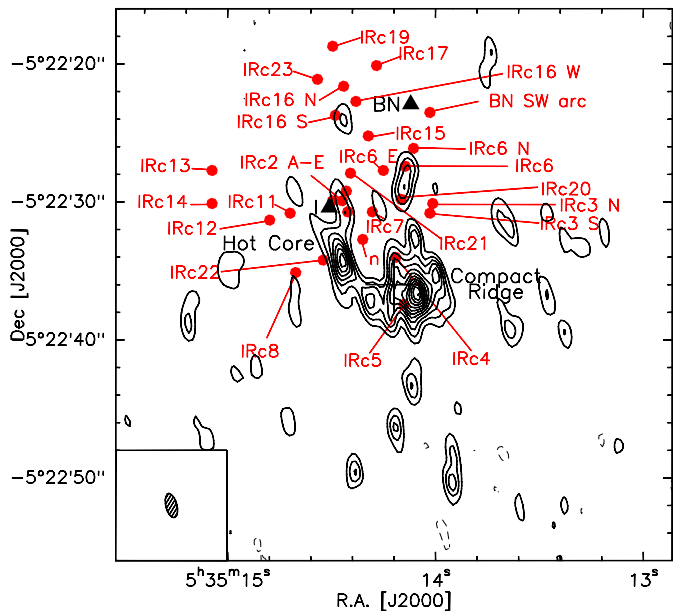}
    \caption{Positions of the IRc sources identified by \citet{Shuping:2004} on the methyl formate integrated intensity map (see Fig. \ref{Fig.vshape}).}
     \label{visuIRshupMF} 
    \end{figure}

Over small distance scales (a few $\arcsec$ or 1000-2000~AU), and with the exception of Parenago 1822/LBLS k at the center of the Compact Ridge (see discussion below), little correlation is seen between the methyl formate (MF) peaks and the stellar objects. Almost no association is obvious as well with the IRc sources, as seen on figure \ref{visuIRshupMF} which shows the position of the IRc sources from \citet{Shuping:2004}  with respect to the methyl formate emission. 
Over larger scales Ê($\ga$10$\arcsec$ or 4500~AU) it seems, as discussed below, that the methyl formate distribution is closely linked to the presence of a few remarkable objects, in particular the \textquotedblleft low-velocity\textquotedblright \ ÊSiO outflow whose origin is attributed to radiosource I by \citet{Goddi:2009} and \citet{Plambeck:2009}, and linked to the matter (traced in CO and H$\rm_{2}$) ejected during the recent ($\sim$ 500~yr) stellar collision or close interaction between the B-type star BN, and the I and n objects (at least) as proposed by e.g. \citet{Zapata:2009}. Long range effects due to heating, photodissociating photons or shocks are also possible from BN itself and source n. Note that our data do not allow us to separate the respective effects of source I from source SMA1 which has recently been identified by \citet{Beuther:2008} as another candidate source for the high-velocity CO and H$\rm_{2}$ outflow. 

%------------------- 
%FIGURE 17
%------------------- 
%------------------- 

%---------------------------------------------------------------
%--HCOOCH3 and "low-velocity" SiO outflow--
%---------------------------------------------------------------
\subsection{ HCOOCH$\rm_{3}$ and the "low-velocity" SiO outflow}

\begin{figure}[h!]
 \centering
\includegraphics[width=8cm,angle=0]{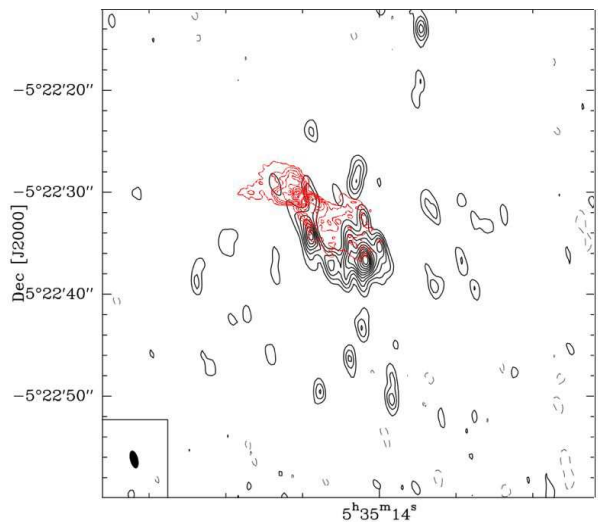}
 Ê\caption{HCOOCH$\rm_{3}$ contours (in black ) overlaid over the SiO J=2--1 v=0 map (red contours) by \citet{Plambeck:2009}. Note the nice match of the SiO outflow with the quiescent and recently shocked gas traced by HCOOCH$\rm_{3}$ as the SiO fills the hole in the methyl formate distribution.}
 ÊÊÊÊÊÊÊÊÊÊÊÊ\label{Fig.SiOMF}%
 ÊÊ\end{figure}

As shown in Fig. \ref{Fig.SiOMF} there is a striking complementarity of the 87~GHz SiO v=0 J=2-1 line map from \citet{Plambeck:2009} with our methyl formate distribution. We consider that this fact is not fortuitous and points to some kind of direct connection between both species. The SiO traces the large \textquotedblleft low-velocity\textquotedblright \ outflow originating from radio source I. As shown by various proper motion studies \citep{Gomez:2005,Rodriguez:2005,Goddi:2010} source I is moving towards the South-East (bottom left).

The HCOOCH$\rm_{3}$ distribution near MF2 (Hot Core-SW) may be due, to a large extent, to the recent sidewise \textquotedblleft encounter\textquotedblright \ of this outflow with quiet dense gas resulting in shocks which in turn disrupt icy grain mantles and release new molecules.ÊThe Hot Core region would then be the prominent result of this sidewise encounter. With this picture in mind one expects a recent major increase in the release of molecules from grains in the Hot Core region and in the eastern side of the large scale V-shaped region mapped in MF in this work. 

The Compact Ridge methyl formate distribution also matches the shape of the SiO outflow. In this area, the interaction is more frontal, the ouflowing matter coming more or less perpendicular to the border of the quiescent gas. The SiO outflow could be older than the stellar collision event. In that case the shock at the Compact Ridge would be older than the collision and be modified only by the recent motion of source I. Alternatively, the outflow could be linked to the stellar collision, or its trajectory might have encountered the Compact Ridge gas because of the recent proper motion of source I. In the first case, the release of molecules from grains would be older than 500~yr, in the latter case it would be more recent. Such a connection between the \textquotedblleft low-velocity\textquotedblright \ outflow and the Compact Ridge has been proposed earlier \citep[see e.g.] []{Irvine:1987}.

A shock from the I outflow can either release CH$\rm_{3}$OH rich ices, followed by gas phase chemistry leading to HCOOCH$\rm_{3}$ formation \citep{Charnley:2005}, or directly release HCOOCH$\rm_{3}$ formed on the grain \citep{Garrod:2006}.
The detailed chemical modelling is beyond the scope of the present study.

%------------------- 
%FIGURE 18
%------------------- 
 Ê
%------------------- 

%---------------------------------------------------------
%------HCOOCH3 and IR 11um map-----
%---------------------------------------------------------
\subsection{HCOOCH$\rm_{3}$ and IR 11~$\mu m$ map}

\begin{figure}[h!]
 \centering
 Ê\includegraphics[width=8cm,angle=0]{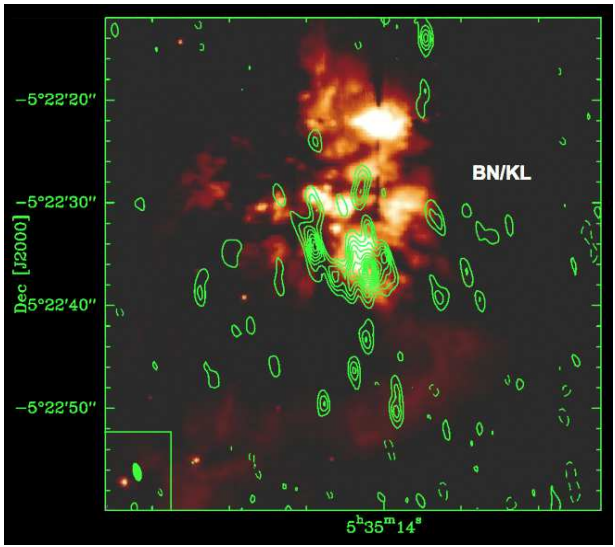}
 Ê\caption{The HCOOCH$\rm_3$ emission contours from MF7 to MF10 follow closely a black zone where 11~$\mu$m IR emission is lacking \citep{Smith:2005}, most probably due to a dense, absorbing foreground gas/dust lane. Such an anticorrelation has been observed for NH$\rm_3$ and interpreted in a similar manner by \citet{Shuping:2004} using the NH$\rm_3$ Êmap ofÊ \citet{Wilson:2000}}.
 ÊÊÊÊÊÊÊÊÊÊÊÊ\label{Fig.SmithMF}%
 ÊÊ\end{figure}

In Fig. \ref{Fig.SmithMF} we compare the methyl formate distribution with the 11~$\mu$m map of \citet{Smith:2005}.
There is a clear trend for Êthe MF emission peaks to lie outside the bright Ê11~$\mu$m regions--- it is especially true for the East branch of the V-shaped MF emission, which encompasses the Hot Core MF2 and goes from ÊMF7 to MF10. A notable apparent exception to this anticorrelation (but see below) is the main source MF1 located in the Compact Ridge.Ê

It has long been noted that there is a tendency for some molecular emission in Orion-KL, in particular NH$\rm_3$, Êto be anticorrelated with mid-IR emission \citep[e.g.] []{Genzel:1989,Wynn-Williams:1984}. Recently \citet{Shuping:2004} reached a similar conclusion from their high resolution mid-IR map and the NH$\rm_{3}$ high resolution map of \citet{Wilson:2000}: they suggest that the NH$\rm_3$ emission comes from a foreground layer of interstellar matter, cold and dense enough to be seen in the mid-IR as an absorbing dark lane against a brighter background. 

We propose that most of the methyl formate emission Êseen in places of low 11~$\mu$m emission is also originating in this Êdark foreground material. In the case of MF1 in the Compact Ridge, there are indications that the methyl formate layer is also in the foreground material, but is thin enough to let the IR background radiation to be seen through it (see discussion on excited H$\rm_2$).

%------------------- 
%FIGURE 19
%------------------- 
 Ê
%------------------- 

%----------------------------------------------------------
%--HCOOCH3 and 2.12mu H2 emission--
%----------------------------------------------------------
\subsection{ HCOOCH$\rm_3$ and 2.12~$\mu m$ H$\rm_2$ emission }

\begin{figure}[h!]
 \centering
 \includegraphics[width=9cm,angle=0]{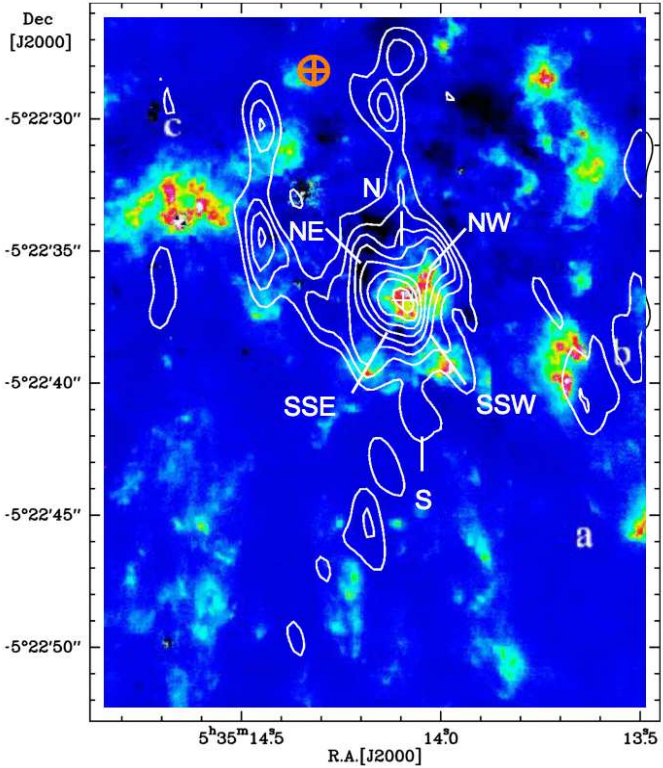}
 Ê\caption{Methyl formate 8.7 ~km~s$^{-1}$ channel map contours overlaid over \citet{Lacombe:2004} 2.12~$\mu$m excited H$\rm_2$ emission showing a good correlation of both tracers toward MF1 (white cross) and around (North West (NW), South South West (SSW) and South South East (SSE)). The North East (NE) region analysis is hampered by the subtraction of strong 2~$\mu$m continuum from IRc4 (see Fig. \ref{visuIRshupMF}) -- which results in an artefact (the zone in black).
Emission toward the North (N) and the South (S) seems of a different nature and is linked to the 9-10~km~s$^{-1}$ velocity gas (see discussion). Note that a fraction only of the H$\rm_2$ \textquotedblleft bullets\textquotedblright \   distribution (those moving towards us and thus closer to the observer) is expected to suffer sufficiently low extinction to be visible. The red circled cross marks the proper motion center where the sources n, I and BN were located 500 years ago \citep{Gomez:2005,Gomez:2008,Rodriguez:2005}.
}
 ÊÊÊÊÊÊÊÊÊÊÊÊ\label{Fig.MFvsLacombe}%
 ÊÊ\end{figure}

%------
%MF1
%------
\paragraph{\textbf{MF1:}}
A most striking spatial correlation is observed between the methyl formate distribution at the MF1 peak and the 2.12~$\mu$m H$\rm_2$ emission imaged by \citet{Lacombe:2004} with adaptative optics (NACO) at the Very Large Telescope. This is shown in Fig. \ref{Fig.MFvsLacombe}. Additional H$\rm_2$ pictures can be found in \citet{Stolovy:1998} and  \citet{Nissen:2007}.

\begin{figure}[h!]
 \centering
 \includegraphics[width=9cm,angle=0]{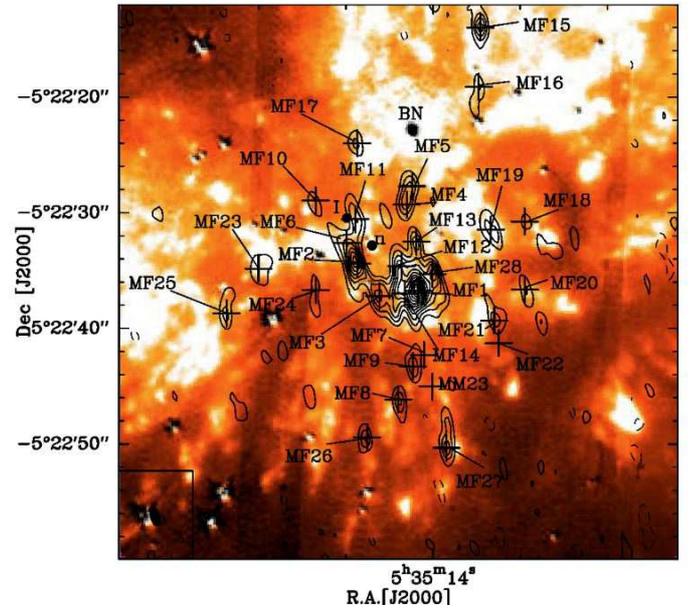}
 Ê\caption{ Map of the integrated methyl formate emission (cf. Fig. 4) overlaid over a Subaru Observatory image of H$\rm_2$ at 2.12~$\mu$m emission (\copyright \ Subaru Telescope, NAOJ. All rights reserved). In this large-scale image, the \textquotedblleft methyl formate - excited H$\rm_2$\textquotedblright \ correlation is evident in several peaks other than the MF1 peak (Fig. \ref{Fig.MFvsLacombe}): e.g. MF19, MF18-SW, MF21 (even more clearly seen in Fig. \ref{Fig.MFvsLacombe}), MF27, MF8, MF23. Note, however, than some HCOOCH$\rm_{3}$ or H$\rm_{2}$ peaks have no counterpart. Possible explanations are given in the text.}
 ÊÊÊÊÊÊÊÊÊÊÊÊ\label{Fig.MFvsLacombeSubaru}%
 ÊÊ\end{figure}

We first note a global overlap of MF1 with a zone of excited H$\rm_2$ emission. Going into details, the MF1 methyl formate distribution departs from a simple elliptical shape, showing several extensions: 1) the MF1 peak corresponds to a local maximum of H$\rm_2$ emission, 2) there is a second peak at  a North West position of the Compact Ridge (labeled NW in Fig. \ref{Fig.MFvsLacombe}) in both tracers, 3) Two extensions of the MF map towards the South South East  and West respectively of the Compact Ridge (labeled SSE and SSW in Fig. \ref{Fig.MFvsLacombe}) correspond to two other peaks. These findings strongly support the association between MF and excited H$\rm_{2}$. At the North East of the Compact Ridge (position labeled NE in Fig. \ref{Fig.MFvsLacombe}), MF overlaps the strong 2~$\mu$m continuum emission which has been Êsubtracted and appears as a black zone in Fig. \ref{Fig.MFvsLacombe}. This prevents a good comparison of the H$\rm_2$ emission in this area. However, inspection of the Subaru map (Fig. Ê\ref{Fig.MFvsLacombeSubaru}) confirms that there is also 2.12~$\mu$m emission there.Ê

The North and South MF extensions have no counterpart at 2.12~$\mu$m (only faint 2.12~$\mu$m emission is present near N). They seem to be related to a different layer of emission, which appears as a North-South bar in the MF channel maps at velocities around or larger than 9~km~s$^{-1}$. Because such a structure does not show up in any optical or IR map, we consider it likely that the gas is on the far side of OMC-1, probably behind Êthe KL nebula itself.

%----------------
%Other peaks
%----------------
\paragraph{\textbf{Other peaks:}}
The methyl formate - excited H$\rm_2$ association observed in the MF1 area is also observed on a broader scale, especially, clockwise from the West, toward MF19, MF18-SW, MF21, MF27, MF8, MF23. However, some MF peaks have no excited H$\rm_{2}$ emission counterpart: for example, and most notably, there is no   H$\rm_{2}$ emission at the MF2 (Hot Core-SW) peak; however excited H$\rm_{2}$ emission is present in its Southern part (Fig. Ê\ref{Fig.MFvsLacombe} and \ref{Fig.MFvsLacombeSubaru}). Conversely bright 2.12~$\mu$m emission areas (e.g. East of the Hot Core, close to star LBLS t) do not exhibit MF emission.

To better understand the diversity of the observations the following explanations are worth considering:
\begin{itemize}
\item MF is seen associated with a 2.12~$\mu$m emission if the shock related to the Orion-KL explosive event passes through interstellar material dense and cold enough so that grains have ice mantles. In that case the correlation observed toward MF1 is explained by the shock-induced release of methyl formate or its progenitor CH$\rm_{3}$OH from ice-coated grain mantles. 
\item The molecular production efficiency related to the shock may also be low in places, or cold grain mantles may be less abundant Ê(e.g. closer to the very luminous Trapezium OB stars located 1$\arcmin$ SE of Orion-KL), so that the MF column density is undetectable.
\item In some other places (e.g. around LBLS t) the 2.12~$\mu$m emission geometry suggests that the emission is linked to the star and thus might be of a different nature. 
\item ÊMF may also peak in regions where it is released or produced by mechanisms (e.g. thermal heating) different from the shock generated by the explosive event. 
\item H$\rm_{2}$ emission may be hidden in some areas by a large column density of foreground dust. 
\end{itemize}

The latter possibility is well illustrated by the comparison of the H$\rm_{2}$ 2.12~$\mu$m map of \citet{Lacombe:2004} (Fig. Ê\ref{Fig.MFvsLacombe}) with the Subaru deep exposure \citep[Fig. Ê\ref{Fig.MFvsLacombeSubaru}, reproduced in extenso in] []{Shuping:2004}. The larger extension of the H$\rm_{2}$ emission seen in the Subaru map indicates that a rather \textquotedblleftÊthinÊ\textquotedblright  \ layer  prevented some H$\rm_{2}$ peaks to be visible in the \citet{Lacombe:2004} map. For a few H$\rm_{2}$ emission spots in this field \citet{Colgan:2007}  estimated indeed a foreground absorption corresponding to A$\rm_v$ =~4-8. In the direction of the Compact Ridge, the NIR sources IRc4 and IRc5 are interpreted as reflection nebulae seen through holes of foreground matter \citep{Shuping:2004, Simpson:2006}; the larger size of these sources (hence of these holes) at 11~$\mu$m compared to 2~$\mu$m (continuum) is an additional indication of the relative thinness of the foreground matter.

On the contrary the MF2 peak would be a case where strong dust absorption hides all H$\rm_{2}$ 2.12~$\mu$m emission. Close to MF2, very large opacities are advocated to explain the non-detection of  the bright source I   \citep[e.g.] []{Greenhill:2004}. Indeed  from our continuum data (Sect. 6) we derive at both MF1 and MF2 peaks large N$\rm_{H\rm_{2}}$ column densities (5 and 3.1~10$^{24}$~cm$^{-2}$ respectively); this corresponds to $A\rm_{v}$ $\ga$ 1000, assuming standard dust opacities. Our interpretation of the contrasted situation at both peaks is that at the Compact Ridge (MF1), a shock hits the front side of a dense clump of matter, whereas at the Hot Core-SW (MF2), it hits the rear side (with respect to the observer). The presence  close to MF2 of a  H$_2$O maser spot \citep[053247.001-052427.81,] []{Gaume:1998}, similar to other spots found in the Compact Ridge, strengthens the hypothesis that a shock is present  there too.

The source of excitation of the H$\rm_{2}$ molecules is most likely linked to the global explosive event.
Most of the H$\rm_{2}$ emission  
can be traced to a common center whose coordinates are given in Table \ref{Objects} (e.g. Zapata et al. 2009 and refs. therein). \citet{Stolovy:1998} analyse in detail two remarkable features which they call the  \textquotedblleft nested arcs\textquotedblright \  and the  \textquotedblleft bullets\textquotedblright \ (see their Fig. 4c),Ê
and suggest that other H$\rm_{2}$ features in their map could be of similar nature but are less easily identified due to projection effects.
If this is indeed the case,  the comparison with these two features favourably seen in a more edge-on configuration sheds some light on the probable geometry of the shocked region at MF1, and how the shock might lead to the release of methyl formate from grain mantles; a detailed modelling at MF1 is  however required.

%------------------- 
%FIGURE 20
%------------------- 
 Ê
%------------------- 

%------------------- 
%FIGURE 21
%------------------- 
 Ê
%------------------- 

%------------------- 
%TABLE 9
%------------------- 
\begin{table}
\begin{minipage}[t]{8.5cm}
\renewcommand{\footnoterule}{} Ê% to avoid a line before footnotes
\caption{Coordinates of noticeable objects or places near the main methyl formate peak (MF1).}
\label{Objects}
\small\addtolength{\tabcolsep}{-2pt}
\begin{center}
\begin{tabular}{lcccc}
\hline\hline
Object/place Ê& R.A. (J2000) & Dec. (J2000) & Precision Ê\\
& 05$^{h}$35$^{m}$ & -05$\degr$22$\arcmin$ & ÊÊ\\
\hline
HCOOCH$\rm_{3}$ MF1 peak \footnote {This study} & ÊÊ05 35 14.09 & ÊÊ-05 22 36.7 & ÊÊ$\pm$0.8$\arcsec$ \\
P1822/LBLS k Ê\footnote {SIMBAD} Ê& ÊÊ05 35 14.092 & ÊÊ-05 22 36.43 & ÊÊÊÊ\\
COUP 600 X source \footnote {\citet{Grosso:2005}} & ÊÊ05 35 14.09 & ÊÊ-05 22 36.4 & ÊÊÊÊ\\
H$\rm_{2}$O supermaser Ê\footnote {\citet{Matveenko:2000} after \citet{Greenhill:2004}} ÊÊ& ÊÊ05 Ê35 14.12 & ÊÊ-5 22 36.27 & ÊÊÊ\\
HC~438 \footnote {\citet{Eisner:2008} } Ê& ÊÊ05 35 14.09 & Ê-5 22 36.6 Ê& ÊÊÊÊ\\ 
%\\
\hline
CO Fingers Outflow Center \footnote {\citet{Stolovy:1998,Zapata:2009}} Ê& Ê05 35 14.37 & Ê-05 22 27.9 & $\pm$1.5$\arcsec$ Ê\\
Proper motion center \footnote {\citet{Gomez:2005,Gomez:2008,Rodriguez:2005}} & Ê05 35 14.35 & Ê-05 22 27.7 & Ê$\pm$1$\arcsec$ Ê\\
\hline
\end{tabular}
\end{center}
\end{minipage}
\end{table}
%------------------- 

%----------------------------------------------------------
%-----Main Methyl Formate peak MF1------
%----------------------------------------------------------
\subsection{Main Methyl Formate peak MF1 (Compact Ridge)}

In addition to the correlation observed between the low-velocity outflow traced by SiO and the excited H$\rm_2$ emission at 2.12~$\mu$m discussed above, some other remarkable objects are observed towards the main MF peak, MF1. The visible star Parenago 1822/LBLS~k \citep{Parenago:1954,Parenago:1997} is situated right in the middle of the MF1 methyl formate peak. ÊIt exhibits 1.3~mm continuum emission which is analyzed as disk emission by \citet{Eisner:2008} under the name HC~438. \citet{Getman:2005} derive an age of 22 000 yr and a small mass of 0.26 M$\rm_{\sun}$. Another nearby remarkable object is the H$\rm_{2}$O  \textquotedblleft Supermaser\textquotedblright \  identified and studied by \citet{Matveenko:2000}, see also \citet{Demichev:2009} and \citet{Matveyenko:2007}. Table \ref{Objects} gives the \textit{absolute} position of this H$\rm_{2}$O source together with our methyl formate position and positions from other studies of nearby objects. \citet{Matveenko:2000} estimate a very small mass of 0.007 M$\rm_{\sun}$ for their H$\rm_{2}$O supermaser central source whose position is distinct from P1822/LBLS~k. However, the proximity and youth of these two objects suggest that they might be a binary still in its forming phase.

P1822/LBLS~k and associated supermaser could perhaps play a role in the observed distribution of methyl formate. Shocks are playing a central role in the Êvicinity of the H$\rm_{2}$O supermaser whose excitation and properties are best explained by molecular collisions in shocked dense gas clumps. These shocks may well have released the methyl formate or its precursor from grain mantles. However the methyl formate clumps cannot be always tightly associated with all the H$\rm_{2}$O maser spots because H$\rm_{2}$O maser emission traces only the hottest and densest pockets of gas. 

We suggest that four phenomena, observed in an area as small as 5$\arcsec$, may play a role in the release of the methyl formate molecule from ice grain mantles: i) bullets ejected 500 years ago (due to the BN-I-n encounter), ii) bipolar outflow from source I, iii) action from the star P1822/LBLS~k, and iv) shocks linked to the excitation of the H$\rm_{2}$O supermaser. From a statistical point of view it seems unlikely that the above suggestions are unrelated. However, the relative importance of these four phenomena in the MF production, their relative timing and their causal relationship deserve further study.

%-----------------------------------------------------------------------------------------
%---------------------------------------CONCLUSIONS----------------
%-----------------------------------------------------------------------------------------

\section{Conclusions}

We have studied the distribution of the complex O-bearing molecule Methyl Formate (MF) HCOOCH$\rm_{3}$ with medium to high angular resolution ($\approx$7$\arcsec$ to 2$\arcsec$) using interferometric data from the IRAM Plateau de Bure Interferometer. 
Our main results and conclusions are the following: 
   \begin{enumerate}
      \item
Our data sets include 21 well detected transitions of MF from E$\rm_{up}$=25~K to E$\rm_{up}$=618~K. A few more lines are present, still usable but partially blended with other lines. Only about 40$\%$ of the lines stronger than the weakest detected line appear free of blend. The line ares optically thin ($\tau$ $<$ 0.1).
      \item 
      We confirm the detection of v$\rm_{t}$=1 transitions. In this study, we used together v$\rm_{t}$=0 and v$\rm_{t}$=1 transitions to derive rotational temperatures. 
      \item 
      We identify at least 28 MF concentrations. The most intense emissions, MF1 and MF2, arise from the Compact Ridge and the Hot Core-SW respectively. The emission toward MF1 is much stronger than toward MF2 for low energy levels whereas the emission at MF2 becomes stronger at high energy levels.
       \item 
       We have determined the MF temperature using rotational diagrams for the five main positions and deduced HCOOCH$\rm_{3}$ column densities assuming LTE. Temperatures cover the range 80 to 170~K, and column densities 1.6$\times$10$^{16}$ to 1.6$\times$10$^{17}$~cm$^{-2}$. 
       \item 
       In the course of the MF data reduction, we had to produce and subtract maps of the continuum emission, using line free channels. We used these continuum maps to identify four major clumps Ca to Cd, for which we derived dust masses in the range 0.01 to 0.05~M$\rm_{\sun}$ using a dust temperature equal to the MF rotational temperature. Assuming a gas-to-dust ratio of 100 the gas masses are in the range 1 to 5~M$\rm_{\sun}$.
       \item 
       MF gas velocities lie in the range 7-8~km~s$^{-1}$. At some places two different velocity components are clearly seen and the high velocity component (around 9-10~km~s$^{-1}$) has a linear North-South structure. We see no gas emission at the 5~km~s$^{-1}$ velocity usually reported for the Hot Core: the gas we observe there is slightly higher in velocity and probably of the same nature as the gas we observe for the Compact Ridge.
       \item 
       We have searched for the two isomers of methyl formate but we did not detect them. We find an abundance ratio of  less than 1/50 for acetic acid CH$\rm_{3}$COOH and less than 1/(200-500) for glycolaldehyde CH$\rm_{2}$OHCHO.
       \item 
We have correlated the MF1/Compact Ridge emission with other gas tracers and with various Orion objects. A very clear association is found with 2.12~$\mu$m excited H$\rm_{2}$ maps. This tends to confirm a scenario of MF production involving the release of molecules from ice mantles, either MF itself or a precursor (CH$\rm_{3}$OH). Four possible origins have been identified for the excited H$\rm_{2}$ emission and the MF production: shock from the \textquotedblleft low-velocity\textquotedblright  outflow from source I, shock from a bullet ejected during the supposed collision-explosion event 500 yr ago \citep{Zapata:2009}, action of the young forming star Parenago 1822/LBLS k , or from the nearby source (and possible companion of the star)  responsible for the H$\rm_{2}$O supermaser \citep{Matveenko:2000}.

We can not conclude yet which of these four likely processes dominates the MF production but all may have contributed. 

To further analyze the Compact Ridge history more radio and IR high angular resolution images are needed. In the future, ALMA and adaptive optics on large ground based telescopes and space telescopes will provide the new data required to progress deeply in the understanding of the Orion-KL region.

   \end{enumerate}

The structure we find in Orion-KL might help to better understand the correlation previously found between cometary and interstellar ices \citep{Bockelee-Morvan:2000}. In this correlation, sources of different  nature {\it a priori},  "hot cores" on one side and a bipolar flow L1157 on the other side, both appeared as a good match of cometary ices. Shocks releasing molecules from icy grain mantles  are in L1157 a major source of molecules in the gas phase . From the correlation we found between methyl formate and excited H$\rm_{2}$ emission, this  might  also be the case in Orion-KL as well as in other hot cores.
   
%===================================================================================================

%-----------------------------------------------------------------------------------------------------------------------------
%-----------------------------------------ACKNOWLEDGEMENTS-----------------------------------
%-----------------------------------------------------------------------------------------------------------------------------

\begin{acknowledgements}
      We thank Laurent Margules and Brian Drouin for their spectroscopic knowledge and advices on HCOOCH$\rm_{3}$.
      We thank Thierry Jacq for his dataset at 80 GHz and 203 GHz, and Daniel Rouan and Nathan Smith for their H$\rm_{2}$ and 11~$\mu$m maps.
      Valuable contributions from David Field are gratetfully acknowledged.
      We also thank Alexandre Faure for the discussion on lines in fundamental and excited torsional states. 
      We thank the IRAM staff in Grenoble for their help to get and reduce the data.
      Finally we thank the anonymous referee for his helpful comments.
      This research has made use of the SIMBAD database and ALADIN software, operated at CDS, Strasbourg, France.
      This work was supported by CNRS national programs PCMI (Physics and Chemistry of the Interstellar Medium) and GDR Exobiology.
\end{acknowledgements}

%===================================================================================================

%-----------------------------------------------------------------------------------------------------------------------------
%-----------------------------------------BIBLIOGRAPHY-----------------------------------------------
%-----------------------------------------------------------------------------------------------------------------------------

\bibliographystyle{aa}
\bibliography{biblio.bib}

\begin{thebibliography}{23}
\expandafter\ifx\csname natexlab\endcsname\relax\def\natexlab#1{#1}\fi

\bibitem[{{Beuther} {et~al.}(2005){Beuther}, {Zhang}, {Greenhill}, {Reid},
  {Wilner}, {Keto}, {Shinnaga}, {Ho}, {Moran}, {Liu}, \&
  {Chang}}]{Beuther:2005}
{Beuther}, H., {Zhang}, Q., {Greenhill}, L.~J., {et~al.} 2005, \apj, 632, 355

\bibitem[{{Blake} {et~al.}(1987){Blake}, {Sutton}, {Masson}, \&
  {Phillips}}]{Blake:1987}
{Blake}, G.~A., {Sutton}, E.~C., {Masson}, C.~R., \& {Phillips}, T.~G. 1987,
  \apj, 315, 621

\bibitem[{{Combes} {et~al.}(1996){Combes}, {Q-Rieu}, \&
  {Wlodarczak}}]{Combes:1996}
{Combes}, F., {Q-Rieu}, N., \& {Wlodarczak}, G. 1996, \aap, 308, 618

\bibitem[{{Comito} {et~al.}(2005){Comito}, {Schilke}, {Phillips}, {Lis},
  {Motte}, \& {Mehringer}}]{Comito:2005}
{Comito}, C., {Schilke}, P., {Phillips}, T.~G., {et~al.} 2005, \apjs, 156, 127

\bibitem[{{de Vicente} {et~al.}(2002){de Vicente}, {Mart{\'{\i}}n-Pintado},
  {Neri}, \& {Rodr{\'{\i}}guez-Franco}}]{de-Vicente:2002}
{de Vicente}, P., {Mart{\'{\i}}n-Pintado}, J., {Neri}, R., \&
  {Rodr{\'{\i}}guez-Franco}, A. 2002, \apjl, 574, L163

\bibitem[{{Draine} \& {Lee}(1984)}]{Draine:1984}
{Draine}, B.~T. \& {Lee}, H.~M. 1984, \apj, 285, 89

\bibitem[{{Eisner} {et~al.}(2008){Eisner}, {Plambeck}, {Carpenter}, {Corder},
  {Qi}, \& {Wilner}}]{Eisner:2008}
{Eisner}, J.~A., {Plambeck}, R.~L., {Carpenter}, J.~M., {et~al.} 2008, \apj,
  683, 304

\bibitem[{{Forster} {et~al.}(1978){Forster}, {Welch}, {Wright}, \&
  {Baudry}}]{Forster:1978}
{Forster}, J.~R., {Welch}, W.~J., {Wright}, M.~C.~H., \& {Baudry}, A. 1978,
  \apj, 221, 137

\bibitem[{{Friedel} \& {Snyder}(2008)}]{Friedel:2008}
{Friedel}, D.~N. \& {Snyder}, L.~E. 2008, \apj, 672, 962

\bibitem[{{Goddi} {et~al.}(2009){Goddi}, {Greenhill}, {Chandler}, {Humphreys},
  {Matthews}, \& {Gray}}]{Goddi:2009}
{Goddi}, C., {Greenhill}, L.~J., {Chandler}, C.~J., {et~al.} 2009, \apj, 698,
  1165

\bibitem[{{G{\'o}mez} {et~al.}(2005){G{\'o}mez}, {Rodr{\'{\i}}guez}, {Loinard},
  {Lizano}, {Poveda}, \& {Allen}}]{Gomez:2005}
{G{\'o}mez}, L., {Rodr{\'{\i}}guez}, L.~F., {Loinard}, L., {et~al.} 2005, \apj,
  635, 1166

\bibitem[{{Gu{\'e}lin} {et~al.}(2008){Gu{\'e}lin}, {Brouillet}, {Cernicharo},
  {Combes}, \& {Wooten}}]{Guelin:2008}
{Gu{\'e}lin}, M., {Brouillet}, N., {Cernicharo}, J., {Combes}, F., \& {Wooten},
  A. 2008, \apss, 313, 45

\bibitem[{{Ilyushin} {et~al.}(2009){Ilyushin}, {Kryvda}, \&
  {Alekseev}}]{Ilyushin:2009}
{Ilyushin}, V., {Kryvda}, A., \& {Alekseev}, E. 2009, Journal of Molecular
  Spectroscopy, 255, 32

\bibitem[{{Kobayashi} {et~al.}(2007){Kobayashi}, {Ogata}, {Tsunekawa}, \&
  {Takano}}]{Kobayashi:2007}
{Kobayashi}, K., {Ogata}, K., {Tsunekawa}, S., \& {Takano}, S. 2007, \apjl,
  657, L17

\bibitem[{{Mathis} {et~al.}(1977){Mathis}, {Rumpl}, \&
  {Nordsieck}}]{Mathis:1977}
{Mathis}, J.~S., {Rumpl}, W., \& {Nordsieck}, K.~H. 1977, \apj, 217, 425

\bibitem[{{Menten} \& {Reid}(1995)}]{Menten:1995}
{Menten}, K.~M. \& {Reid}, M.~J. 1995, \apjl, 445, L157

\bibitem[{{Menten} {et~al.}(2007){Menten}, {Reid}, {Forbrich}, \&
  {Brunthaler}}]{Menten:2007}
{Menten}, K.~M., {Reid}, M.~J., {Forbrich}, J., \& {Brunthaler}, A. 2007, \aap,
  474, 515

\bibitem[{{Pickett} {et~al.}(1998){Pickett}, {Poynter}, {Cohen}, {Delitsky},
  {Pearson}, \& {Muller}}]{Pickett:1998}
{Pickett}, H.~M., {Poynter}, I.~R.~L., {Cohen}, E.~A., {et~al.} 1998, Journal
  of Quantitative Spectroscopy and Radiative Transfer, 60, 883

\bibitem[{{Plambeck} {et~al.}(1995){Plambeck}, {Wright}, {Mundy}, \&
  {Looney}}]{Plambeck:1995}
{Plambeck}, R.~L., {Wright}, M.~C.~H., {Mundy}, L.~G., \& {Looney}, L.~W. 1995,
  \apjl, 455, L189+

\bibitem[{{Rodr{\'{\i}}guez} {et~al.}(2005){Rodr{\'{\i}}guez}, {Poveda},
  {Lizano}, \& {Allen}}]{Rodriguez:2005}
{Rodr{\'{\i}}guez}, L.~F., {Poveda}, A., {Lizano}, S., \& {Allen}, C. 2005,
  \apjl, 627, L65

\bibitem[{{Turner}(1991)}]{Turner:1991}
{Turner}, B.~E. 1991, \apjs, 76, 617

\bibitem[{{Wilson} {et~al.}(2000){Wilson}, {Gaume}, {Gensheimer}, \&
  {Johnston}}]{Wilson:2000}
{Wilson}, T.~L., {Gaume}, R.~A., {Gensheimer}, P., \& {Johnston}, K.~J. 2000,
  \apj, 538, 665

\bibitem[{{Wynn-Williams} {et~al.}(1984){Wynn-Williams}, {Genzel}, {Becklin},
  \& {Downes}}]{Wynn-Williams:1984}
{Wynn-Williams}, C.~G., {Genzel}, R., {Becklin}, E.~E., \& {Downes}, D. 1984,
  \apj, 281, 172

\end{thebibliography}


\begin{thebibliography}{95}
\expandafter\ifx\csname natexlab\endcsname\relax\def\natexlab#1{#1}\fi

\bibitem[{{Belloche} {et~al.}(2008){Belloche}, {Menten}, {Comito},
  {M{\"u}ller}, {Schilke}, {Ott}, {Thorwirth}, \& {Hieret}}]{Belloche:2008a}
{Belloche}, A., {Menten}, K.~M., {Comito}, C., {et~al.} 2008, \aap, 482, 179

\bibitem[{{Beuther} \& {Nissen}(2008)}]{Beuther:2008}
{Beuther}, H. \& {Nissen}, H.~D. 2008, \apjl, 679, L121

\bibitem[{{Beuther} {et~al.}(2004){Beuther}, {Zhang}, {Greenhill}, {Reid},
  {Wilner}, {Keto}, {Marrone}, {Ho}, {Moran}, {Rao}, {Shinnaga}, \&
  {Liu}}]{Beuther:2004}
{Beuther}, H., {Zhang}, Q., {Greenhill}, L.~J., {et~al.} 2004, \apjl, 616, L31

\bibitem[{{Beuther} {et~al.}(2005){Beuther}, {Zhang}, {Greenhill}, {Reid},
  {Wilner}, {Keto}, {Shinnaga}, {Ho}, {Moran}, {Liu}, \&
  {Chang}}]{Beuther:2005}
{Beuther}, H., {Zhang}, Q., {Greenhill}, L.~J., {et~al.} 2005, \apj, 632, 355

\bibitem[{{Beuther} {et~al.}(2006){Beuther}, {Zhang}, {Reid}, {Hunter},
  {Gurwell}, {Wilner}, {Zhao}, {Shinnaga}, {Keto}, {Ho}, {Moran}, \&
  {Liu}}]{Beuther:2006}
{Beuther}, H., {Zhang}, Q., {Reid}, M.~J., {et~al.} 2006, \apj, 636, 323

\bibitem[{{Bisschop} {et~al.}(2007){Bisschop}, {J{\o}rgensen}, {van Dishoeck},
  \& {de Wachter}}]{Bisschop:2007}
{Bisschop}, S.~E., {J{\o}rgensen}, J.~K., {van Dishoeck}, E.~F., \& {de
  Wachter}, E.~B.~M. 2007, \aap, 465, 913

\bibitem[{{Blake} {et~al.}(1996){Blake}, {Mundy}, {Carlstrom}, {Padin},
  {Scott}, {Scoville}, \& {Woody}}]{Blake:1996}
{Blake}, G.~A., {Mundy}, L.~G., {Carlstrom}, J.~E., {et~al.} 1996, \apjl, 472,
  L49+

\bibitem[{{Blake} {et~al.}(1987){Blake}, {Sutton}, {Masson}, \&
  {Phillips}}]{Blake:1987}
{Blake}, G.~A., {Sutton}, E.~C., {Masson}, C.~R., \& {Phillips}, T.~G. 1987,
  \apj, 315, 621

\bibitem[{{Blake} {et~al.}(1984){Blake}, {Sutton}, {Masson}, {Phillips},
  {Herbst}, {Plummer}, \& {De Lucia}}]{Blake:1984}
{Blake}, G.~A., {Sutton}, E.~C., {Masson}, C.~R., {et~al.} 1984, \apj, 286, 586

\bibitem[{{Bockel{\'e}e-Morvan} {et~al.}(2000){Bockel{\'e}e-Morvan}, {Lis},
  {Wink}, {Despois}, {Crovisier}, {Bachiller}, {Benford}, {Biver}, {Colom},
  {Davies}, {G{\'e}rard}, {Germain}, {Houde}, {Mehringer}, {Moreno}, {Paubert},
  {Phillips}, \& {Rauer}}]{Bockelee-Morvan:2000}
{Bockel{\'e}e-Morvan}, D., {Lis}, D.~C., {Wink}, J.~E., {et~al.} 2000, \aap,
  353, 1101

\bibitem[{{Bonnarel} {et~al.}(2000){Bonnarel}, {Fernique}, {Bienaym{\'e}},
  {Egret}, {Genova}, {Louys}, {Ochsenbein}, {Wenger}, \&
  {Bartlett}}]{Bonnarel:2000}
{Bonnarel}, F., {Fernique}, P., {Bienaym{\'e}}, O., {et~al.} 2000, \aaps, 143,
  33

\bibitem[{{Boucher} {et~al.}(1980){Boucher}, {Burie}, {Bauer}, {Dubrulle}, \&
  {Demaison}}]{Boucher:1980}
{Boucher}, D., {Burie}, J., {Bauer}, A., {Dubrulle}, A., \& {Demaison}, J.
  1980, Journal of Physical and Chemical Reference Data, 9, 659

\bibitem[{{Charnley} \& {Rodgers}(2005)}]{Charnley:2005}
{Charnley}, S.~B. \& {Rodgers}, S.~D. 2005, in IAU Symposium, Vol. 231,
  Astrochemistry: Recent Successes and Current Challenges, ed. {D.~C.~Lis,
  G.~A.~Blake, \& E.~Herbst}, 237--246

\bibitem[{{Colgan} {et~al.}(2007){Colgan}, {Schultz}, {Kaufman}, {Erickson}, \&
  {Hollenbach}}]{Colgan:2007}
{Colgan}, S.~W.~J., {Schultz}, A.~S.~B., {Kaufman}, M.~J., {Erickson}, E.~F.,
  \& {Hollenbach}, D.~J. 2007, \apj, 671, 536

\bibitem[{{Combes} {et~al.}(1996){Combes}, {Q-Rieu}, \&
  {Wlodarczak}}]{Combes:1996}
{Combes}, F., {Q-Rieu}, N., \& {Wlodarczak}, G. 1996, \aap, 308, 618

\bibitem[{{Comito} {et~al.}(2005){Comito}, {Schilke}, {Phillips}, {Lis},
  {Motte}, \& {Mehringer}}]{Comito:2005}
{Comito}, C., {Schilke}, P., {Phillips}, T.~G., {et~al.} 2005, \apjs, 156, 127

\bibitem[{{de Vicente} {et~al.}(2002){de Vicente}, {Mart{\'{\i}}n-Pintado},
  {Neri}, \& {Rodr{\'{\i}}guez-Franco}}]{de-Vicente:2002}
{de Vicente}, P., {Mart{\'{\i}}n-Pintado}, J., {Neri}, R., \&
  {Rodr{\'{\i}}guez-Franco}, A. 2002, \apjl, 574, L163

\bibitem[{{Demichev} \& {Matveenko}(2009)}]{Demichev:2009}
{Demichev}, V.~A. \& {Matveenko}, L.~I. 2009, Astronomy Reports, 53, 79

\bibitem[{{Draine}(2010)}]{Draine:2010}
{Draine}, B.~T. 2010, {Physics of the Interstellar and Intergalactic Medium},
  ed. {Draine, B.~T.}

\bibitem[{{Draine} \& {Lee}(1984)}]{Draine:1984}
{Draine}, B.~T. \& {Lee}, H.~M. 1984, \apj, 285, 89

\bibitem[{{Eisner} {et~al.}(2008){Eisner}, {Plambeck}, {Carpenter}, {Corder},
  {Qi}, \& {Wilner}}]{Eisner:2008}
{Eisner}, J.~A., {Plambeck}, R.~L., {Carpenter}, J.~M., {et~al.} 2008, \apj,
  683, 304

\bibitem[{{Faure} \& {Josselin}(2008)}]{Faure:2008}
{Faure}, A. \& {Josselin}, E. 2008, \aap, 492, 257

\bibitem[{{Friedel} \& {Snyder}(2008)}]{Friedel:2008}
{Friedel}, D.~N. \& {Snyder}, L.~E. 2008, \apj, 672, 962

\bibitem[{{Garrod} \& {Herbst}(2006)}]{Garrod:2006}
{Garrod}, R.~T. \& {Herbst}, E. 2006, \aap, 457, 927

\bibitem[{{Garrod} {et~al.}(2008){Garrod}, {Weaver}, \& {Herbst}}]{Garrod:2008}
{Garrod}, R.~T., {Weaver}, S.~L.~W., \& {Herbst}, E. 2008, \apj, 682, 283

\bibitem[{{Gaume} {et~al.}(1998){Gaume}, {Wilson}, {Vrba}, {Johnston}, \&
  {Schmid-Burgk}}]{Gaume:1998}
{Gaume}, R.~A., {Wilson}, T.~L., {Vrba}, F.~J., {Johnston}, K.~J., \&
  {Schmid-Burgk}, J. 1998, \apj, 493, 940

\bibitem[{{Genzel} \& {Stutzki}(1989)}]{Genzel:1989}
{Genzel}, R. \& {Stutzki}, J. 1989, \araa, 27, 41

\bibitem[{{Getman} {et~al.}(2005){Getman}, {Flaccomio}, {Broos}, {Grosso},
  {Tsujimoto}, {Townsley}, {Garmire}, {Kastner}, {Li}, {Harnden}, {Wolk},
  {Murray}, {Lada}, {Muench}, {McCaughrean}, {Meeus}, {Damiani}, {Micela},
  {Sciortino}, {Bally}, {Hillenbrand}, {Herbst}, {Preibisch}, \&
  {Feigelson}}]{Getman:2005}
{Getman}, K.~V., {Flaccomio}, E., {Broos}, P.~S., {et~al.} 2005, \apjs, 160,
  319

\bibitem[{{Goddi} {et~al.}(2009){Goddi}, {Greenhill}, {Chandler}, {Humphreys},
  {Matthews}, \& {Gray}}]{Goddi:2009}
{Goddi}, C., {Greenhill}, L.~J., {Chandler}, C.~J., {et~al.} 2009, \apj, 698,
  1165

\bibitem[{{Goddi} {et~al.}(2010){Goddi}, {Humphreys}, {Greenhill}, {Chandler},
  \& {Matthews}}]{Goddi:2010}
{Goddi}, C., {Humphreys}, E.~M.~L., {Greenhill}, L.~J., {Chandler}, C.~J., \&
  {Matthews}, L.~D. 2010, ArXiv e-prints

\bibitem[{{G{\'o}mez} {et~al.}(2008){G{\'o}mez}, {Rodr{\'{\i}}guez}, {Loinard},
  {Lizano}, {Allen}, {Poveda}, \& {Menten}}]{Gomez:2008}
{G{\'o}mez}, L., {Rodr{\'{\i}}guez}, L.~F., {Loinard}, L., {et~al.} 2008, \apj,
  685, 333

\bibitem[{{G{\'o}mez} {et~al.}(2005){G{\'o}mez}, {Rodr{\'{\i}}guez}, {Loinard},
  {Lizano}, {Poveda}, \& {Allen}}]{Gomez:2005}
{G{\'o}mez}, L., {Rodr{\'{\i}}guez}, L.~F., {Loinard}, L., {et~al.} 2005, \apj,
  635, 1166

\bibitem[{{Greenhill} {et~al.}(2004){Greenhill}, {Gezari}, {Danchi}, {Najita},
  {Monnier}, \& {Tuthill}}]{Greenhill:2004}
{Greenhill}, L.~J., {Gezari}, D.~Y., {Danchi}, W.~C., {et~al.} 2004, \apjl,
  605, L57

\bibitem[{{Grosso} {et~al.}(2005){Grosso}, {Feigelson}, {Getman}, {Townsley},
  {Broos}, {Flaccomio}, {McCaughrean}, {Micela}, {Sciortino}, {Bally}, {Smith},
  {Muench}, {Garmire}, \& {Palla}}]{Grosso:2005}
{Grosso}, N., {Feigelson}, E.~D., {Getman}, K.~V., {et~al.} 2005, \apjs, 160,
  530

\bibitem[{{Gu{\'e}lin} {et~al.}(2008){Gu{\'e}lin}, {Brouillet}, {Cernicharo},
  {Combes}, \& {Wooten}}]{Guelin:2008}
{Gu{\'e}lin}, M., {Brouillet}, N., {Cernicharo}, J., {Combes}, F., \& {Wooten},
  A. 2008, \apss, 313, 45

\bibitem[{{Halfen} {et~al.}(2006){Halfen}, {Apponi}, {Woolf}, {Polt}, \&
  {Ziurys}}]{Halfen:2006}
{Halfen}, D.~T., {Apponi}, A.~J., {Woolf}, N., {Polt}, R., \& {Ziurys}, L.~M.
  2006, \apj, 639, 237

\bibitem[{{Hillenbrand} \& {Carpenter}(2000)}]{Hillenbrand:2000}
{Hillenbrand}, L.~A. \& {Carpenter}, J.~M. 2000, \apj, 540, 236

\bibitem[{{Hollis} {et~al.}(2000){Hollis}, {Lovas}, \& {Jewell}}]{Hollis:2000}
{Hollis}, J.~M., {Lovas}, F.~J., \& {Jewell}, P.~R. 2000, \apjl, 540, L107

\bibitem[{{Hollis} {et~al.}(2003){Hollis}, {Pedelty}, {Snyder}, {Jewell},
  {Lovas}, {Palmer}, \& {Liu}}]{Hollis:2003}
{Hollis}, J.~M., {Pedelty}, J.~A., {Snyder}, L.~E., {et~al.} 2003, \apj, 588,
  353

\bibitem[{{Ilyushin} {et~al.}(2008){Ilyushin}, {Kleiner}, \&
  {Lovas}}]{Ilyushin:2008}
{Ilyushin}, V., {Kleiner}, I., \& {Lovas}, F.~J. 2008, Journal of Physical and
  Chemical Reference Data, 37, 97

\bibitem[{{Ilyushin} {et~al.}(2009){Ilyushin}, {Kryvda}, \&
  {Alekseev}}]{Ilyushin:2009}
{Ilyushin}, V., {Kryvda}, A., \& {Alekseev}, E. 2009, Journal of Molecular
  Spectroscopy, 255, 32

\bibitem[{{Irvine} {et~al.}(1987){Irvine}, {Goldsmith}, \&
  {Hjalmarson}}]{Irvine:1987}
{Irvine}, W.~M., {Goldsmith}, P.~F., \& {Hjalmarson}, A. 1987, Astrophysics and
  Space Science Library, Vol. 134, {Chemical abundances in molecular clouds},
  ed. {D.~J.~Hollenbach \& H.~A.~Thronson Jr.}, 561--609

\bibitem[{{Johansson} {et~al.}(1984){Johansson}, {Andersson}, {Ellder},
  {Friberg}, {Hjalmarson}, {Hoglund}, {Irvine}, {Olofsson}, \&
  {Rydbeck}}]{Johansson:1984}
{Johansson}, L.~E.~B., {Andersson}, C., {Ellder}, J., {et~al.} 1984, \aap, 130,
  227

\bibitem[{{Kobayashi} {et~al.}(2007){Kobayashi}, {Ogata}, {Tsunekawa}, \&
  {Takano}}]{Kobayashi:2007}
{Kobayashi}, K., {Ogata}, K., {Tsunekawa}, S., \& {Takano}, S. 2007, \apjl,
  657, L17

\bibitem[{{Lacombe} {et~al.}(2004){Lacombe}, {Gendron}, {Rouan}, {Cl{\'e}net},
  {Field}, {Lemaire}, {Gustafsson}, {Lagrange}, {Mouillet}, {Rousset}, {Fusco},
  {Rousset-Rouvi{\`e}re}, {Servan}, {Marlot}, \& {Feautrier}}]{Lacombe:2004}
{Lacombe}, F., {Gendron}, E., {Rouan}, D., {et~al.} 2004, \aap, 417, L5

\bibitem[{{Lada} {et~al.}(2004){Lada}, {Muench}, {Lada}, \&
  {Alves}}]{Lada:2004}
{Lada}, C.~J., {Muench}, A.~A., {Lada}, E.~A., \& {Alves}, J.~F. 2004, \aj,
  128, 1254

\bibitem[{{Lattelais} {et~al.}(2009){Lattelais}, {Pauzat}, {Ellinger}, \&
  {Ceccarelli}}]{Lattelais:2009}
{Lattelais}, M., {Pauzat}, F., {Ellinger}, Y., \& {Ceccarelli}, C. 2009, \apjl,
  696, L133

\bibitem[{{Leurini} {et~al.}(2004){Leurini}, {Schilke}, {Menten}, {Flower},
  {Pottage}, \& {Xu}}]{Leurini:2004}
{Leurini}, S., {Schilke}, P., {Menten}, K.~M., {et~al.} 2004, \aap, 422, 573

\bibitem[{{Liu} {et~al.}(2002){Liu}, {Girart}, {Remijan}, \&
  {Snyder}}]{Liu:2002}
{Liu}, S., {Girart}, J.~M., {Remijan}, A., \& {Snyder}, L.~E. 2002, \apj, 576,
  255

\bibitem[{{Lonsdale} {et~al.}(1982){Lonsdale}, {Becklin}, {Lee}, \&
  {Stewart}}]{Lonsdale:1982}
{Lonsdale}, C.~J., {Becklin}, E.~E., {Lee}, T.~J., \& {Stewart}, J.~M. 1982,
  \aj, 87, 1819

\bibitem[{{Mathis} {et~al.}(1977){Mathis}, {Rumpl}, \&
  {Nordsieck}}]{Mathis:1977}
{Mathis}, J.~S., {Rumpl}, W., \& {Nordsieck}, K.~H. 1977, \apj, 217, 425

\bibitem[{{Matthews} {et~al.}(2010){Matthews}, {Greenhill}, {Goddi},
  {Chandler}, {Humphreys}, \& {Kunz}}]{Matthews:2010}
{Matthews}, L.~D., {Greenhill}, L.~J., {Goddi}, C., {et~al.} 2010, \apj, 708,
  80

\bibitem[{{Matveenko} {et~al.}(2000){Matveenko}, {Diamond}, \&
  {Graham}}]{Matveenko:2000}
{Matveenko}, L.~I., {Diamond}, P.~J., \& {Graham}, D.~A. 2000, Astronomy
  Reports, 44, 592

\bibitem[{{Matveyenko} {et~al.}(2007){Matveyenko}, {Graham}, \&
  {Demichev}}]{Matveyenko:2007}
{Matveyenko}, L.~I., {Graham}, D.~A., \& {Demichev}, V.~A. 2007, Memorie della
  Societa Astronomica Italiana, 78, 419

\bibitem[{{Menten} {et~al.}(2007){Menten}, {Reid}, {Forbrich}, \&
  {Brunthaler}}]{Menten:2007}
{Menten}, K.~M., {Reid}, M.~J., {Forbrich}, J., \& {Brunthaler}, A. 2007, \aap,
  474, 515

\bibitem[{{Menten} {et~al.}(1988){Menten}, {Walmsley}, {Henkel}, \&
  {Wilson}}]{Menten:1988}
{Menten}, K.~M., {Walmsley}, C.~M., {Henkel}, C., \& {Wilson}, T.~L. 1988,
  \aap, 198, 253

\bibitem[{{Mezger} {et~al.}(1990){Mezger}, {Wink}, \& {Zylka}}]{Mezger:1990}
{Mezger}, P.~G., {Wink}, J.~E., \& {Zylka}, R. 1990, \aap, 228, 95

\bibitem[{{Muench} {et~al.}(2002){Muench}, {Lada}, {Lada}, \&
  {Alves}}]{Muench:2002}
{Muench}, A.~A., {Lada}, E.~A., {Lada}, C.~J., \& {Alves}, J. 2002, \apj, 573,
  366

\bibitem[{{M{\"u}ller} {et~al.}(2005){M{\"u}ller}, {Schl{\"o}der}, {Stutzki},
  \& {Winnewisser}}]{Muller:2005}
{M{\"u}ller}, H.~S.~P., {Schl{\"o}der}, F., {Stutzki}, J., \& {Winnewisser}, G.
  2005, Journal of Molecular Structure, 742, 215

\bibitem[{{M{\"u}ller} {et~al.}(2001){M{\"u}ller}, {Thorwirth}, {Roth}, \&
  {Winnewisser}}]{Muller:2001}
{M{\"u}ller}, H.~S.~P., {Thorwirth}, S., {Roth}, D.~A., \& {Winnewisser}, G.
  2001, \aap, 370, L49

\bibitem[{{Nissen} {et~al.}(2007){Nissen}, {Gustafsson}, {Lemaire},
  {Cl{\'e}net}, {Rouan}, \& {Field}}]{Nissen:2007}
{Nissen}, H.~D., {Gustafsson}, M., {Lemaire}, J.~L., {et~al.} 2007, \aap, 466,
  949

\bibitem[{{O'dell}(2001)}]{ODell:2001}
{O'dell}, C.~R. 2001, in Revista Mexicana de Astronomia y Astrofisica, vol. 27,
  Vol.~10, Revista Mexicana de Astronomia y Astrofisica Conference Series, ed.
  {J.~Cant{\'o} \& L.~F.~Rodr{\'{\i}}guez}, 1--8

\bibitem[{{O'Dell} \& {Henney}(2008)}]{ODell:2008}
{O'Dell}, C.~R. \& {Henney}, W.~J. 2008, \aj, 136, 1566

\bibitem[{{O'Dell} {et~al.}(2009){O'Dell}, {Henney}, {Abel}, {Ferland}, \&
  {Arthur}}]{ODell:2009}
{O'Dell}, C.~R., {Henney}, W.~J., {Abel}, N.~P., {Ferland}, G.~J., \& {Arthur},
  S.~J. 2009, \aj, 137, 367

\bibitem[{{O'dell} {et~al.}(1993){O'dell}, {Wen}, \& {Hu}}]{ODell:1993}
{O'dell}, C.~R., {Wen}, Z., \& {Hu}, X. 1993, \apj, 410, 696

\bibitem[{{Ossenkopf} \& {Henning}(1994)}]{Ossenkopf:1994}
{Ossenkopf}, V. \& {Henning}, T. 1994, \aap, 291, 943

\bibitem[{{Parenago}(1954)}]{Parenago:1954}
{Parenago}, P.~P. 1954, Trudy Gosudarstvennogo Astronomicheskogo Instituta, 25,
  3

\bibitem[{{Parenago}(1997)}]{Parenago:1997}
{Parenago}, P.~P. 1997, VizieR Online Data Catalog, 2171, 0

\bibitem[{{Persson} {et~al.}(2007){Persson}, {Olofsson}, {Koning}, {Bergman},
  {Bernath}, {Black}, {Frisk}, {Geppert}, {Hasegawa}, {Hjalmarson}, {Kwok},
  {Larsson}, {Lecacheux}, {Nummelin}, {Olberg}, {Sandqvist}, \&
  {Wirstr{\"o}m}}]{Persson:2007}
{Persson}, C.~M., {Olofsson}, A.~O.~H., {Koning}, N., {et~al.} 2007, \aap, 476,
  807

\bibitem[{{Pickett}(1991)}]{Pickett:1991}
{Pickett}, H.~M. 1991, Journal of Molecular Spectroscopy, 148, 371

\bibitem[{{Pickett} {et~al.}(1998){Pickett}, {Poynter}, {Cohen}, {Delitsky},
  {Pearson}, \& {Muller}}]{Pickett:1998}
{Pickett}, H.~M., {Poynter}, I.~R.~L., {Cohen}, E.~A., {et~al.} 1998, Journal
  of Quantitative Spectroscopy and Radiative Transfer, 60, 883

\bibitem[{{Plambeck} \& {Wright}(1988)}]{Plambeck:1988}
{Plambeck}, R.~L. \& {Wright}, M.~C.~H. 1988, \apjl, 330, L61

\bibitem[{{Plambeck} {et~al.}(2009){Plambeck}, {Wright}, {Friedel}, {Widicus
  Weaver}, {Bolatto}, {Pound}, {Woody}, {Lamb}, \& {Scott}}]{Plambeck:2009}
{Plambeck}, R.~L., {Wright}, M.~C.~H., {Friedel}, D.~N., {et~al.} 2009, \apjl,
  704, L25

\bibitem[{{Remijan} {et~al.}(2003){Remijan}, {Snyder}, {Friedel}, {Liu}, \&
  {Shah}}]{Remijan:2003}
{Remijan}, A., {Snyder}, L.~E., {Friedel}, D.~N., {Liu}, S., \& {Shah}, R.~Y.
  2003, \apj, 590, 314

\bibitem[{{Robberto} {et~al.}(2005){Robberto}, {Beckwith}, {Panagia}, {Patel},
  {Herbst}, {Ligori}, {Custo}, {Boccacci}, \& {Bertero}}]{Robberto:2005}
{Robberto}, M., {Beckwith}, S.~V.~W., {Panagia}, N., {et~al.} 2005, \aj, 129,
  1534

\bibitem[{{Rodr{\'{\i}}guez} {et~al.}(2005){Rodr{\'{\i}}guez}, {Poveda},
  {Lizano}, \& {Allen}}]{Rodriguez:2005}
{Rodr{\'{\i}}guez}, L.~F., {Poveda}, A., {Lizano}, S., \& {Allen}, C. 2005,
  \apjl, 627, L65

\bibitem[{{Rohlfs} \& {Wilson}(2000)}]{Rohlfs:2000}
{Rohlfs}, K. \& {Wilson}, T.~L. 2000, {Book Review: Tools of radio astronomy.-
  3rd rev. ed. Springer, 2000}, Vol. 120, 289

\bibitem[{{Schilke} {et~al.}(1997){Schilke}, {Groesbeck}, {Blake}, \&
  {Phillips}}]{Schilke:1997}
{Schilke}, P., {Groesbeck}, T.~D., {Blake}, G.~A., \& {Phillips}, T.~G. 1997,
  \apjs, 108, 301

\bibitem[{{Shiao} {et~al.}(2010){Shiao}, {Looney}, {Remijan}, {Snyder}, \&
  {Friedel}}]{Shiao:2010}
{Shiao}, Y., {Looney}, L.~W., {Remijan}, A.~J., {Snyder}, L.~E., \& {Friedel},
  D.~N. 2010, \apj, 716, 286

\bibitem[{{Shuping} {et~al.}(2004){Shuping}, {Morris}, \&
  {Bally}}]{Shuping:2004}
{Shuping}, R.~Y., {Morris}, M., \& {Bally}, J. 2004, \aj, 128, 363

\bibitem[{{Simpson} {et~al.}(2006){Simpson}, {Colgan}, {Erickson}, {Burton}, \&
  {Schultz}}]{Simpson:2006}
{Simpson}, J.~P., {Colgan}, S.~W.~J., {Erickson}, E.~F., {Burton}, M.~G., \&
  {Schultz}, A.~S.~B. 2006, \apj, 642, 339

\bibitem[{{Smith} {et~al.}(2005){Smith}, {Bally}, {Shuping}, {Morris}, \&
  {Kassis}}]{Smith:2005}
{Smith}, N., {Bally}, J., {Shuping}, R.~Y., {Morris}, M., \& {Kassis}, M. 2005,
  \aj, 130, 1763

\bibitem[{{Snyder} {et~al.}(2005){Snyder}, {Lovas}, {Hollis}, {Friedel},
  {Jewell}, {Remijan}, {Ilyushin}, {Alekseev}, \& {Dyubko}}]{Snyder:2005}
{Snyder}, L.~E., {Lovas}, F.~J., {Hollis}, J.~M., {et~al.} 2005, \apj, 619, 914

\bibitem[{{Stolovy} {et~al.}(1998){Stolovy}, {Burton}, {Erickson}, {Kaufman},
  {Chrysostomou}, {Young}, {Colgan}, {Axon}, {Thompson}, {Rieke}, \&
  {Schneider}}]{Stolovy:1998}
{Stolovy}, S.~R., {Burton}, M.~G., {Erickson}, E.~F., {et~al.} 1998, \apjl,
  492, L151+

\bibitem[{{Tang} {et~al.}(2010){Tang}, {Ho}, {Koch}, \& {Rao}}]{Tang:2010}
{Tang}, Y., {Ho}, P.~T.~P., {Koch}, P.~M., \& {Rao}, R. 2010, \apj, 717, 1262

\bibitem[{{Tielens}(2005)}]{Tielens:2005}
{Tielens}, A.~G.~G.~M. 2005, {The Physics and Chemistry of the Interstellar
  Medium}, ed. {Tielens, A.~G.~G.~M.}

\bibitem[{{Turner}(1991)}]{Turner:1991}
{Turner}, B.~E. 1991, \apjs, 76, 617

\bibitem[{{van der Tak} {et~al.}(2000){van der Tak}, {van Dishoeck}, \&
  {Caselli}}]{van-der-Tak:2000}
{van der Tak}, F.~F.~S., {van Dishoeck}, E.~F., \& {Caselli}, P. 2000, \aap,
  361, 327

\bibitem[{{Wang} {et~al.}(2010){Wang}, {Kuan}, {Liu}, \&
  {Charnley}}]{Wang:2010}
{Wang}, K., {Kuan}, Y., {Liu}, S., \& {Charnley}, S.~B. 2010, \apj, 713, 1192

\bibitem[{{Wang} {et~al.}(2011){Wang}, {Bergin}, {Crockett}, {Goldsmith},
  {Lis}, {Pearson}, {Schilke}, {Bell}, {Comito}, {Blake}, {Caux}, {Ceccarelli},
  {Cernicharo}, {Daniel}, {Dubernet}, {Emprechtinger}, {Encrenaz}, {Gerin},
  {Giesen}, {Goicoechea}, {Gupta}, {Herbst}, {Joblin}, {Johnstone}, {Langer},
  {Latter}, {Lord}, {Maret}, {Martin}, {Melnick}, {Menten}, {Morris}, {Muller},
  {Murphy}, {Neufeld}, {Ossenkopf}, {Perault}, {Phillips}, {Plume}, {Qin},
  {Schlemmer}, {Stutzki}, {Trappe}, {van der Tak}, {Vastel}, {Yorke}, {Yu}, \&
  {Zmuidzinas}}]{Wang:2011}
{Wang}, S., {Bergin}, E.~A., {Crockett}, N.~R., {et~al.} 2011, ArXiv e-prints

\bibitem[{{Wilson} {et~al.}(2000){Wilson}, {Gaume}, {Gensheimer}, \&
  {Johnston}}]{Wilson:2000}
{Wilson}, T.~L., {Gaume}, R.~A., {Gensheimer}, P., \& {Johnston}, K.~J. 2000,
  \apj, 538, 665

\bibitem[{{Wynn-Williams} {et~al.}(1984){Wynn-Williams}, {Genzel}, {Becklin},
  \& {Downes}}]{Wynn-Williams:1984}
{Wynn-Williams}, C.~G., {Genzel}, R., {Becklin}, E.~E., \& {Downes}, D. 1984,
  \apj, 281, 172

\bibitem[{{Zapata} {et~al.}(2009){Zapata}, {Schmid-Burgk}, {Ho},
  {Rodr{\'{\i}}guez}, \& {Menten}}]{Zapata:2009}
{Zapata}, L.~A., {Schmid-Burgk}, J., {Ho}, P.~T.~P., {Rodr{\'{\i}}guez}, L.~F.,
  \& {Menten}, K.~M. 2009, \apjl, 704, L45

\bibitem[{{Zapata} {et~al.}(2010){Zapata}, {Schmid-Burgk}, \&
  {Menten}}]{Zapata:2010b}
{Zapata}, L.~A., {Schmid-Burgk}, J., \& {Menten}, K.~M. 2010, ArXiv e-prints

\bibitem[{{Ziurys} \& {McGonagle}(1993)}]{Ziurys:1993}
{Ziurys}, L.~M. \& {McGonagle}, D. 1993, \apjs, 89, 155

\end{thebibliography}

%===================================================================================================

%-----------------------------------------------------------------------------------------------------------------------------
%---------------------------------------------------TABLE MF1 ------------------------------------------------
%-----------------------------------------------------------------------------------------------------------------------------

%------------------- 
%MF1 TABLE
%------------------- 
\longtabL{10}{
\begin{landscape}
\begin{longtable}{clllccccccccccl}
\caption{\label{MF1} Transitions of methyl formate observed with the Plateau de Bure Interferometer toward position MF1 in Orion-KL}\\
\hline\hline
N$\degr$ & Set & Frequency &Transition & E$\rm_{up}$ & S$\mu$$^{2}$ & sigma & v & $\Delta$v$\rm_{1/2}$ & T$\rm_{B}$ & W & N$\rm_{up}$/g$\rm_{up}$& \multicolumn{2}{c}{Source size}  &Comments \\
  & &(MHz) &  & (K) & (D$^{2}$) & (K) & (~km~s$^{-1}$) & (~km~s$^{-1}$) & (K) & (K~km~s$^{-1}$) & (10$^{12}$~cm$^{-2}$)  & ($\arcsec$ x $\arcsec$) &PA($\degr$) & \\
(1) & (2) & (3) & (4) &  (5) &  (6) &  (7) &  (8) &  (9) &  (10) &  (11) &  (12)&  (13) &  (14) &(15) \\
\hline
\hline
\endfirsthead
\caption{continued.}\\
\hline\hline
N$\degr$ & Set& Frequency &Transition & E$\rm_{up}$ & S$\mu$$^{2}$ & sigma & v & $\Delta$v$\rm_{1/2}$ & T$\rm_{B}$ & W & N$\rm_{up}$/g$\rm_{up}$& \multicolumn{2}{c}{Source size}  &Comments \\
  & &(MHz) &  & (K) & (D$^{2}$) & (K) & (~km~s$^{-1}$) & (~km~s$^{-1}$) & (K) & (K~km~s$^{-1}$) & (10$^{12}$~cm$^{-2}$)  & ($\arcsec$ x $\arcsec$) &PA($\degr$) & \\
(1) & (2) & (3) & (4) &  (5) &  (6) &  (7) &  (8) &  (9) &  (10) &  (11) &  (12)&  (13) &  (14) &(15) \\
\hline
\endhead
\hline
\endfoot
1 & 1& 80531.669(0.030) & $9_{2,8} - 9_{0,9}$ E, (v$\rm_{t}$=1) & 216 & 0.6 & 0.05 &- & - & $\la$0.15 &$\la$0.6 &$\la$1.04& -&-& not detected \\
2 & 1&80565.210(0.010) & $10_{2,8} - 9_{3,7}$ E & 36 & 1.1 & 0.05 & 7.9(0.1)& 3.8(0.3)& 0.51 & 2.0(0.1)& 1.88 & 10.0x7.0 & 10 &detected \\
3 &1& 80572.589(0.010) & $10_{2,8} - 9_{3,7}$ A & 36 & 1.1 & 0.05 & 8.0(0.5) & 4.3(1.5) & 0.47 & 2.2(0.5) & 2.07& 10.0x8.0 & 45 &detected\\
4  &2& 80604.508(0.030) & $12_{9,3} - 13_{8,5}$ E, (v$\rm_{t}$=1) & 288 & 0.2 & 0.09 & - & - & $\la$0.31 & $\la$1.3 &$\la$6.73& -&-& blend \\
5 & 2&80652.521(0.010) & $15_{4,11} - 14_{5,10}$ A & 83 & 1.0 & 0.09 & - & - & $\la$0.74 & $\la$3.0 &$\la$3.10& -&-& blend \\
6 & 3&101202.806(0.010) &  $9_{1,9} - 8_{0,8}$ A, (v$\rm_{t}$=1) & 213 & 4.2 & 0.04 & 7.6(0.1) & 3.5(0.2) & 1.02 & 3.8(0.2) & 0.75& -&-& partial blend \\
&&&&&&&&&&&&&&with U-line\\
7 & 3&101210.208(0.004)& $24_{3,21} - 25_{2,24}$ A&187 &0.04 & 0.04 & - & - & $\la$0.12 & $\la$0.4 &$\la$8.68& -&-& not detected \\
8  &3& 101279.006(0.010) & $9_{1,9} - 8_{0,8}$ E, (v$\rm_{t}$=1) & 212 & 4.0 & 0.04 & 7.8(0.1)& 3.4(0.1)& 1.18 & 4.3(0.1) & 0.89 & 5.5x3.5 & 84 & detected \\
9  &3& 101289.471(0.010) & $29_{9,20} - 28_{10,19}$ A, (v$\rm_{t}$=1) & 498 & 1.8 & 0.04 & - & - & $\la$0.12 & $\la$0.4 &$\la$0.18& -&-& not detected \\
10 &3& 101302.159(0.010) & $25_{6,19} - 25_{5,20}$ A & 219 & 11.1 & 0.04 & 7.6(0.1) & 3.9(0.4) & 3.11 & 12.8(1.0) & 0.95 & 5.0x3.5 & 26 & detected \\
11 &3& 101305.506(0.010) & $25_{6,19} -25 _{5,20}$ E & 219 & 11.1 & 0.04 & 7.8(0.2) & 3.7(0.4) & 2.65 & 10.4(1.0)& 0.77& 5.5x3.5 & 20 & detected \\
12 &3& 101318.966(0.010) & $13_{3,11} - 13_{2,12}$ E, (v$\rm_{t}$=1) & 246 & 3.2 & 0.04 & 7.8(0.1) & 3.1(0.3) & 0.70 & 2.3(0.2) &0.59 & 5.5x3.5 & 5 & detected \\
13 & 3&101356.788(0.010) & $15_{2,13} - 15_{1,14}$ A, (v$\rm_{t}$=1) & 263 & 3.5 & 0.04 & - & - & $\la$0.59 & $\la$2.1&$\la$0.49& -&-& blend \\
14 & 3&101370.505(0.010) & $13_{3,11} - 13_{2,12}$ E & 60 & 3.2 & 0.04 & 7.9(0.1)& 3.7(0.1)& 3.78 & 14.7(0.1)& 3.78& 5.0x3.5 & 5 & detected \\
15 & 3&101414.746(0.010) & $13_{3,11} - 13_{2,12}$ A & 60 & 3.2 & 0.04 & 7.8(0.1) & 3.6(0.1) & 3.73 & 14.3(0.1) & 3.68& 5.0x3.5 & 5&  detected  \\
16 & 3&101418.330(0.001)& $5_{3,3} - 5_{1,4}$ A, (v$\rm_{t}$=1) & 203 &0.04 & 0.04 & - & - & $\la$0.12 & $\la$0.4 &$\la$7.37& -&-& not detected \\
17 &3& 101477.421(0.010) & $18_{3,15} - 18_{3,16}$ E & 111 & 1.8 & 0.04 & - & - &$\la$9.31 & $\la$33.2&$\la$15.17& -&-& blend with H$\rm_{2}$CS \\
18 &3& 101545.453(0.010) & $18_{3,15} - 18_{3,16}$ A & 111 & 1.8 & 0.04 & 7.8(0.1) & 3.6(0.1) &1.80 & 6.9(0.2) &3.15 & 5.0x3.5 & 5 &  detected \\
19 & 3&101626.884(0.010) & $9_{1,9} - 8_{0,8}$ E & 25 & 4.1 & 0.04 & - & - & $\la$5.96 &$\la$21.2 &$\la$4.25& -&-& blend with   \\
&&&&&&&&&&&&&&HCOOCH$\rm_{3}$-A\\
20 & 3&101628.149(0.010) & $9_{1,9} - 8_{0,8}$ A & 25 & 4.1 & 0.04 & - & - & $\la$5.47 & $\la$19.5 &$\la$3.91& -&-& blend with  \\
&&&&&&&&&&&&&&HCOOCH$\rm_{3}$-E\\
21 &4& 105663.117(0.030) & $19_{13,7} - 20_{12,9}$ E, (v$\rm_{t}$=1) & 412 & 0.4 &0.02 &- & - & $\la$0.06 & $\la$0.1 &$\la$0.20& -&-& not detected \\
22 & 4&105722.672(0.030)  & $10_{4,6} - 10_{3,8}$ E, (v$\rm_{t}$=1) & 230 & 0.5 &0.02 & - & - & $\la$0.11& $\la$0.2 &$\la$0.32& -&-& blend \\
23 & 5&105810.803(0.030)  & $14_{11,4} - 15_{10,6}$ E & 142 & 0.2 &0.03 & - & - & $\la$0.11 &$\la$0.2 &$\la$0.79& -&-& blend \\
24 & 5&105815.953(0.030)  & $3_{3,1} - 2_{2,0}$ E & 10 & 0.2 &0.03 & 7.6(0.1) & 2.2(0.2) & 0.39 & 0.9(0.1)& 3.55& 10.5x7.5 & 6 &detected \\
25 & 5&105832.067(0.030)  & $14_{11,3} - 15_{10,5}$ E & 142 & 0.2 & 0.03& 7.9(0.1) & 1.6(0.2) & 0.20 & 0.4(0.1) & 1.58& 10.5x7.5 & 6 & detected \\
26 & 6&110153.652(0.010) & $10_{1,10} - 9_{1,9}$ A, (v$\rm_{t}$=1) & 218 & 25.1& 0.12 & - & - & $\la$8.08 & $\la$9.7 &$\la$0.29 & -&-& blend with NH$\rm_{2}$D\\
27 &7& 203349.158(0.006)& $24_{19,5} - 25_{18,7}$ E & 416 &0.2 & 0.09 &- &- &$\la$1.73 &$\la$2.8 &$\la$5.77 &-&- & blend\\ 
28 &7& 203378.812(0.007)& $24_{19,6} - 25_{18,8}$ E & 416 &0.2 & 0.09 &- &- &$\la$5.64 &$\la$9.2 &$\la$18.94 &-&- & blend\\ 
29 &8& 203427.575(0.100) & $21_{8,13} - 21_{7,14}$ A, (v$\rm_{t}$=1) & 366 & 6.2 & 0.09 & 7.9(0.4) & 1.7(0.8) & 1.40 & 2.5(1.0) & 0.17& 4.5x2.5 & 45 & detected \\
30 & 8&203435.554(0.100) & $19_{8,11} - 19_{7,12}$ A & 155 & 5.3 & 0.09 & 7.6(0.9) & 1.6(0.9) & 6.40 & 10.6(1.6) &0.82 & 5.0x3.5 & 65 & detected  \\
31 &8& 203471.840(0.05) & $19_{8,11} - 19_{7,12}$ E & 155 & 4.9 & 0.09 & 7.5(0.1)& 1.6(0.1)& 5.28 & 9.2(2.0)& 0.77& 5.0x3.5 & 40 & detected  \\
32 & 9& 223435.416(0.001)& $11_{4,8} - 10_{3,7}$ A, (v$\rm_{t}$=1) & 237 & 1.8 & 0.17 & - & - &$\la$3.66 &$\la$6.7 &$\la$1.38 &- &- & blend \\ 
33 & 9&223465.340(0.050) & $11_{4,8} - 10_{3,7}$ E & 50 & 2.1 & 0.17 & 7.6(0.1)& 2.0(0.2)& 10.74 &22.5(1.9) & 4.00 & 3.0x2.0& 0 &detected \\
34 &9& 223500.463(0.001) & $11_{4,8} - 10_{3,7}$ A & 50 & 1.8 & 0.17 & 7.4(0.1)&1.8(0.1) & 11.49 &21.8(1.7)& 4.61& 3.0x1.5& 0 &detected \\
35 & 9&223534.727(0.100) & $18_{5,14} - 17_{5,13}$ E, (v$\rm_{t}$=1) & 305 & 41.6 & 0.17 & 7.4(0.1) & 1.7(0.1)& 10.25 & 18.5(0.8) & 0.17& 3.5x2.0& 0 & detected\\
36 & 9&223538.511(0.100) & $43_{8,35} - 43_{7,36}$ E & 618 & 13.9 & 0.17 & - & - & $\la$0.51 &$\la$0.9 &$\la$0.02& -&-& not detected \\
37 & 9&223592.231(0.100) & $43_{8,35} - 43_{7,36}$ A & 618 & 13.9 &0.17 & - & - & $\la$0.91 & $\la$1.7&$\la$0.05& -&-& blend \\
38 & 9&223618.470(0.007)& $27_{9,19} - 27_{8,19}$ E, (v$\rm_{t}$=1) & 464 &0.4 & 0.17 & - & - & $\la$0.51 &$\la$0.9 &$\la$0.90& -&-& not detected \\ 
39 & 9&223624.496(0.100) & $37_{8,30} - 37_{7,31}$ E & 463 & 11.7 & 0.17& 7.7(0.3)& 1.8(0.6) & 0.68  & 1.3(0.4) &0.04 & -& - & partial blend \\
40 & 9&223634.916(0.100) & $37_{8,30} - 37_{7,31}$ A & 463 & 11.7 & 0.17& - & - &$\la$2.06 & $\la$3.8 &$\la$0.12& -&-& blend \\
41 & 9&223642.180(0.001)  & $12_{5,7} - 12_{3,10}$ E & 63 & 0.1 & 0.17 & - &-& $\la$4.08 & $\la$7.4& $\la$32.97&-&-& blend \\
42 & 9& 223645.342(0.001)& $12_{5,7} - 12_{3,10}$ A & 63 & 0.1 & 0.17 & - & - &$\la$1.66 &$\la$3.0 &$\la$11.23 &- &- & blend \\ 
43 &9&  223651.592(0.007)& $33_{9,24} - 33_{8,26}$ E, (v$\rm_{t}$=1) & 573 &0.7 & 0.17 & - & - &$\la$1.35 &$\la$2.5 &$\la$1.25 &- &- & blend \\ 
44 & 9& 223676.472(0.006)& $35_{7,29} - 35_{5,30}$ A, (v$\rm_{t}$=1) & 593 &3.0 & 0.17 & - & - &$\la$2.09 &$\la$3.8 &$\la$0.47 &- &- & blend \\ 
45 & 9& 223782.081(0.007)& $33_{6,28} - 33_{5,29}$ E, (v$\rm_{t}$=1) & 543 &7.0 & 0.17 & - & - & $\la$0.51 &$\la$0.9 &$\la$0.05& -&-& not detected \\ 
46 & 9& 223821.523(0.002)& $35_{7,29} - 35_{6,30}$ E & 409 &8.2 & 0.17 & - & - &$\la$2.31 &$\la$4.2 &$\la$0.19 &- &- & blend \\ 
47 &9& 223854.201(0.100) & $35_{7,29} - 35_{6,30}$ A & 409 & 10.1 & 0.17& - & - & $\la$4.63 & $\la$8.4 &$\la$0.31& -&-& blend \\
48 & 10&225689.371(0.008)& $34_{5,29} - 34_{5,30}$ A, (v$\rm_{t}$=1) & 562 &3.0 & 0.13 &- & -& $\la$12.20& $\la$12.6&$\la$1.57& -&-& blend \\
49  &10& 225696.842(0.100) & $20_{1,19} - 19_{1,18}$ E, (v$\rm_{t}$=1) & 307 & 50.2 & 0.13& - & - & $\la$16.8 & $\la$17.4 &$\la$0.13& -&-& blend  \\
50 &10& 225702.857(0.100) & $19_{2,17} - 18_{2,16}$ A, (v$\rm_{t}$=1)& 304 & 46.8 &0.13 & - & - & $\la$20.09 & $\la$20.8 &$\la$0.16& -&-& blend \\
51 &10& 225727.506(0.100) & $6_{6,1} - 5_{5,0}$ A, (v$\rm_{t}$=1) & 224 & 3.1 &0.13 & - & - & $\la$4.04 & $\la$4.2 &$\la$0.50& -&-& blend \\
52 &10& 225727.506(0.100) & $6_{6,0} - 5_{5,1}$ A, (v$\rm_{t}$=1) & 224 & 3.1 &0.13 & - & - & $\la$4.04 & $\la$4.2 &$\la$0.50& -&-& blend \\
53  &11& 225855.505(0.100) & $6_{6,1} - 5_{5,1}$ E & 36 & 3.1 &0.10 & 7.6(0.1)&1.2(0.1)& 6.84 & 9.1(0.1) &1.08 & 5.5x4.5 & 5 & detected\\
54 &11& 225900.684(0.100) & $6_{6,0} - 5_{5,0}$ E & 36 & 3.1 & 0.10& - & - &$\la$12.74 & $\la$13.2 &$\la$1.57& -&-& blend with HDO \\
55 &11& 225928.659(0.100) & $6_{6,0} - 5_{5,1}$ A & 36 & 3.1 & 0.10& - & - & $\la$11.42 & $\la$11.8 &$\la$1.41& -&-& blend  \\
56 & 11&225928.659(0.100) & $6_{6,1} - 5_{5,0}$ A & 36 & 3.1 & 0.10& - & - & $\la$11.42 & $\la$11.8 &$\la$1.41& -&-& blend\\
57 & 12&225999.145(0.100) & $30_{7,23} - 29_{8,22}$ E & 312 & 2.2 &0.11 & 7.5(0.1) & 0.9(0.3) & 0.70 & 0.6(0.2)& 0.10& 5.0x3.5 & 0 & detected \\
58 & 12&226061.796(0.100) & $20_{3,17} - 19_{4,16}$ E, (v$\rm_{t}$=1) & 321 & 4.0 & 0.11 & 7.6(0.2)& 1.0(0.5) & 0.72 & 0.8(0.3) & 0.07& 5.5x3.5 & 25 & detected \\
59 & 12& 226077.920(0.001)& $10_{3,7} - 9_{1,8}$ E & 39 &0.3 & 0.11 &- & -& $\la$2.62& $\la$2.7&$\la$3.53& -&-& blend \\
60 & 12& 226081.175(0.002)& $30_{7,23} - 29_{8,22}$ A & 312 &1.8 & 0.11 &- & -& $\la$0.71&$\la$0.7 &$\la$0.14& -&-& blend \\
61 & 12& 226087.718(0.007)& $29_{4,26} - 29_{3,27}$ A, (v$\rm_{t}$=1) & 450 &4.3 & 0.11 &- & -& $\la$0.82& $\la$0.8&$\la$0.07& -&-& blend \\
62 & 12&226090.301(0.100) & $19_{2,17} - 18_{2,16}$ E, (v$\rm_{t}$=1) & 303 & 47.0 & 0.11 & - & - & $\la$9.62 & $\la$9.6 &$\la$0.08& -&-& blend \\
63 &12&  226112.470(0.007)& $29_{4,26} - 29_{2,27}$ A, (v$\rm_{t}$=1) & 450 &2.1 & 0.11 &- & -& $\la$0.33& $\la$0.3&$\la$0.05& -&-& not detected \\
64 & 12& 226125.600(0.001)& $10_{3,7} - 9_{1,8}$ A & 39 &0.3 & 0.11 &- & -& $\la$4.04& $\la$4.2&$\la$5.49& -&-& blend \\
\end{longtable}
\flushleft
Notes: (1) Numbering of the observed transitions with E$\rm_{upper}$ $\la$ 650 K. (2) Reference number of the corresponding data set (see Table \ref{Table.dataset_parameters}). (3) Frequencies and uncertainties taken from Ilyushin et al. (2009) and from the JPL database (http://spec.jpl.nasa.gov/). (4) Transitions in ground state (v$\rm_{t}$=0). Transitions in torsional levels (v$\rm_{t}$=1) are specified. (5) Energy of the upper level. (6) Line strength calculated by Ilyushin et al. (2009). For the transitions coming from the JPL database, the line strength is calculated from the formulae of \citet{Pickett:1998}. (7) Noise estimated from spectrum (1$\sigma$). (8) Lines with number 30, 31, 33, 34, 35, 41, 53 and 57 are decomposed. Only the first component is given here, other components are given in Table \ref{Table.MF1bis}, except for lines 29, 39 and 58 because of the blend with the second component. (8), (9), (10) and (11) Velocity, linewidth at half intensity, brightness temperature and integrated intensities. The uncertainties are estimated with the CLASS software. (12) Column densities of the upper state of the transition with respect to the rotational degeneracy (2J$\rm_{up}$+1). (13) and (14) Source size estimated at half flux density from detected transitions with spatial resolution allowing to isolate the source. (15) Blended lines are specified. 
\end{landscape}
} %End \longtabL
%------------------- 

%===================================================================================================

%-----------------------------------------------------------------------------------------------------------------------------
%------------------------------------TABLES - ONLINE APPENDICES---------------------------
%-----------------------------------------------------------------------------------------------------------------------------

\Online

\begin{appendix} %First online appendix
\section{Transitions of methyl formate observed with the Plateau de Bure Interferometer towards positions MF2 to MF5 in Orion-KL.}

The following tables summarize the line parameters for all detected, blended or not detected transitions of the methyl formate molecule (HCOOCH$\rm_{3}$) in all our PdBI data sets towards the emission peaks MF2 to MF5.
%------------------- 
%MF2 TABLE
%-------------------
\longtabL{1}{
\begin{landscape}
\begin{longtable}{clllccccccccccl}
\caption{\label{MF2} Transitions of methyl formate observed with the Plateau de Bure Interferometer toward position MF2 in Orion-KL}\\
\hline\hline
N$\degr$ &Set& Frequency &Transition & E$\rm_{up}$ & S$\mu$$^{2}$ & sigma & v & $\Delta$v$\rm_{1/2}$ & T$\rm_{B}$ & W &N$\rm_{up}$/g$\rm_{up}$& \multicolumn{2}{c}{Source size}  &Comments \\
  && (MHz) &  & (K) & (D$^{2}$) & (K) & (~km~s$^{-1}$) & (~km~s$^{-1}$) & (K) & (K~km~s$^{-1}$) &(10$^{12}$~cm$^{-2}$)  & ($\arcsec$ x $\arcsec$) &PA($\degr$) & \\
(1) & (2) & (3) & (4) &  (5) &  (6) &  (7) &  (8) &  (9) &  (10) &  (11) &  (12)&  (13) &  (14)  & (15)\\
\hline
\hline
\endfirsthead
\caption{continued.}\\
\hline\hline
N$\degr$ &Set& Frequency &Transition & E$\rm_{up}$ & S$\mu$$^{2}$ & sigma & v & $\Delta$v$\rm_{1/2}$ & T$\rm_{B}$ & W &N$\rm_{up}$/g$\rm_{up}$& \multicolumn{2}{c}{Source size}  &Comments \\
  && (MHz) &  & (K) & (D$^{2}$) & (K) & (~km~s$^{-1}$) & (~km~s$^{-1}$) & (K) & (K~km~s$^{-1}$) &(10$^{12}$~cm$^{-2}$)  & ($\arcsec$ x $\arcsec$) &PA($\degr$) & \\
(1) & (2) & (3) & (4) &  (5) &  (6) &  (7) &  (8) &  (9) &  (10) &  (11) &  (12)&  (13) &  (14)  & (15)\\
\hline
\endhead
\hline
\endfoot
1&1 & 80531.669(0.030) & $9_{2,8} - 9_{0,9}$ E, (v$\rm_{t}$=1) & 216 & 0.6 & 0.03 & - & - & $\la$0.09 & $\la$0.4 &$\la$0.69& -&-& not detected \\
2&1  & 80565.210(0.010) & $10_{2,8} - 9_{3,7}$ E & 36 & 1.1 & 0.03 & 7.9(0.3) & 4.4(0.5) & 0.24 & 1.1(0.1)& 1.04& -&-& detected \\
3 &1 & 80572.589(0.010) & $10_{2,8} - 9_{3,7}$ A & 36 & 1.1 & 0.03 & - & - & $\la$0.30 & $\la$1.3 &$\la$1.22& -&-& blend \\
4 &2  & 80604.508(0.030) & $12_{9,3} - 13_{8,5}$ E, (v$\rm_{t}$=1) & 288 & 0.2 & 0.10 & - & - &$\la$8.03 &$\la$35.3 &$\la$182.73& -&-& blend \\
5 &2 & 80652.521(0.010) & $15_{4,11} - 14_{5,10}$ A & 83 & 1.0 & 0.10 & - & - &$\la$3.92 &$\la$17.2 &$\la$17.80& -&-& blend \\
6 &3 & 101202.806(0.010) &  $9_{1,9} - 8_{0,8}$ A, (v$\rm_{t}$=1) & 213 & 4.2 & 0.04 & - & - &$\la$0.64 &$\la$2.6 &$\la$0.51& -&-&blend  \\
7 &3 & 101210.208(0.004)& $24_{3,21} - 25_{2,24}$ A&187 &0.04 & 0.04& - & - &$\la$0.34 &$\la$1.4 &$\la$30.36& -&-&blend  \\
8  &3 & 101279.006(0.010) & $9_{1,9} - 8_{0,8}$ E, (v$\rm_{t}$=1) & 212 & 4.0 & 0.04 & 7.6(0.7)&4.1(2.0) & 0.64&2.8(1.0) & 0.58& 4.0x2.5 & 20 &detected \\
9   &3& 101289.471(0.010) & $29_{9,20} - 28_{10,19}$ A, (v$\rm_{t}$=1) & 498 & 1.8 & 0.04 & - &- &$\la$0.12 &$\la$0.5 &$\la$0.23& -&-&not detected \\
10  &3& 101302.159(0.010) & $25_{6,19} - 25_{5,20}$ A & 219 & 11.1 & 0.04 & 7.1(0.4) & 4.3(1.0) & 2.09 & 9.6(1.9) &0.71 &- &- & partial blend \\
11  &3& 101305.506(0.010) & $25_{6,19} -25 _{5,20}$ E & 219 & 11.1 & 0.04 & 7.5(0.8)& 4.4(2.2) & 1.58 & 7.3(3.0)& 0.54 & 5.0x2.5 & 20 & detected \\
12  &3& 101318.966(0.010) & $13_{3,11} - 13_{2,12}$ E, (v$\rm_{t}$=1) & 246 & 3.2 & 0.04 &7.7(0.5) &3.8(1.2) & 0.50& 2.0(0.5) & 0.51& 5.5x2.5 & 35 &detected \\
13  &3& 101356.788(0.010) & $15_{2,13} - 15_{1,14}$ A, (v$\rm_{t}$=1) & 263 & 3.5 & 0.04 & -&- &$\la$0.56 &$\la$2.2 &$\la$0.52& -&-& blend \\
14 &3 & 101370.505(0.010) & $13_{3,11} - 13_{2,12}$ E & 60 & 3.2 & 0.04 & 7.5(0.1) & 3.8(0.1)& 1.66 & 6.6(0.2)  & 1.70 & 4.5x2.5 & 35 & detected \\
15 &3 & 101414.746(0.010) & $13_{3,11} - 13_{2,12}$ A & 60 & 3.2 &  0.04 & 7.5(0.1) &4.1(0.3) & 1.49 & 6.6(0.4)&1.70 & 4.5x2.5 & 35 & detected  \\
16 &3 & 101418.330(0.001)& $5_{3,3} - 5_{1,4}$ A, (v$\rm_{t}$=1) & 203 &0.04 & 0.04 & - &- &$\la$0.12 &$\la$0.5 &$\la$9.22& -&-&not detected \\
17 &3 & 101477.421(0.010) & $18_{3,15} - 18_{3,16}$ E & 111 & 1.8 & 0.04 & -& -& $\la$9.17& $\la$36.7 &$\la$16.77& -&-& blend with H$\rm_{2}$CS \\
18  &3& 101545.453(0.010) & $18_{3,15} - 18_{3,16}$ A & 111 & 1.8 & 0.04 & 7.6(0.3) & 3.5(0.6) & 1.03 & 3.8(0.6)& 1.73& 4.5x2.5 & 35 &detected \\
19  &3& 101626.884(0.010) & $9_{1,9} - 8_{0,8}$ E & 25 & 4.1 & 0.04 & - & - & $\la$2.80 &$\la$11.2 &$\la$2.24& -&-& blend with   \\
&&&&&&&&&&&&&&HCOOCH$\rm_{3}$-A\\
20 &3 & 101628.149(0.010) & $9_{1,9} - 8_{0,8}$ A & 25 & 4.1 & 0.04 & - & - & $\la$2.97 & $\la$11.9 &$\la$2.38& -&-& blend with  \\
&&&&&&&&&&&&&&HCOOCH$\rm_{3}$-E\\
21  &4& 105663.117(0.030) & $19_{13,7} - 20_{12,9}$ E, (v$\rm_{t}$=1) & 412 & 0.4 &0.02 & -& - &$\la$0.19 & $\la$0.6 &$\la$1.18& -&-& blend  \\
22  &4& 105722.672(0.030)  & $10_{4,6} - 10_{3,8}$ E, (v$\rm_{t}$=1) & 230 & 0.5 &0.02 & - & - & $\la$0.06 & $\la$0.2 &$\la$0.32& -&-& not detected \\
23 &5 & 105810.803(0.030)  & $14_{11,4} - 15_{10,6}$ E & 142 & 0.2 & 0.02 &- &- &$\la$0.09 & $\la$0.3 &$\la$1.18& -&-& blend \\
24 &5 & 105815.953(0.030)  & $3_{3,1} - 2_{2,0}$ E & 10 & 0.2  &0.02 &7.7(0.2) & 3.0(0.9)& 0.17& 0.5(0.1) & 1.97& - & - & detected \\
25 &5 & 105832.067(0.030)  & $14_{11,3} - 15_{10,5}$ E & 142 & 0.2 &0.02 & - &-  & $\la$0.23& $\la$0.7 &$\la$2.76& -&-& blend \\
26 &6 & 110153.652(0.010) & $10_{1,10} - 9_{1,9}$ A, (v$\rm_{t}$=1) & 218 & 25.1&0.10 & - & - & $\la$8.63 & $\la$9.5 &$\la$0.29& -&-&blend with NH$\rm_{2}$D\\
27 &7 & 203349.158(0.006)& $24_{19,5} - 25_{18,7}$ E & 416 &0.2 & 0.07 &- &- &$\la$3.35 &$\la$8.3 &$\la$17.09 &-&- & blend\\ 
28&7  & 203378.812(0.007)& $24_{19,6} - 25_{18,8}$ E & 416 &0.2 & 0.07 &- &- &$\la$20.42 &$\la$50.4 &$\la$103.77&-&- & blend\\ 
29 &8 & 203427.575(0.100) & $21_{8,13} - 21_{7,14}$ A, (v$\rm_{t}$=1) & 366 & 6.2 &0.07 & 7.8(0.4) & 2.3(1.0) &1.14 & 2.7(1.1)& 0.18 & 6.5x3.0 & 35 & detected\\
30 &8& 203435.554(0.100) & $19_{8,11} - 19_{7,12}$ A & 155 & 5.3 &0.07 & 7.8(0.2)&2.6(0.6) & 4.24 &11.8(2.0)  & 0.91 & 6.0x2.5 & 25 & detected \\
31&8 & 203471.840(0.050) & $19_{8,11} - 19_{7,12}$ E & 155 & 4.9 & 0.07& 7.6(0.5)& 2.5(1.3)& 3.22 & 8.5(3.4)& 0.71& 6.0x2.5 & 25 & detected \\
32 &9&  223435.416(0.001)& $11_{4,8} - 10_{3,7}$ A, (v$\rm_{t}$=1) & 237 &1.8 & 0.25 & - & - &$\la$5.75 &$\la$13.9&$\la$2.86 &- &- & blend\\ 
33 &9& 223465.340(0.050) & $11_{4,8} - 10_{3,7}$ E & 50 & 2.1 &0.25 & 7.8(0.1)&2.2(0.2) & 5.62 & 12.9(1.3)& 2.29& 2.5x1.2 & 10 & detected \\
34 &9& 223500.463(0.001) & $11_{4,8} - 10_{3,7}$ A & 50 & 1.8 & 0.25& 7.5(0.2)& 2.3(0.5)& 5.71& 13.7(2.8)& 2.90& 2.5x1.2 & 10 & detected \\
35 &9& 223534.727(0.100) & $18_{5,14} - 17_{5,13}$ E, (v$\rm_{t}$=1) & 305 & 41.6 &0.25 & 7.5(0.1) &2.6(0.3) & 12.67 & 35.5(3.5)& 0.32& -&-& partial blend\\
36 &9& 223538.511(0.100) & $43_{8,35} - 43_{7,36}$ E & 618 & 13.9 & 0.25& - &- &$\la$1.23 & $\la$3.0 &$\la$0.08& -&-& blend \\
37 &9& 223592.231(0.100) & $43_{8,35} - 43_{7,36}$ A & 618 & 13.9 &0.25 & 7.8(0.9)& 2.2(1.8)& 1.22& 2.8(2.2) &0.08 & 2.0x1.2 & 20 &detected \\
38 &9& 223618.470(0.007)& $27_{9,19} - 27_{8,19}$ E, (v$\rm_{t}$=1) & 464 &0.4 & 0.25& - & - & $\la$0.75 &$\la$1.8&$\la$1.80& -&-& not detected \\ 
39 &9& 223624.496(0.100) & $37_{8,30} - 37_{7,31}$ E & 463 & 11.7 &0.25 & - & -& $\la$0.92 &$\la$2.2 &$\la$0.07& -&-& blend \\
40 &9& 223634.916(0.100) & $37_{8,30} - 37_{7,31}$ A & 463 & 11.7 & 0.25& -& -& $\la$1.83& $\la$4.2 &$\la$0.13& -&-& blend\\
41 &9& 223642.180(0.001) & $12_{5,7} - 12_{3,10}$ E & 63 & 0.1 & 0.25& - & -&$\la$3.58 &$\la$8.7 &$\la$38.77 & - & - &blend  \\
42 &9&  223645.342(0.001)& $12_{5,7} - 12_{3,10}$ A & 63 &0.1 & 0.25 & - & - &$\la$0.83 &$\la$2.0&$\la$7.49 &- &- & blend\\
43&9 &  223651.592(0.007)& $33_{9,24} - 33_{8,26}$ E, (v$\rm_{t}$=1) & 573 &0.7 & 0.25 & - & - &$\la$1.88 &$\la$4.5&$\la$2.25 &- &- & blend\\
44&9 &  223676.472(0.006)& $35_{7,29} - 35_{5,30}$ A, (v$\rm_{t}$=1) & 593 &3.0 & 0.25 & - & - &$\la$3.42 &$\la$8.3&$\la$1.02 &- &- & blend\\ 
45&9 &  223782.081(0.007)& $33_{6,28} - 33_{5,29}$ E, (v$\rm_{t}$=1) & 543 &7.0 & 0.25& - & - & $\la$0.75 &$\la$1.8&$\la$0.10& -&-& not detected \\ 
46 &9&  223821.523(0.002)& $35_{7,29} - 35_{6,30}$ E & 409 &8.2  & 0.25 & - & - &$\la$2.06 &$\la$5.0&$\la$0.23 &- &- & blend\\ 
47&9 & 223854.201(0.100) & $35_{7,29} - 35_{6,30}$ A & 409 & 10.1 & 0.25& -& -&$\la$2.33 &$\la$5.6 &$\la$0.21& -&-& blend \\
48&10 & 225689.371(0.008)& $34_{5,29} - 34_{5,30}$ A, (v$\rm_{t}$=1) & 562 &3.0 & 0.10 &- & -& $\la$17.26& $\la$19.0&$\la$2.37& -&-& blend \\
49 &10 & 225696.842(0.100) & $20_{1,19} - 19_{1,18}$ E, (v$\rm_{t}$=1) & 307 & 50.2 & 0.10& -& - & $\la$24.50& $\la$27.0&$\la$0.20& -&-& blend \\
50 &10& 225702.857(0.100) & $19_{2,17} - 18_{2,16}$ A, (v$\rm_{t}$=1)& 304 & 46.8 & 0.10 &- &- &$\la$43.23 &$\la$47.6 &$\la$0.38& -&-& blend \\
51 &10& 225727.506(0.100) & $6_{6,1} - 5_{5,0}$ A, (v$\rm_{t}$=1) & 224 & 3.1 &0.10 &- &- & $\la$4.61& $\la$5.1&$\la$0.61& -&-& blend \\
52&10 & 225727.506(0.100) & $6_{6,0} - 5_{5,1}$ A, (v$\rm_{t}$=1) & 224 & 3.1& 0.10 &- &- & $\la$4.61& $\la$5.1&$\la$0.61& -&-& blend \\
53 &11 & 225855.505(0.100) & $6_{6,1} - 5_{5,1}$ E & 36 & 3.1 &0.11 & - & - & $\la$3.68&$\la$4.0 &$\la$0.48& -&-&blend \\
54&11 & 225900.684(0.100) & $6_{6,0} - 5_{5,0}$ E & 36 & 3.1&0.11  &- & -& $\la$19.23& $\la$21.2 &$\la$2.53& -&-& blend with HDO \\
55&11 & 225928.659(0.100) & $6_{6,0} - 5_{5,1}$ A & 36 & 3.1& 0.11 & - & - & $\la$5.80& $\la$6.4 &$\la$0.76& -&-& blend  \\
56&11 & 225928.659(0.100) & $6_{6,1} - 5_{5,0}$ A & 36 & 3.1 &0.11 & - & - & $\la$5.80& $\la$6.4&$\la$0.76& -&-& blend \\
57&12 & 225999.145(0.100) & $30_{7,23} - 29_{8,22}$ E & 312 & 2.2 & 0.10 & - & -& $\la$0.30& $\la$0.3&$\la$0.05& -&-& not detected \\
58 &12& 226061.796(0.100) & $20_{3,17} - 19_{4,16}$ E, (v$\rm_{t}$=1) & 321 & 4.0 &0.10 & 7.8(0.3)& 1.1(0.6)& 0.42 & 0.5(0.2) & 0.05& -&-&partial blend  \\
59 &12&  226077.920(0.001)& $10_{3,7} - 9_{1,8}$ E & 39 &0.3 & 0.10 &-&- &$\la$2.52 & $\la$2.8&$\la$3.66& -&-&blend  \\
60 &12&  226081.175(0.002)& $30_{7,23} - 29_{8,22}$ A & 312 &1.8 & 0.10 &-&- &$\la$0.42 & $\la$0.5&$\la$0.10& -&-&blend  \\
61&12 &  226087.718(0.007)& $29_{4,26} - 29_{3,27}$ A, (v$\rm_{t}$=1) & 450 &4.3 & 0.10 &-&- &$\la$0.73 & $\la$0.8&$\la$0.07& -&-&blend  \\
62&12 & 226090.301(0.100) & $19_{2,17} - 18_{2,16}$ E, (v$\rm_{t}$=1) & 303 & 47.0 &0.10 & -&- &$\la$7.67 & $\la$8.4 &$\la$0.07& -&-&blend  \\
63 &12&  226112.470(0.007)& $29_{4,26} - 29_{2,27}$ A, (v$\rm_{t}$=1) & 450 &2.1 & 0.10 &-&- &$\la$0.53 & $\la$0.6&$\la$0.11& -&-&blend  \\
64 &12&  226125.600(0.001)& $10_{3,7} - 9_{1,8}$ A & 39 &0.3 & 0.10 &-&- &$\la$3.22 & $\la$3.5&$\la$4.58& -&-&blend  \\
\end{longtable}
\flushleft
Notes: (1) Numbering of the observed transitions with E$\rm_{upper}$ $\la$ 650 K. (2) Reference number of the corresponding data set (see Table \ref{Table.dataset_parameters}). (3) Frequencies and uncertainties taken from Ilyushin et al. (2009) and from the JPL database (http://spec.jpl.nasa.gov/). (4) Transitions in ground state (v$\rm_{t}$=0). Transitions in torsional levels (v$\rm_{t}$=1) are specified. (5) Energy of the upper level. (6) Line strength calculated by Ilyushin et al. (2009). For the transitions coming from the JPL database, the line strength is calculated from the formulae of \citet{Pickett:1998}. (7) Noise estimated from spectrum (1$\sigma$). (8), (9), (10) and (11) Velocity, linewidth at half intensity, brightness temperature and integrated intensities. The uncertainties are estimated with the CLASS software. (12) Column densities of the upper state of the transition with respect to the rotational degeneracy (2J$\rm_{up}$+1). (13) and (14) Source size estimated at half flux density from detected transitions with spatial resolution allowing to isolate the source. (15) Blended lines are specified. 
\end{landscape}
} %End \longtabL
%-------------------

%------------------- 
%MF3 TABLE
%-------------------
\longtabL{2}{
\begin{landscape}
\begin{longtable}{clllccccccccccl}
\caption{\label{MF3} Transitions of methyl formate observed with the Plateau de Bure Interferometer toward position MF3 in Orion-KL}\\
\hline\hline
N$\degr$ &Set& Frequency &Transition & E$\rm_{up}$ & S$\mu$$^{2}$ & sigma & v & $\Delta$v$\rm_{1/2}$ & T$\rm_{B}$ & W & N$\rm_{up}$/g$\rm_{up}$& \multicolumn{2}{c}{Source size}  &Comments \\
  & & (MHz) &  & (K) & (D$^{2}$) & (K) & (~km~s$^{-1}$) & (~km~s$^{-1}$) & (K) & (K~km~s$^{-1}$) & (10$^{12}$~cm$^{-2}$) & ($\arcsec$ x $\arcsec$) &PA($\degr$) & \\
(1) & (2) & (3) & (4) &  (5) &  (6) &  (7) &  (8) &  (9) &  (10) &  (11) &  (12)&  (13) &  (14) &(15) \\
\hline
\hline
\endfirsthead
\caption{continued.}\\
\hline\hline
N$\degr$ &Set& Frequency &Transition & E$\rm_{up}$ & S$\mu$$^{2}$ & sigma & v & $\Delta$v$\rm_{1/2}$ & T$\rm_{B}$ & W & N$\rm_{up}$/g$\rm_{up}$& \multicolumn{2}{c}{Source size}  &Comments \\
  & & (MHz) &  & (K) & (D$^{2}$) & (K) & (~km~s$^{-1}$) & (~km~s$^{-1}$) & (K) & (K~km~s$^{-1}$) & (10$^{12}$~cm$^{-2}$) & ($\arcsec$ x $\arcsec$) &PA($\degr$) & \\
(1) & (2) & (3) & (4) &  (5) &  (6) &  (7) &  (8) &  (9) &  (10) &  (11) &  (12)&  (13) &  (14) &(15) \\
\hline
\endhead
\hline
\endfoot
1  &1 & 80531.669(0.030) & $9_{2,8} - 9_{0,9}$ E, (v$\rm_{t}$=1) & 216 & 0.6 & 0.02 & - & -& $\la$0.06& $\la$0.2 &$\la$0.35& -&-& not detected \\
2  &1 & 80565.210(0.010) & $10_{2,8} - 9_{3,7}$ E & 36 & 1.1 & 0.02 & 7.8(0.1) &3.8(0.2) &0.43 & 1.8(0.1) & 1.69& 10.0x7.0 & 10 & detected \\
3 &1  & 80572.589(0.010) & $10_{2,8} - 9_{3,7}$ A & 36 & 1.1 & 0.02 & - & - & $\la$0.39 & $\la$1.5 &$\la$1.41& -&-& blend \\
4  &2  & 80604.508(0.030) & $12_{9,3} - 13_{8,5}$ E, (v$\rm_{t}$=1) & 288 & 0.2 & 0.06 & - & - &$\la$3.20 &$\la$12.2 &$\la$63.15& -&-& blend \\
5 &2  & 80652.521(0.010) & $15_{4,11} - 14_{5,10}$ A & 83 & 1.0 &0.06 & - & - & $\la$1.94 &$\la$7.4 &$\la$7.66& -&-& blend  \\
6 &3  & 101202.806(0.010) &  $9_{1,9} - 8_{0,8}$ A, (v$\rm_{t}$=1) & 213 & 4.2 & 0.04 & 8.0(0.1)& 4.0(0.2)& 0.51&2.2(0.1) & 0.43& -&-&partial blend \\
7 &3& 101210.208(0.004)& $24_{3,21} - 25_{2,24}$ A&187 &0.04 & 0.04  & - & -  &$\la$0.12 &$\la$0.5&$\la$10.84& -&-&not detected \\
8 &3 & 101279.006(0.010) & $9_{1,9} - 8_{0,8}$ E, (v$\rm_{t}$=1) & 212 & 4.0 & 0.04  & 7.5(0.4)&3.5(0.9) & 0.55& 2.0(0.4) & 0.41& -&-&detected \\
9  &3& 101289.471(0.010) & $29_{9,20} - 28_{10,19}$ A, (v$\rm_{t}$=1) & 498 & 1.8 & 0.04  & - & -  &$\la$0.12 &$\la$0.5&$\la$0.23& -&-&not detected \\
10&3 & 101302.159(0.010) & $25_{6,19} - 25_{5,20}$ A & 219 & 11.1 &  0.04 & 7.4(0.4)& 4.1(1.2) &  1.76 & 7.6(1.8)& 0.56& -&-& partial blend \\
11 &3& 101305.506(0.010) & $25_{6,19} -25 _{5,20}$ E & 219 & 11.1 &  0.04 & 7.8(0.2) &4.0(0.6) & 1.33 & 5.7(0.6) &0.42 & -&-& detected \\
12 &3& 101318.966(0.010) & $13_{3,11} - 13_{2,12}$ E, (v$\rm_{t}$=1) & 246 & 3.2 &  0.04 & 7.5(0.3) & 3.2(0.8)& 0.28 & 1.0(0.2) & 0.26& -&-&detected \\
13 &3& 101356.788(0.010) & $15_{2,13} - 15_{1,14}$ A, (v$\rm_{t}$=1) & 263 & 3.5 &  0.04 & - & - & $\la$0.30 &$\la$1.1 &$\la$0.26& -&-&blend \\
14 &3& 101370.505(0.010) & $13_{3,11} - 13_{2,12}$ E & 60 & 3.2 &  0.04 & 7.7(0.1) & 3.8(0.1) &1.77 &7.2(0.1) & 1.85& -&-&detected \\
15 &3& 101414.746(0.010) & $13_{3,11} - 13_{2,12}$ A & 60 & 3.2 &  0.04 & 7.7(0.1)&3.7(0.1) & 1.83& 7.2(0.2) & 1.85& -&-& detected \\
16 &3& 101418.330(0.001)& $5_{3,3} - 5_{1,4}$ A, (v$\rm_{t}$=1) & 203 &0.04 & 0.04 & - & -  &$\la$0.12 &$\la$0.5&$\la$9.22& -&-&not detected \\
17&3 & 101477.421(0.010) & $18_{3,15} - 18_{3,16}$ E & 111 & 1.8 & 0.04  & - & - &$\la$10.05 &$\la$37.8 &$\la$17.27& -&-& blend with H$\rm_{2}$CS \\
18&3 & 101545.453(0.010) & $18_{3,15} - 18_{3,16}$ A & 111 & 1.8 & 0.04  &  7.9(0.1)& 3.8(0.2)& 0.83 & 3.4(0.1) & 1.55& -&-&detected \\
19&3 & 101626.884(0.010) & $9_{1,9} - 8_{0,8}$ E & 25 & 4.1 &  0.04 & -  & - & $\la$ 3.21 & $\la$12.1&$\la$2.42& -&-& blend with  \\
&&&&&&&&&&&&&&HCOOCH$\rm_{3}$-A\\
20 &3& 101628.149(0.010) & $9_{1,9} - 8_{0,8}$ A & 25 & 4.1 & 0.04  & - & - &  $\la$3.23 & $\la$12.2 &$\la$2.44& -&-&blend with \\
&&&&&&&&&&&&&&HCOOCH$\rm_{3}$-E\\
21 &4& 105663.117(0.030) & $19_{13,7} - 20_{12,9}$ E, (v$\rm_{t}$=1) & 412 & 0.4 &0.02 & 7.6(0.3)& 2.1(0.8)& 0.09 & 0.2(0.1)& 0.39& 11x8.5 & 6 & detected \\
22 &4& 105722.672(0.030)  & $10_{4,6} - 10_{3,8}$ E, (v$\rm_{t}$=1) & 230 & 0.5 & 0.02 & -& - & $\la$0.11& $\la$0.2 &$\la$0.32& -&-& blend \\
23&5 & 105810.803(0.030)  & $14_{11,4} - 15_{10,6}$ E & 142 & 0.2 & 0.02 & - &- & $\la$0.12 & $\la$0.2&$\la$0.79& -&-& blend  \\
24 &5 & 105815.953(0.030)  & $3_{3,1} - 2_{2,0}$ E & 10 & 0.2 & 0.02 & 7.6(0.1)&2.1(0.3) & 0.35& 0.8(0.1)& 3.15& 10.5x8.5 & 6 & detected\\
25 &5 & 105832.067(0.030)  & $14_{11,3} - 15_{10,5}$ E & 142 & 0.2 & 0.02 & 7.7(0.1)& 1.5(0.2)& 0.17 & 0.3(0.1) & 1.18& 10.0x7.5 & 6 &detected \\
26&6  & 110153.652(0.010) & $10_{1,10} - 9_{1,9}$ A, (v$\rm_{t}$=1) & 218 & 25.1& 0.12 & - & - & $\la$6.40 &$\la$12.2 &$\la$0.37 & -&-& blend with NH$\rm_{2}$D\\
27 &7 & 203349.158(0.006)& $24_{19,5} - 25_{18,7}$ E & 416 &0.2 & 0.20 & - & - & $\la$0.60 &$\la$1.0 &$\la$2.06& -&-&not detected \\
28&7  & 203378.812(0.007)& $24_{19,6} - 25_{18,8}$ E & 416 &0.2 & 0.20 & - & - & $\la$2.94 &$\la$5.0 &$\la$10.29& -&-&blend \\
29&8  & 203427.575(0.100) & $21_{8,13} - 21_{7,14}$ A, (v$\rm_{t}$=1) & 366 & 6.2 & 0.20 & - & - & $\la$0.60&$\la$1.0 &$\la$0.07& -&-&not detected \\
30 &8 & 203435.554(0.100) & $19_{8,11} - 19_{7,12}$ A & 155 & 5.3 & 0.20 & 7.7(0.1) & 1.7(0.1)& 3.33 & 6.1(0.3) &0.47 & -&-& detected  \\
31&8  & 203471.840(0.050) & $19_{8,11} - 19_{7,12}$ E & 155 & 4.9 & 0.20 & 7.6(0.1) & 1.7(0.3) & 2.73 & 4.8(0.8) &0.40 & -&-& detected \\
32&9  &  223435.416(0.001)& $11_{4,8} - 10_{3,7}$ A, (v$\rm_{t}$=1) & 237 &1.8 & 0.30 &- &- &$\la$0.90 &$\la$1.4 &$\la$0.29& -&-&not detected \\ 
33 &9 & 223465.340(0.050) & $11_{4,8} - 10_{3,7}$ E & 50 & 2.1 & 0.30 & 7.7(0.1)& 1.7(0.3) &3.24 &5.8(1.2) &1.03 & -&-&partial blend \\
34&9  & 223500.463(0.001) & $11_{4,8} - 10_{3,7}$ A & 50 &1.8 & 0.30 & 7.7(0.1) &1.2(0.1) &4.02 &5.2(0.3) & 1.10& 3.0x1.5 & 0 &detected \\
35&9  & 223534.727(0.100) & $18_{5,14} - 17_{5,13}$ E, (v$\rm_{t}$=1) & 305 & 41.6 & 0.30 & 7.7(0.1)&1.6(0.1) &3.49 &6.0(0.4) &0.05 & 2.5x1.5 & 0 &detected \\
36&9  & 223538.511(0.100) & $43_{8,35} - 43_{7,36}$ E & 618 & 13.9 & 0.30 &- & -&$\la$0.90 &$\la$1.4 &$\la$0.04& -&-&not detected \\
37 &9 & 223592.231(0.100) & $43_{8,35} - 43_{7,36}$ A & 618 & 13.9 & 0.30 &- & -&$\la$0.90 &$\la$1.4 &$\la$0.04& -&-&not detected \\
38 &9 & 223618.470(0.007)& $27_{9,19} - 27_{8,19}$ E, (v$\rm_{t}$=1) & 464 &0.4 & 0.30&- &- &$\la$0.90 &$\la$1.4 &$\la$1.40& -&-&not detected \\ 
39 &9 & 223624.496(0.100) & $37_{8,30} - 37_{7,31}$ E & 463 & 11.7 & 0.30 &- & -&$\la$0.90 &$\la$1.4&$\la$0.04& -&-&not detected \\
40&9  & 223634.916(0.100) & $37_{8,30} - 37_{7,31}$ A & 463 & 11.7 & 0.30 &- & -&$\la$0.90 &$\la$1.4 &$\la$0.04& -&-&not detected  \\
41&9 & 223642.180(0.001) & $12_{5,7} - 12_{3,10}$ E & 63 &0.1& 0.30 & - & - & $\la$1.13 & $\la$1.7 & $\la$7.57& -&-& blend \\
42 &9 &  223645.342(0.001)& $12_{5,7} - 12_{3,10}$ A & 63 &0.1 & 0.30 &- &- &$\la$0.90 &$\la$1.4&$\la$5.24& -&-&not detected \\ 
43 &9 &  223651.592(0.007)& $33_{9,24} - 33_{8,26}$ E, (v$\rm_{t}$=1) & 573 &0.7 & 0.30 &- &- &$\la$0.90 &$\la$1.4&$\la$0.70& -&-&not detected \\ 
44 &9 &  223676.472(0.006)& $35_{7,29} - 35_{5,30}$ A, (v$\rm_{t}$=1) & 593 &3.0 & 0.30 &- &- &$\la$0.90 &$\la$1.4&$\la$0.17& -&-&not detected \\ 
45 &9 &  223782.081(0.007)& $33_{6,28} - 33_{5,29}$ E, (v$\rm_{t}$=1) & 543 &7.0 & 0.30 &- &- &$\la$0.90 &$\la$1.4 &$\la$0.08& -&-&not detected \\ 
46 &9 &  223821.523(0.002)& $35_{7,29} - 35_{6,30}$ E & 409 &8.2 & 0.30 &- &- &$\la$0.90 &$\la$1.4 &$\la$0.06& -&-&not detected \\ 
47 &9 & 223854.201(0.100) & $35_{7,29} - 35_{6,30}$ A & 409 & 10.1 & 0.30 & - & - &$\la$0.90 &$\la$1.4 &$\la$0.05& -&-&not detected  \\
48 &10 & 225689.371(0.008)& $34_{5,29} - 34_{5,30}$ A, (v$\rm_{t}$=1) & 562 &3.0 & 0.14&- & -& $\la$13.06& $\la$22.2&$\la$2.77& -&-& blend \\
49 &10 & 225696.842(0.100) & $20_{1,19} - 19_{1,18}$ E, (v$\rm_{t}$=1) & 307 & 50.2 & 0.14 & - & - & $\la$22.85 & $\la$38.8 &$\la$0.29& -&-& blend  \\
50 &10& 225702.857(0.100) & $19_{2,17} - 18_{2,16}$ A, (v$\rm_{t}$=1)& 304 & 46.8 &0.14 & - & - & $\la$25.75 & $\la$43.8 &$\la$0.35& -&-& blend \\
51 &10& 225727.506(0.100) & $6_{6,1} - 5_{5,0}$ A, (v$\rm_{t}$=1) & 224 & 3.1 & 0.14 & - & - & $\la$3.20 & $\la$5.4&$\la$0.64& -&-& blend \\
52 &10& 225727.506(0.100) & $6_{6,0} - 5_{5,1}$ A, (v$\rm_{t}$=1) & 224 & 3.1 & 0.14 & - & - & $\la$3.20 & $\la$5.4 &$\la$0.64& -&-& blend \\
53 &11 & 225855.505(0.100) & $6_{6,1} - 5_{5,1}$ E & 36 & 3.1 & 0.09 & - &-  & $\la$5.27 & $\la$9.0&$\la$1.07& -&-& blend \\
54 &11& 225900.684(0.100) & $6_{6,0} - 5_{5,0}$ E & 36 & 3.1&0.09  & - &-  & $\la$11.08 & $\la$18.8 &$\la$2.24& -&-& blend with HDO \\
55&11 & 225928.659(0.100) & $6_{6,0} - 5_{5,1}$ A & 36 & 3.1 &0.09  & - &-  & $\la$7.14 & $\la$12.1&$\la$1.44& -&-&blend \\
56 &11& 225928.659(0.100) & $6_{6,1} - 5_{5,0}$ A & 36 & 3.1 &0.09  & - &-  & $\la$7.14 & $\la$12.1&$\la$1.44& -&-&blend \\
57&12 & 225999.145(0.100) & $30_{7,23} - 29_{8,22}$ E & 312 & 2.2 &0.12 & - & - & $\la$0.36 &$\la$0.6 &$\la$0.10& -&-&not detected  \\
58 &12& 226061.796(0.100) & $20_{3,17} - 19_{4,16}$ E, (v$\rm_{t}$=1) & 321 & 4.0 &0.12 &  - &-  & $\la$0.46 & $\la$0.8 &$\la$0.07& -&-& blend \\
59&12 &  226077.920(0.001)& $10_{3,7} - 9_{1,8}$ E & 39 &0.3 & 0.12 &-&- &$\la$2.48 & $\la$4.2&$\la$5.50& -&-&blend  \\
60 &12&  226081.175(0.002)& $30_{7,23} - 29_{8,22}$ A & 312 &1.8 & 0.12 &-&- &$\la$0.36  & $\la$0.6&$\la$0.12& -&-&not detected \\
61 &12&  226087.718(0.007)& $29_{4,26} - 29_{3,27}$ A, (v$\rm_{t}$=1) & 450 &4.3 & 0.12 &-&- &$\la$0.46 & $\la$0.8&$\la$0.07& -&-&blend  \\
62 &12& 226090.301(0.100) & $19_{2,17} - 18_{2,16}$ E, (v$\rm_{t}$=1) & 303 & 47.0 & 0.12& - &-  & $\la$6.10 & $\la$10.4 &$\la$0.08& -&-& blend \\
63 &12&  226112.470(0.007)& $29_{4,26} - 29_{2,27}$ A, (v$\rm_{t}$=1) & 450 &2.1 & 0.12 &-&- &$\la$0.36  & $\la$0.6&$\la$0.11& -&-&not detected \\
64 &12&  226125.600(0.001)& $10_{3,7} - 9_{1,8}$ A & 39 &0.3 & 0.12 &-&- &$\la$1.83 & $\la$3.1&$\la$4.05& -&-&blend  \\
\end{longtable}
\flushleft
Notes: (1) Numbering of the observed transitions with E$\rm_{upper}$ $\la$ 650 K. (2) Reference number of the corresponding data set (see Table \ref{Table.dataset_parameters}). (3) Frequencies and uncertainties taken from Ilyushin et al. (2009) and from the JPL database (http://spec.jpl.nasa.gov/). (4) Transitions in ground state (v$\rm_{t}$=0). Transitions in torsional levels (v$\rm_{t}$=1) are specified. (5) Energy of the upper level. (6) Line strength calculated by Ilyushin et al. (2009). For the transitions coming from the JPL database, the line strength is calculated from the formulae of \citet{Pickett:1998}. (7) Noise estimated from spectrum (1$\sigma$). (8), (9), (10) and (11) Velocity, linewidth at half intensity, brightness temperature and integrated intensities. The uncertainties are estimated with the CLASS software. (12) Column densities of the upper state of the transition with respect to the rotational degeneracy (2J$\rm_{up}$+1). (13) and (14) Source size estimated at half flux density from detected transitions with spatial resolution allowing to isolate the source. (15) Blended lines are specified. 
\end{landscape}
} %End \longtabL
%------------------- 

%------------------- 
%MF4 TABLE
%-------------------
\longtabL{3}{
\begin{landscape}
\begin{longtable}{clllccccccccccl}
\caption{\label{MF4} Transitions of methyl formate observed with the Plateau de Bure Interferometer toward position MF4 in Orion-KL}\\
\hline\hline
N$\degr$ &Set& Frequency &Transition & E$\rm_{up}$ & S$\mu$$^{2}$ & sigma & v & $\Delta$v$\rm_{1/2}$ & T$\rm_{B}$  & W & N$\rm_{up}$/g$\rm_{up}$& \multicolumn{2}{c}{Source size}  &Comments \\
  & &(MHz) &  & (K) & (D$^{2}$) & (K) & (~km~s$^{-1}$) & (~km~s$^{-1}$) & (K) & (K~km~s$^{-1}$) &(10$^{12}$~cm$^{-2}$)& ($\arcsec$ x $\arcsec$) &PA($\degr$) & \\
(1) & (2) & (3) & (4) &  (5) &  (6) &  (7) &  (8) &  (9) &  (10) &  (11) &  (12)&  (13) &  (14)  &(15)\\
\hline
\hline
\endfirsthead
\caption{continued.}\\
\hline\hline
N$\degr$ &Set& Frequency &Transition & E$\rm_{up}$ & S$\mu$$^{2}$ & sigma & v & $\Delta$v$\rm_{1/2}$ & T$\rm_{B}$  & W & N$\rm_{up}$/g$\rm_{up}$& \multicolumn{2}{c}{Source size}  &Comments \\
  & &(MHz) &  & (K) & (D$^{2}$) & (K) & (~km~s$^{-1}$) & (~km~s$^{-1}$) & (K) & (K~km~s$^{-1}$) &(10$^{12}$~cm$^{-2}$)& ($\arcsec$ x $\arcsec$) &PA($\degr$) & \\
(1) & (2) & (3) & (4) &  (5) &  (6) &  (7) &  (8) &  (9) &  (10) &  (11) &  (12)&  (13) &  (14)  &(15)\\
\hline
\endhead
\hline
\endfoot
1 &1& 80531.669(0.030) & $9_{2,8} - 9_{0,9}$ E, (v$\rm_{t}$=1) & 216 & 0.6 & 0.02 &- & -&$\la$0.06 &$\la$0.2 &$\la$0.35& -&-&not detected \\
2  &1& 80565.210(0.010) & $10_{2,8} - 9_{3,7}$ E & 36 & 1.1 &0.02  &8.1(0.2) & 4.0(0.4)& 0.28& 1.2(0.1)& 1.13& -&-&detected \\
3  &1& 80572.589(0.010) & $10_{2,8} - 9_{3,7}$ A & 36 & 1.1 &0.02  & - &- &$\la$0.33 &$\la$1.3 &$\la$1.22& -&-&blend \\
4  &2 & 80604.508(0.030) & $12_{9,3} - 13_{8,5}$ E, (v$\rm_{t}$=1) & 288 & 0.2 & 0.12& - & -&$\la$6.26 &$\la$25.0 &$\la$129.41& -&-& blend \\
5  &2& 80652.521(0.010) & $15_{4,11} - 14_{5,10}$ A & 83 & 1.0 &0.12 & - & -& $\la$2.33&$\la$9.3 &$\la$9.62& -&-&blend \\
6 &3 & 101202.806(0.010) &  $9_{1,9} - 8_{0,8}$ A, (v$\rm_{t}$=1) & 213 & 4.2 & 0.03& - & -&$\la$0.42 &$\la$1.8 &$\la$0.35& -&-&blend \\
7 &3 & 101210.208(0.004)& $24_{3,21} - 25_{2,24}$ A&187 &0.04 & 0.03 & - & - & $\la$0.37& $\la$1.6 &$\la$34.70& -&-& blend \\
8&3   & 101279.006(0.010) & $9_{1,9} - 8_{0,8}$ E, (v$\rm_{t}$=1) & 212 & 4.0 & 0.03& - & - & $\la$0.42& $\la$1.8 &$\la$0.37& -&-& blend \\
9  &3 & 101289.471(0.010) & $29_{9,20} - 28_{10,19}$ A, (v$\rm_{t}$=1) & 498 & 1.8 &0.03 & - & - & $\la$0.09 &$\la$0.4 &$\la$0.18& -&-&not detected  \\
10 &3 & 101302.159(0.010) & $25_{6,19} - 25_{5,20}$ A & 219 & 11.1 &0.03 & 8.0(0.1) &4.3(0.2) & 1.07 & 4.9(0.3)& 0.36& -&-&partial blend \\
11 &3 & 101305.506(0.010) & $25_{6,19} -25 _{5,20}$ E & 219 & 11.1 & 0.03&8.1(0.2) & 4.5(0.6)&1.04 &4.9(0.6) &0.36 & 5.0x2.7 & 0 &detected \\
12 &3 & 101318.966(0.010) & $13_{3,11} - 13_{2,12}$ E, (v$\rm_{t}$=1) & 246 & 3.2 & 0.03& 8.4(0.3)& 3.9(0.6)& 0.23&0.9(0.1) & 0.23& 4.5x2.7 & 20 &detected \\
13 &3 & 101356.788(0.010) & $15_{2,13} - 15_{1,14}$ A, (v$\rm_{t}$=1) & 263 & 3.5 &0.03 & - & - & $\la$0.29& $\la$1.2&$\la$0.28& -&-& blend \\
14&3  & 101370.505(0.010) & $13_{3,11} - 13_{2,12}$ E & 60 & 3.2 & 0.03& 7.9(0.1)& 4.0(0.1)& 1.27&5.4(0.1) & 1.39& -&-&partial blend \\
15&3  & 101414.746(0.010) & $13_{3,11} - 13_{2,12}$ A & 60 & 3.2 & 0.03& 8.2(0.1)& 4.6(0.1) & 1.36 & 6.6(0.2) & 1.70& 5.5x3.0 & 0 & detected \\
16 &3 & 101418.330(0.001)& $5_{3,3} - 5_{1,4}$ A, (v$\rm_{t}$=1) & 203 &0.04& 0.03 & - & - & $\la$0.09&$\la$0.4 &$\la$7.37& -&-&not detected  \\
17 &3 & 101477.421(0.010) & $18_{3,15} - 18_{3,16}$ E & 111 & 1.8 &0.03 & - &- &$\la$7.79 & $\la$32.8&$\la$14.99& -&-& blend with H$\rm_{2}$CS \\
18&3  & 101545.453(0.010) & $18_{3,15} - 18_{3,16}$ A & 111 & 1.8 & 0.03& 8.7(0.3)& 4.0(0.6)& 0.57& 2.4(0.3) & 1.10& 5.5x2.7 & 0 &detected \\
19&3  & 101626.884(0.010) & $9_{1,9} - 8_{0,8}$ E & 25 & 4.1 & 0.03& -& -& $\la$2.88 & $\la$12.1&$\la$2.42& -&-& blend with  \\
&&&&&&&&&&&&&&HCOOCH$\rm_{3}$-A\\
20&3  & 101628.149(0.010) & $9_{1,9} - 8_{0,8}$ A & 25 & 4.1 & 0.03& - & - & $\la$2.22 &$\la$9.4 &$\la$1.88& -&-& blend with  \\
&&&&&&&&&&&&&&HCOOCH$\rm_{3}$-E\\
21 &4 & 105663.117(0.030) & $19_{13,7} - 20_{12,9}$ E, (v$\rm_{t}$=1) & 412 & 0.4 &0.02& - &- & $\la$0.10&$\la$0.3 &$\la$0.59& -&-&blend \\
22 &4 & 105722.672(0.030)  & $10_{4,6} - 10_{3,8}$ E, (v$\rm_{t}$=1) & 230 & 0.5 &0.02 &- & - & $\la$0.06 &$\la$0.2 &$\la$0.32& -&-& not detected \\
23 &5 & 105810.803(0.030)  & $14_{11,4} - 15_{10,6}$ E & 142 & 0.2 & 0.02 &- & - & $\la$0.06 &$\la$0.2 &$\la$0.79& -&-& not detected \\
24&5 & 105815.953(0.030)  & $3_{3,1} - 2_{2,0}$ E & 10 & 0.2 & 0.02 & 7.9(0.1)& 2.5(0.3)& 0.15 &0.4(0.1) &1.58 & -&-&partial blend \\
25 &5& 105832.067(0.030)  & $14_{11,3} - 15_{10,5}$ E & 142 & 0.2 &0.02 & 7.6(0.2) & 2.5(0.4) & 0.14& 0.4(0.1)&1.58 & -&-& detected \\
26 &6& 110153.652(0.010) & $10_{1,10} - 9_{1,9}$ A, (v$\rm_{t}$=1) & 218 & 25.1& 0.11& - & - &$\la$3.32 &$\la$6.6 &$\la$0.20& -&-& blend with NH$\rm_{2}$D\\
27&7 & 203349.158(0.006)& $24_{19,5} - 25_{18,7}$ E & 416 &0.2 & 0.09& - &- &$\la$1.15 &$\la$2.3 &$\la$4.74& -&-& blend\\
28 &7& 203378.812(0.007)& $24_{19,6} - 25_{18,8}$ E & 416 &0.2 & 0.09 &- &- &$\la$16.84 &$\la$33.7 &$\la$69.38& -&-& blend\\
29&8 & 203427.575(0.100) & $21_{8,13} - 21_{7,14}$ A, (v$\rm_{t}$=1) & 366 & 6.2 & 0.09&- & - & $\la$0.27&$\la$0.5 &$\la$0.03& -&-& not detected \\
30&8 & 203435.554(0.100) & $19_{8,11} - 19_{7,12}$ A & 155 & 5.3 & 0.09&7.8(0.1) &2.0(0.3) & 2.13& 4.6(0.5) & 0.36& -&-&partial blend \\
31&8 & 203471.840(0.050) & $19_{8,11} - 19_{7,12}$ E & 155 & 4.9 & 0.09&- &- &$\la$1.09 &$\la$2.2 &$\la$0.18& -&-& blend\\
32 &9&  223435.416(0.001)& $11_{4,8} - 10_{3,7}$ A, (v$\rm_{t}$=1) & 237 &1.8 & 0.20 & - & - & $\la$0.60 &$\la$1.2 &$\la$0.25& -&-&not detected  \\
33 &9& 223465.340(0.050) & $11_{4,8} - 10_{3,7}$ E & 50 & 2.1 &0.20 & 8.0(0.1)&1.9(0.3) &3.83 & 7.7(2.5)& 1.37& 2.0x1.5 & 60 &  detected \\
34 &9& 223500.463(0.001) & $11_{4,8} - 10_{3,7}$ A & 50 &1.8  &0.20& 8.0(0.1)&2.2(0.2) &1.90& 4.4(0.4) & 0.93&  2.0x1.5 &  60 &detected \\
35&9& 223534.727(0.100) &$18_{5,14} - 17_{5,13}$ E, (v$\rm_{t}$=1) &305 &  41.6 &0.20 & 8.7(0.3) & 3.0(0.4)& 1.78& 5.8(0.8) &0.05 &3.0x1.5 &  60& detected \\
36 &9& 223538.511(0.100) & $43_{8,35} - 43_{7,36}$ E & 618 & 13.9 &0.20 & - & - & $\la$0.60 &$\la$1.2 &$\la$0.03& -&-&not detected  \\
37 &9& 223592.231(0.100) & $43_{8,35} - 43_{7,36}$ A & 618 & 13.9 & 0.20& - & - & $\la$0.60 &$\la$1.2 &$\la$0.03& -&-&not detected  \\
38&9 & 223618.470(0.007)& $27_{9,19} - 27_{8,19}$ E, (v$\rm_{t}$=1) & 464 &0.4& 0.20 & - & - & $\la$0.60&$\la$1.2 &$\la$1.20& -&-&not detected  \\
39 &9& 223624.496(0.100) & $37_{8,30} - 37_{7,31}$ E & 463 & 11.7 & 0.20&  - & - & $\la$0.60 &$\la$1.2 &$\la$0.04& -&-&not detected  \\
40 &9& 223634.916(0.100) & $37_{8,30} - 37_{7,31}$ A & 463 & 11.7 & 0.20& - & - & $\la$0.60 &$\la$1.2 &$\la$0.04& -&-&not detected  \\
41 &9& 223642.180(0.001) & $12_{5,7} - 12_{3,10}$ E & 63 & 0.1 & 0.20& - & - & $\la$0.60 &$\la$1.2 &$\la$5.35& -&-&not detected  \\
42 &9&  223645.342(0.001)& $12_{5,7} - 12_{3,10}$ A & 63 &0.1 & 0.20 & - & - & $\la$0.60&$\la$1.2 &$\la$4.49& -&-&not detected  \\
43 &9&  223651.592(0.007)& $33_{9,24} - 33_{8,26}$ E, (v$\rm_{t}$=1) & 573 &0.7 & 0.20 & - & - & $\la$0.60&$\la$1.2 &$\la$0.60& -&-&not detected  \\
44 &9&  223676.472(0.006)& $35_{7,29} - 35_{5,30}$ A, (v$\rm_{t}$=1) & 593 &3.0 & 0.20 &- &- &$\la$2.20 &$\la$4.5 &$\la$0.55 &- &-& blend\\ 
45&9&  223782.081(0.007)& $33_{6,28} - 33_{5,29}$ E, (v$\rm_{t}$=1) & 543 &7.0 & 0.20 & - & - & $\la$0.60 &$\la$1.2 &$\la$0.06& -&-&not detected  \\
46&9 &  223821.523(0.002)& $35_{7,29} - 35_{6,30}$ E & 409 &8.2 & 0.20 & - & - & $\la$0.60&$\la$1.2 &$\la$0.05& -&-&not detected  \\
47 &9& 223854.201(0.100) & $35_{7,29} - 35_{6,30}$ A & 409 & 10.1 & 0.20& - & - & $\la$0.60 &$\la$1.2 &$\la$0.04& -&-&not detected  \\
48 &10 & 225689.371(0.008)& $34_{5,29} - 34_{5,30}$ A, (v$\rm_{t}$=1) & 562 &3.0 & 0.12 &- & -& $\la$20.83& $\la$41.7&$\la$5.19& -&-& blend \\
49  &10 & 225696.842(0.100) & $20_{1,19} - 19_{1,18}$ E, (v$\rm_{t}$=1) & 307 & 50.2 & 0.12 & - & -& $\la$10.57& $\la$21.1&$\la$0.16& -&-& strongly blend \\
50  &10& 225702.857(0.100) & $19_{2,17} - 18_{2,16}$ A, (v$\rm_{t}$=1)& 304 & 46.8 & 0.12 &- & -& $\la$14.30& $\la$28.6 &$\la$0.23& -&-&  blend \\
51  &10& 225727.506(0.100) & $6_{6,1} - 5_{5,0}$ A, (v$\rm_{t}$=1) & 224 & 3.1 &0.12 &- & -& $\la$0.71& $\la$1.4 &$\la$0.17& -&-&  blend \\
52  &10& 225727.506(0.100) & $6_{6,0} - 5_{5,1}$ A, (v$\rm_{t}$=1) & 224 & 3.1 & 0.12 &- & -& $\la$0.71& $\la$1.4&$\la$0.17& -&-&  blend \\
53  &11 & 225855.505(0.100) & $6_{6,1} - 5_{5,1}$ E & 36 & 3.1 & 0.07& - & -&$\la$1.02 &$\la$2.0 &$\la$0.24& -&-&blend\\
54 &11& 225900.684(0.100) & $6_{6,0} - 5_{5,0}$ E & 36 & 3.1& 0.07&- & -&$\la$5.59 &$\la$11.2 &$\la$1.33& -&-&blend with HDO \\
55 &11& 225928.659(0.100) & $6_{6,0} - 5_{5,1}$ A & 36 & 3.1& 0.07& -& -&$\la$1.74 & $\la$3.5 &$\la$0.42& -&-&blend  \\
56 &11& 225928.659(0.100) & $6_{6,1} - 5_{5,0}$ A & 36 & 3.1 & 0.07&-& -&$\la$1.74 & $\la$3.5 &$\la$0.42& -&-&blend  \\
57 &12& 225999.145(0.100) & $30_{7,23} - 29_{8,22}$ E & 312 & 2.2 & 0.09& - & - & $\la$0.27 &$\la$0.5 &$\la$0.08& -&-&not detected  \\
58  &12& 226061.796(0.100) & $20_{3,17} - 19_{4,16}$ E, (v$\rm_{t}$=1) & 321 & 4.0 & 0.09&- & - & $\la$0.27 &$\la$0.5 &$\la$0.05& -&-&not detected  \\
59  &12&  226077.920(0.001)& $10_{3,7} - 9_{1,8}$ E & 39 &0.3 &0.09 &-&- &$\la$0.27& $\la$0.5&$\la$0.65& -&-&not detected \\
60 &12 &  226081.175(0.002)& $30_{7,23} - 29_{8,22}$ A & 312 &1.8 & 0.09 &-&- &$\la$0.27  & $\la$0.5&$\la$0.10& -&-&not detected \\
61 &12 &  226087.718(0.007)& $29_{4,26} - 29_{3,27}$ A, (v$\rm_{t}$=1) & 450 &4.3 & 0.09 &- &- & $\la$0.84&$\la$1.7 &$\la$0.15& -&-& blend \\
62 &12 & 226090.301(0.100) & $19_{2,17} - 18_{2,16}$ E, (v$\rm_{t}$=1) & 303 & 47.0 & 0.09&- &- & $\la$2.57&$\la$5.1 &$\la$0.04& -&-& blend \\
63  &12&  226112.470(0.007)& $29_{4,26} - 29_{2,27}$ A, (v$\rm_{t}$=1) & 450 &2.1 & 0.09 &-&- &$\la$0.27  & $\la$0.5&$\la$0.09& -&-&not detected \\
64 &12 &  226125.600(0.001)& $10_{3,7} - 9_{1,8}$ A & 39 &0.3 & 0.09 &- &- & $\la$0.32&$\la$0.6&$\la$0.78& -&-& blend \\
\end{longtable}
\flushleft
Notes: (1) Numbering of the observed transitions with E$\rm_{upper}$ $\la$ 650 K. (2) Reference number of the corresponding data set (see Table \ref{Table.dataset_parameters}). (3) Frequencies and uncertainties taken from Ilyushin et al. (2009) and from the JPL database (http://spec.jpl.nasa.gov/). (4) Transitions in ground state (v$\rm_{t}$=0). Transitions in torsional levels (v$\rm_{t}$=1) are specified. (5) Energy of the upper level. (6) Line strength calculated by Ilyushin et al. (2009). For the transitions coming from the JPL database, the line strength is calculated from the formulae of \citet{Pickett:1998}. (7) Noise estimated from spectrum (1$\sigma$). (8) Lines with number 33 and 34 are decomposed. Only the first component is given here, other components are given in Table \ref{Table.MF4bis}, except for line 30 because of the blend with the second component. (8), (9), (10) and (11) Velocity, linewidth at half intensity, brightness temperature and integrated intensities. The uncertainties are estimated with the CLASS software. (12) Column densities of the upper state of the transition with respect to the rotational degeneracy (2J$\rm_{up}$+1). (13) and (14) Source size estimated at half flux density from detected transitions with spatial resolution allowing to isolate the source. (15) Blended lines are specified. 
\end{landscape}
} %End \longtabL
%------------------- 

%------------------- 
%MF5 TABLE
%-------------------
\longtabL{4}{
\begin{landscape}
\begin{longtable}{clllccccccccccl}
\caption{\label{MF5} Transitions of methyl formate observed with the Plateau de Bure Interferometer toward position MF5 in Orion-KL}\\
\hline\hline
N$\degr$ &Set& Frequency &Transition & E$\rm_{up}$ & S$\mu$$^{2}$ & sigma & v & $\Delta$v$\rm_{1/2}$ & T$\rm_{B}$  & W & N$\rm_{up}$/g$\rm_{up}$& \multicolumn{2}{c}{Source size}  &Comments \\
  && (MHz) &  & (K) & (D$^{2}$) & (K) & (~km~s$^{-1}$) & (~km~s$^{-1}$) & (K) & (K~km~s$^{-1}$) & (10$^{12}$~cm$^{-2}$) & ($\arcsec$ x $\arcsec$) &PA($\degr$) & \\
(1) & (2) & (3) & (4) &  (5) &  (6) &  (7) &  (8) &  (9) &  (10) &  (11) &  (12)&  (13) &  (14) &(15)  \\
\hline
\hline
\endfirsthead
\caption{continued.}\\
\hline\hline
N$\degr$ &Set& Frequency &Transition & E$\rm_{up}$ & S$\mu$$^{2}$ & sigma & v & $\Delta$v$\rm_{1/2}$ & T$\rm_{B}$  & W & N$\rm_{up}$/g$\rm_{up}$& \multicolumn{2}{c}{Source size}  &Comments \\
  && (MHz) &  & (K) & (D$^{2}$) & (K) & (~km~s$^{-1}$) & (~km~s$^{-1}$) & (K) & (K~km~s$^{-1}$) & (10$^{12}$~cm$^{-2}$) & ($\arcsec$ x $\arcsec$) &PA($\degr$) & \\
(1) & (2) & (3) & (4) &  (5) &  (6) &  (7) &  (8) &  (9) &  (10) &  (11) &  (12)&  (13) &  (14) &(15)  \\
\hline
\endhead
\hline
\endfoot
1 &1& 80531.669(0.030) & $9_{2,8} - 9_{0,9}$ E, (v$\rm_{t}$=1) & 216 & 0.6 & 0.03& -& -& $\la$0.09&$\la$0.4&$\la$0.69& -&-&not detected \\
2 &1& 80565.210(0.010) & $10_{2,8} - 9_{3,7}$ E & 36 & 1.1 &0.03 & -& -& $\la$0.21 &$\la$0.8 &$\la$0.75& -&-&blend \\
3 &1& 80572.589(0.010) & $10_{2,8} - 9_{3,7}$ A & 36 & 1.1 & 0.03 & -& -& $\la$0.28 &$\la$1.1 &$\la$1.04& -&-&blend \\
4 &2 & 80604.508(0.030) & $12_{9,3} - 13_{8,5}$ E, (v$\rm_{t}$=1) & 288 & 0.2 &0.11 & -& -&$\la$5.48 &$\la$21.9 &$\la$113.37& -&-&blend \\
5 &2& 80652.521(0.010) & $15_{4,11} - 14_{5,10}$ A & 83 & 1.0 & 0.11& -& -& $\la$1.96&$\la$7.8 &$\la$8.07& -&-&blend \\
6 &3 & 101202.806(0.010) &  $9_{1,9} - 8_{0,8}$ A, (v$\rm_{t}$=1) & 213 & 4.2 & 0.04 & 7.7(0.2)&4.5(0.4)&0.43 & 2.1(0.2) &0.41 & -&-&partial blend \\
7 &3 & 101210.208(0.004)& $24_{3,21} - 25_{2,24}$ A&187 &0.04 & 0.04 &- &- &$\la$0.41 &$\la$1.9 &$\la$41.21& -&-&blend \\
8 &3  & 101279.006(0.010) & $9_{1,9} - 8_{0,8}$ E, (v$\rm_{t}$=1) & 212 & 4.0 & 0.04 & 7.5(0.5)&4.6(1.2) &0.49 &2.4(0.5) &0.49 & 5.0x2.7 & 0 &detected \\
9  &3 & 101289.471(0.010) & $29_{9,20} - 28_{10,19}$ A, (v$\rm_{t}$=1) & 498 & 1.8 & 0.04 & -& -& $\la$0.12&$\la$0.5 &$\la$0.23& -&-&not detected  \\
10&3  & 101302.159(0.010) & $25_{6,19} - 25_{5,20}$ A & 219 & 11.1 & 0.04 & 7.7(0.5)& 5.5(1.4)& 1.16& 6.8(1.1)& 0.50& -&-& partial blend \\
11&3  & 101305.506(0.010) & $25_{6,19} -25 _{5,20}$ E & 219 & 11.1 & 0.04 &7.7(0.3) & 4.2(0.6) &1.11 &5.0(0.6)  & 0.37& 5.0x2.7 & 0 &detected \\
12 &3 & 101318.966(0.010) & $13_{3,11} - 13_{2,12}$ E, (v$\rm_{t}$=1) & 246 & 3.2 & 0.04 & 7.7(0.2)&4.5(0.5) & 0.30& 1.4(0.1) & 0.36& 4.5x2.7 & 20 & detected\\
13 &3 & 101356.788(0.010) & $15_{2,13} - 15_{1,14}$ A, (v$\rm_{t}$=1) & 263 & 3.5 &0.04  &- &- &$\la$0.19 &$\la$0.9 &$\la$0.21& -&-&blend \\
14 &3 & 101370.505(0.010) & $13_{3,11} - 13_{2,12}$ E & 60 & 3.2 & 0.04 & 7.9(0.1)&4.7(0.1) &1.42 &7.1(0.1)  & 1.83& 5.0x2.7 & 0 &detected \\
15 &3 & 101414.746(0.010) & $13_{3,11} - 13_{2,12}$ A & 60 & 3.2 & 0.04 & 7.8(0.1)&4.8(0.1) &1.42 &7.2(0.1)  & 1.85& 5.5x3.0 & 0 & detected \\
16 &3 & 101418.330(0.001)& $5_{3,3} - 5_{1,4}$ A, (v$\rm_{t}$=1) & 203 &0.04& 0.04 & -& -& $\la$0.12&$\la$0.5 &$\la$9.22& -&-&not detected  \\ 
17 &3 & 101477.421(0.010) & $18_{3,15} - 18_{3,16}$ E & 111 & 1.8 &0.04  &- &- &$\la$7.56 &$\la$34.4 &$\la$15.72& -&-&blend with H$\rm_{2}$CS \\
18 &3 & 101545.453(0.010) & $18_{3,15} - 18_{3,16}$ A & 111 & 1.8 & 0.04 & -& -&$\la$0.58 &$\la$2.6 &$\la$1.19& -&-&blend \\
19 &3 & 101626.884(0.010) & $9_{1,9} - 8_{0,8}$ E & 25 & 4.1 & 0.04 &- & -& $\la$2.89 &$\la$13.1 &$\la$2.62& -&-& blend with \\
&&&&&&&&&&&&&&HCOOCH$\rm_{3}$-A\\
20 &3 & 101628.149(0.010) & $9_{1,9} - 8_{0,8}$ A & 25 & 4.1 &0.04  & -&-  &$\la$2.42 &$\la$11.0 &$\la$2.20& -&-& blend with  \\
&&&&&&&&&&&&&&HCOOCH$\rm_{3}$-E\\
21 &4 & 105663.117(0.030) & $19_{13,7} - 20_{12,9}$ E, (v$\rm_{t}$=1) & 412 & 0.4 & 0.02&-& -& $\la$0.08 & $\la$0.2 &$\la$0.39& -&-&blend  \\
22 &4 & 105722.672(0.030)  & $10_{4,6} - 10_{3,8}$ E, (v$\rm_{t}$=1) & 230 & 0.5 &0.02&-& -& $\la$0.06&$\la$0.1&$\la$0.16& -&-&not detected \\
23&5  & 105810.803(0.030)  & $14_{11,4} - 15_{10,6}$ E & 142 & 0.2 & 0.02&-& -& $\la$0.06&$\la$0.1&$\la$0.39& -&-&not detected \\
24 &5 & 105815.953(0.030)  & $3_{3,1} - 2_{2,0}$ E & 10 & 0.2 & 0.02& 7.9(0.4)&2.6(0.3) &0.13 & 0.3(0.1)& 1.18& -&-& partial blend\\
25 &5 & 105832.067(0.030)  & $14_{11,3} - 15_{10,5}$ E & 142 & 0.2 & 0.02& 7.5(0.2)&2.1(0.3) &0.12 &0.3(0.1) &1.18 & -&-&detected \\
26 &6 & 110153.652(0.010) & $10_{1,10} - 9_{1,9}$ A, (v$\rm_{t}$=1) & 218 & 25.1& 0.10 &- &-  &$\la$2.75 & $\la$7.2&$\la$0.22& -&-& blend with NH$\rm_{2}$D\\
27&7  & 203349.158(0.006)& $24_{19,5} - 25_{18,7}$ E & 416 &0.2 & 0.16 & -& -&$\la$1.44 &$\la$2.9  &$\la$5.97& -&-&blend \\
28&7  & 203378.812(0.007)& $24_{19,6} - 25_{18,8}$ E & 416 &0.2 & 0.16 & -& -&$\la$4.30 &$\la$8.6 &$\la$17.71& -&-&blend \\
29&8  & 203427.575(0.100) & $21_{8,13} - 21_{7,14}$ A, (v$\rm_{t}$=1) & 366 & 6.2 &0.16 & -& -& $\la$0.48&$\la$1.0&$\la$0.07& -&-&not detected \\
30 &8 & 203435.554(0.100) & $19_{8,11} - 19_{7,12}$ A & 155 & 5.3 & 0.16& -& -&$\la$1.85 &$\la$3.7  &$\la$0.29& -&-&blend \\
31 &8 & 203471.840(0.050) & $19_{8,11} - 19_{7,12}$ E & 155 & 4.9 & 0.16&-& -&$\la$1.25 &$\la$2.5  &$\la$0.21& -&-&blend \\
32&9  &  223435.416(0.001)& $11_{4,8} - 10_{3,7}$ A, (v$\rm_{t}$=1) & 237 &1.8 & 0.27 & -& -& $\la$0.81&$\la$1.7&$\la$0.35& -&-&not detected \\
33 &9&  223465.340(0.050) & $11_{4,8} - 10_{3,7}$ E&50& 2.1&0.27 &7.8(0.1) &4.3(0.3)&2.2 &10.1(0.6)& 1.79&  2.0x1.5&60& detected \\
34 &9& 223500.463(0.001)&  $11_{4,8} - 10_{3,7}$ A & 50& 1.8& 0.27&7.6(0.1) &4.2(0.2)&2.4 &10.7(0.5) & 2.26& 2.0x1.5 &  60& detected\\
35&9& 223534.727(0.100) &  $18_{5,14} - 17_{5,13}$ E, (v$\rm_{t}$=1) & 305& 41.6 & 0.27&8.1(0.1) & 5.7(0.3)& 2.5 &15.1(0.8) &0.14&3.0x1.5 & 70& detected \\
36 &9& 223538.511(0.100) & $43_{8,35} - 43_{7,36}$ E & 618 & 13.9 &0.27 &-& -& $\la$0.81&$\la$1.7&$\la$0.05& -&-&not detected \\
37 &9& 223592.231(0.100) & $43_{8,35} - 43_{7,36}$ A & 618 & 13.9 & 0.27& -& -& $\la$0.81&$\la$1.7&$\la$0.05& -&-&not detected \\
38 &9& 223618.470(0.007)& $27_{9,19} - 27_{8,19}$ E, (v$\rm_{t}$=1) & 464 &0.4 & 0.27 & -& -& $\la$0.81&$\la$1.7&$\la$1.70& -&-&not detected \\
39 &9& 223624.496(0.100) & $37_{8,30} - 37_{7,31}$ E & 463 & 11.7 &0.27 &-& -& $\la$0.81&$\la$1.7&$\la$0.05& -&-&not detected \\
40 &9& 223634.916(0.100) & $37_{8,30} - 37_{7,31}$ A & 463 & 11.7 &0.27 & -& -& $\la$0.81&$\la$1.7&$\la$0.05& -&-&not detected \\
41 &9& 223642.180(0.001) & $12_{5,7} - 12_{3,10}$ E & 63 & 0.1& 0.27&  -& -& $\la$0.81&$\la$1.7&$\la$7.57& -&-&not detected \\
42&9 &  223645.342(0.001)& $12_{5,7} - 12_{3,10}$ A & 63 &0.1 & 0.27 & -& -& $\la$0.81&$\la$1.7&$\la$6.37& -&-&not detected \\
43&9 &  223651.592(0.007)& $33_{9,24} - 33_{8,26}$ E, (v$\rm_{t}$=1) & 573 &0.7 & 0.27 & -& -&$\la$0.81&$\la$1.7&$\la$0.85& -&-&not detected \\
44&9 &  223676.472(0.006)& $35_{7,29} - 35_{5,30}$ A, (v$\rm_{t}$=1) & 593 &3.0 & 0.27 & -& -&$\la$0.81&$\la$1.7&$\la$0.21& -&-&not detected \\
45 &9&  223782.081(0.007)& $33_{6,28} - 33_{5,29}$ E, (v$\rm_{t}$=1) & 543 &7.0 & 0.27 & -& -& $\la$0.81&$\la$1.7&$\la$0.09& -&-&not detected \\
46 &9&  223821.523(0.002)& $35_{7,29} - 35_{6,30}$ E & 409 &8.2 & 0.27 & -& -& $\la$0.81&$\la$1.7&$\la$0.08& -&-&not detected \\
47 &9& 223854.201(0.100) & $35_{7,29} - 35_{6,30}$ A & 409 & 10.1 & 0.27&-& -& $\la$0.81&$\la$1.7&$\la$0.06& -&-&not detected \\
48 &10 & 225689.371(0.008)& $34_{5,29} - 34_{5,30}$ A, (v$\rm_{t}$=1) & 562 &3.0 & 0.13 &- & -& $\la$15.43& $\la$30.9&$\la$3.85& -&-& blend \\
49 &10 & 225696.842(0.100) & $20_{1,19} - 19_{1,18}$ E, (v$\rm_{t}$=1) & 307 & 50.2 &0.13 &- & -& $\la$13.25& $\la$26.5&$\la$0.20& -&-& blend \\
50 &10 & 225702.857(0.100) & $19_{2,17} - 18_{2,16}$ A, (v$\rm_{t}$=1)& 304 & 46.8 & 0.13& -& -&$\la$7.71 &$\la$15.4&$\la$0.12& -&-&blend \\
51 &10 & 225727.506(0.100) & $6_{6,1} - 5_{5,0}$ A, (v$\rm_{t}$=1) & 224 & 3.1 & 0.13& -& -& $\la$0.90&$\la$1.8 &$\la$0.21& -&-&blend \\
52 &10 & 225727.506(0.100) & $6_{6,0} - 5_{5,1}$ A, (v$\rm_{t}$=1) & 224 & 3.1 &0.13& -& -& $\la$0.90&$\la$1.8 &$\la$0.21& -&-&blend \\
53 &11  & 225855.505(0.100) & $6_{6,1} - 5_{5,1}$ E & 36 & 3.1 & 0.08&- &- & $\la$1.20& $\la$2.4&$\la$0.29& -&-& blend\\
54&11  & 225900.684(0.100) & $6_{6,0} - 5_{5,0}$ E & 36 & 3.1& 0.08& -& -&$\la$3.16 &$\la$6.3 &$\la$0.75& -&-&blend with HDO \\
55&11  & 225928.659(0.100) & $6_{6,0} - 5_{5,1}$ A & 36 & 3.1& 0.08& -&-& $\la$1.53 &$\la$3.1 &$\la$0.37& -&-&blend  \\
56&11  & 225928.659(0.100) & $6_{6,1} - 5_{5,0}$ A & 36 & 3.1 & 0.08& -&-& $\la$1.53 &$\la$3.1 &$\la$0.37& -&-&blend  \\
57&12  & 225999.145(0.100) & $30_{7,23} - 29_{8,22}$ E & 312 & 2.2 &0.10 & -& -& $\la$0.30&$\la$0.6&$\la$0.10& -&-&not detected \\
58&12  & 226061.796(0.100) & $20_{3,17} - 19_{4,16}$ E, (v$\rm_{t}$=1) & 321 & 4.0 & 0.10&-& -& $\la$0.30&$\la$0.6&$\la$0.06& -&-&not detected \\
59 &12 &  226077.920(0.001)& $10_{3,7} - 9_{1,8}$ E & 39 &0.3 & 0.10 &-& -& $\la$0.30&$\la$0.6&$\la$0.79& -&-&not detected \\
60&12  &  226081.175(0.002)& $30_{7,23} - 29_{8,22}$ A & 312 &1.8 & 0.10&-& -& $\la$0.30&$\la$0.6&$\la$0.12& -&-&not detected \\
61&12  &  226087.718(0.007)& $29_{4,26} - 29_{3,27}$ A, (v$\rm_{t}$=1) & 450 &4.3 & 0.10 & -&-& $\la$0.79 &$\la$1.6&$\la$0.14& -&-&blend  \\
62 &12 & 226090.301(0.100) & $19_{2,17} - 18_{2,16}$ E, (v$\rm_{t}$=1) & 303 & 47.0 &0.10 & -&-& $\la$1.00 &$\la$2.0&$\la$0.02& -&-&blend  \\
63 &12 &  226112.470(0.007)& $29_{4,26} - 29_{2,27}$ A, (v$\rm_{t}$=1) & 450 &2.1 & 0.10&-& -& $\la$0.30&$\la$0.6&$\la$0.11& -&-&not detected \\
64 &12 &  226125.600(0.001)& $10_{3,7} - 9_{1,8}$ A & 39 &0.3 & 0.10 & -&-& $\la$0.59 &$\la$1.2 &$\la$1.57& -&-&blend  \\
\end{longtable}
\flushleft
Notes: (1) Numbering of the observed transitions with E$\rm_{upper}$ $\la$ 650 K. (2) Reference number of the corresponding data set (see Table \ref{Table.dataset_parameters}). (3) Frequencies and uncertainties taken from Ilyushin et al. (2009) and from the JPL database (http://spec.jpl.nasa.gov/). (4) Transitions in ground state (v$\rm_{t}$=0). Transitions in torsional levels (v$\rm_{t}$=1) are specified. (5) Energy of the upper level. (6) Line strength calculated by Ilyushin et al. (2009). For the transitions coming from the JPL database, the line strength is calculated from the formulae of \citet{Pickett:1998}. (7) Noise estimated from spectrum (1$\sigma$). (8), (9), (10) and (11) Velocity, linewidth at half intensity, brightness temperature and integrated intensities. The uncertainties are estimated with the CLASS software. (12) Column densities of the upper state of the transition with respect to the rotational degeneracy (2J$\rm_{up}$+1). (13) and (14) Source size estimated at half flux density from detected transitions with spatial resolution allowing to isolate the source. (15) Blended lines are specified. 
\end{landscape}
} %End \longtabL
%-------------------
\end{appendix}

%===================================================================================================
%
\end{document}